\documentclass[twocolumn, superscriptaddress,amsmath,amssymb,aps, prb,
showkeys,showpacs,nofootinbib] {revtex4-1}
\usepackage{graphicx}
\usepackage{color}
\usepackage{bm}
\usepackage{amsmath}
\usepackage{amssymb}
\usepackage{bm}
\usepackage{hyperref}
\usepackage{dcolumn}
\usepackage{slashed}
\usepackage{float}
\usepackage{yfonts}

\begin{document}
\title{Quantum-limited Parametric Amplification with Josephson Circuits in the Regime of Pump Depletion}

\author{Ananda Roy}
\email{roy@physik.rwth-aachen.de}
\affiliation{Institute for Quantum Information, RWTH Aachen University, 52056 Aachen, Germany}

\author{Michel Devoret}
\email{michel.devoret@yale.edu}
\affiliation{Department of Applied Physics, Yale University, PO BOX 208284, New Haven, CT 06511}

\begin{abstract}
Linear parametric amplification is a key operation in information processing. Our interest here is quantum-limited parametric amplification, $i.e.$, amplification of quantum signals while adding the minimum amount of noise allowed by quantum mechanics, which is essential for any viable implementation of quantum information processing.  We describe parametric amplifiers based on the dispersive nonlinearity of Josephson junctions driven with appropriate tones playing the role of pumps. We discuss two defining characteristics in the architecture of these amplifiers: the number of modes occupied by the signal, idler and pump waves and the number of independent ports through which these waves enter into the circuit. The  scattering properties of these amplifiers is also reviewed. The main focus of this work are the computations of the dynamic range and phase-space distributions of the fluctuations of the modes of the amplifiers. 
\end{abstract}

\maketitle

\section{Introduction}

Photons of microwave radiation in the band $3-12\ \rm{GHz}$ (25-100 mm wavelength)
have an  energy approximately 10$^{5}$ smaller than those of visible light.
Yet, at a temperature 2$\times $10$^{4}$ smaller than room temperature, now
routinely achievable with commercial dilution refrigerators, it is possible to
detect and process signals whose energy is equivalent to that of single
microwave photons. \cite{Devoret_Schoelkopf_2013} There are three advantages of single photon microwave
electronics when compared with quantum optics. First, signal envelopes with a relative bandwidth of a few
percent at carrier frequencies of a few GHz can be controlled with much greater relative precision than their
equivalent at few hundreds of THz. This is because microwave generators tend to 
have better short term stability than lasers, and also because microwave
components are mechanically very stable, particularly when cooled, compared
with traditional optical components. Second, on-chip circuitry of single-photon microwave electronics can be well in the lumped
element regime and consequently, the control of spatial mode structure is more easily
achieved than in the optical domain. Third, there exists a simple, robust,
non-dissipative component, the Josephson tunnel junction, whose
non-linearity can dominate over the linear characteristics of the circuit at
the single photon level. Many quantum signal processing functions have
thus been realized, both digital and analog. In this work, we concentrate on analog Josephson
amplifying devices pumped with one or several microwave tones. In particular, only devices
that demonstrate linear amplification with added noise at the level of the
standard quantum limit \cite{Caves_1982} are considered here. These novel devices have taken
the work pioneered by B. Yurke at the Bell labs thirty years ago \cite{Yurke_Simon_1988, Yurke_Whittaker_1989} to the point
where new original experiments can be performed successfully owing  to
Josephson amplifiers as the first link in the chain of measurements. \cite{Vijay_Siddiqi_2011, Hatridge_Devoret_2013}

The main desired characteristics of a Josephson amplifier are: i) {\it{low added noise}}: the noise added by the amplifier should be no larger than the minimum imposed by quantum mechanics, ii) {\it{high gain}}: the gain of the amplifier should be large enough (in practice, 20 dB or more)
 to beat the noise added by the subsequent stages in the amplifier chain, iii) {\it{large bandwidth}}: the amplifier should have a constant gain over a bandwidth that is large enough for the desired application, ranging from several MHz-s to several GHz-s, iv) {\it{large dynamic range}}: the amplifier
should function as a linear device with output signal power proportional to the input signal power for a wide range of input signal power. This range of power, known as the dynamic range of the amplifier, should be large enough so that more than just a few incident photons can be reliably detected, v) {\it{unidirectional}}: the amplifier should, ideally, amplify only signals incident from the system being probed and deamplify signals coming from subsequent devices in the amplification chain. This is necessary to prevent spurious noise in other parts of the setup to propagate back into the system under measurement, vi) {\it{ease of operation}}: the necessary energy for the amplification process should be delivered to the amplifier in a manner that is as simple and robust as possible, without very precise tuning, vii) {\it{ease of construction}}: the circuit implementing the amplifier should have a minimal number of parts, and the parts should not require too delicate tolerances.


The aim of this article is to discuss the physics behind these Josephson parametric amplifiers in the high-gain regime where pump depletion effects can cause a reduction of the dynamic range of the amplifiers [item (iv) in the list above]. In particular, we provide a self-consistent, mean-field analysis of the dynamic range and stability of these amplifiers in this pump depletion regime. Furthermore, we analyze, in the same regime, the phase space distribution of the modes participating in the amplification process, building an understanding of the manner in which quantum fluctuations modify the parametric instability, the dark side of the phenomenon of amplification. This analysis is done within the framework of the Fokker-Planck equation. Although the focus of the paper is rather narrow compared to the above list of desired amplifier characteristics, for the sake of pedagogical clarity and to make the text self-contained, we review the fundamental amplifier characteristics like gain and bandwidth, which have been calculated elsewhere~\cite{Kamal_Devoret_2009, Bergeal_Devoret_2010, Abdo_Devoret_2013_b, Roy2016}. 
We also compare several circuit realizations of these amplifiers, classified according to their number of modes and access ports.

 The article is organized as follows.  In Sec. \ref{sec_4}, the notion of effective parametric amplifier Hamiltonian is introduced. The important distinction between the degenerate and non-degenerate  amplifiers that arises from a fundamental difference in the nature and number of degrees of freedom of the two kinds of devices is discussed. The linear scattering theory of these amplifiers is presented in this section. In practical amplifiers, the linear scattering theory ceases to be a sufficient description when the amplitude of the signal being amplified ceases to be infinitesimal compared to that of the pump tone giving rise to amplification. Then, the amplified output signal power is no longer a linear function of the input signal power. 
This more involved topic behind reduction of dynamic range, {\it i.e.} pump depletion effects leading to a reduction of  gain of the device, is discussed in Sec. \ref{sec_5}. 
Phase-space fluctuations of the signals are calculated in Sec. \ref{sec_6}. Several practical implementations of amplifiers are given in Sec. \ref{sec_7}. A concluding summary is provided in Sec. \ref{sec_8}. For pedagogical reasons and to make the article self-sufficient, we have also included several appendices that familiarize the reader with the formalism required for theoretical analysis of the topics discussed in the main text. Appendix \ref{appQuantumSignals} extends the concept of classical signals to the quantum domain in the field of microwave electronics. Appendix \ref{appQLE} describes the important theoretical tool of the Quantum Langevin Equation (QLE) and input-output theory. Using this theory, the amplifier characteristics can be calculated, starting from the circuit Hamiltonian and the coupling parameters of its ports. We keep our treatments of the concepts in the appendices sufficient general and present them beyond the usual rotating wave approximation (RWA).  Details of some of the calculations of the results presented in the main text are given in Appendices \ref{output_power_calc} and \ref{dyn_range_deg_paramp}. 

\section{Model amplifiers}
\label{sec_4}
\begin{figure}
\centering
\includegraphics[width = 0.5\textwidth]{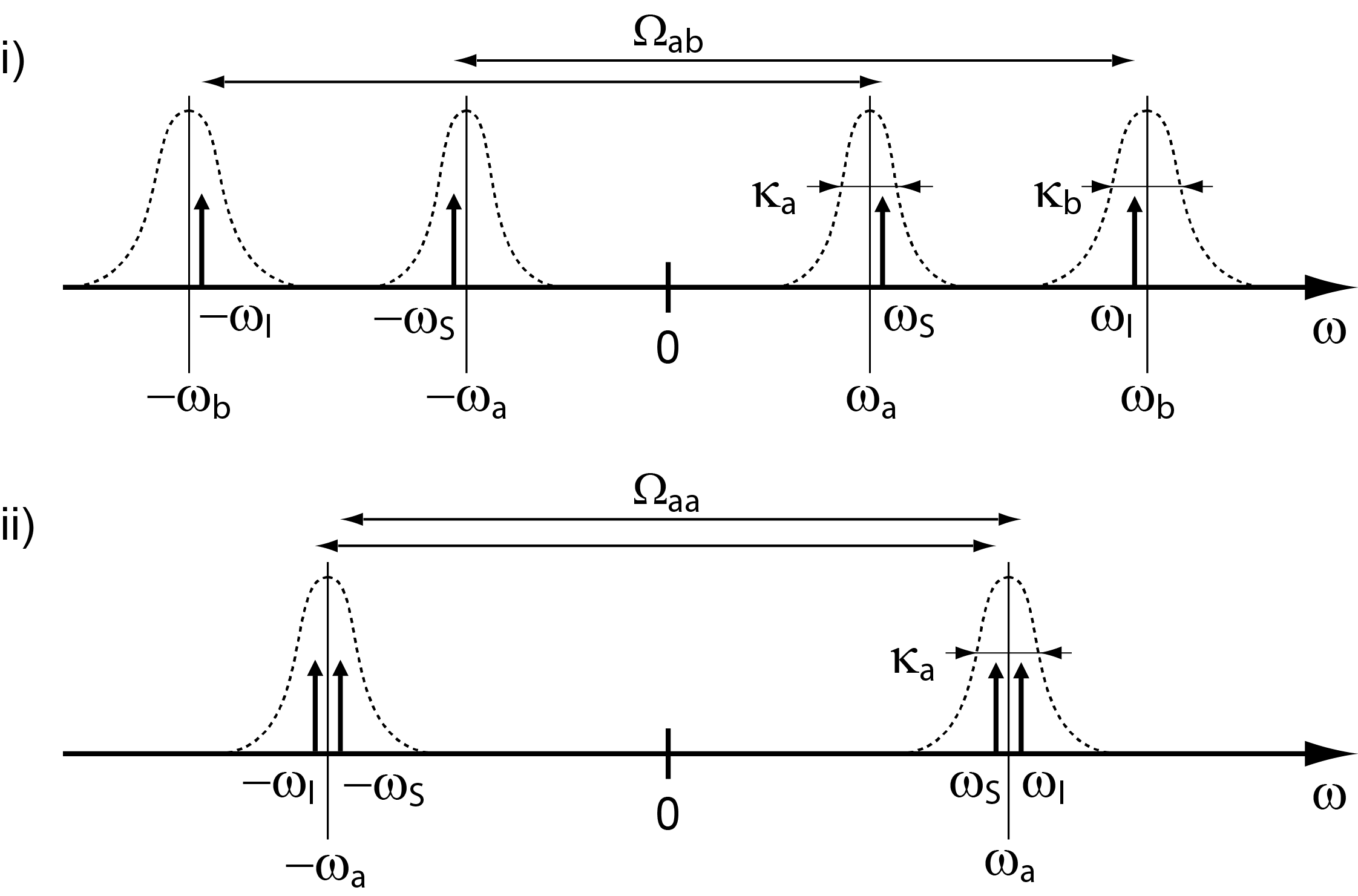}
\caption{\label{Fig_5} Frequency landscape for the non-degenerate (i) and degenerate (ii)
parametric amplifiers. The dashed lines corresponds to response curves of
each mode, as measured by a probe tone injected in the circuit elements of the mode. The vertical arrows
correspond to the spectral densities of the signal and the idler tones arriving in the circuit through its ports. The
horizontal arrows denote the frequency translations between signal and idler
operated by the parametrical modulation induced by the pump tone.}
\end{figure}

An amplifying circuit can be conveniently described as a collection of simple harmonic modes with time-dependent couplings. First, we address, using this model, the question of how a linear amplification function can arise from the Hamiltonian. Thus, let us consider the time-dependent effective quadratic Hamiltonian%
\begin{align}
\frac{H}{\hbar }&=\sum_{m}\omega _{m}a_{m}^{\dag }a_{m}+i\sum_{m\leq
p}g_{mp}^{\rm{eff}}\big\{ a_{m}a_{p}e^{i\left( \Omega _{mp}t+\theta _{mp}\right)
}\nonumber\\&\qquad-\rm{H.c.}\big\},
\end{align}
where $a_m$ are bosonic mode annihilation operators and $m, p$ are circuit mode indices. The real, positive
parameters $\omega_{m}$ and $g_{ml}^{\rm{eff}}$ are, in general, functions
of elementary parameters of the circuit combined with the values of
time-dependent driving fields imposed from the outside and treated
classically. We will see later in Sec. \ref{sec_7} how these effective parameters arise. The driving fields excite the circuit, thus providing energy
for the amplification process and are often nicknamed ``pumps". Another important ingredient of the model is that there are ports that couple the modes $m$ to outside circuitry. In the
case of one port per mode, this coupling is described by constants $\kappa
_{m}$, the rate of excitation decay of mode $m$ through the port. The phase factors $e^{i\theta _{mp}}$ depend on the details of the
excitation, while the drive frequencies $\Omega _{mp}$ are in the vicinity
of $\omega _{m}+\omega _{p}$ (or sometimes $\left\vert \omega
_{m}-\omega _{p}\right\vert $; this corresponds to photon conversion 
and we do not consider this case here). By vicinity, we mean within the bandwidth determined by the port
coupling constants: $\left\vert \Omega _{mp}-\omega _{m}-\omega
_{p}\right\vert  \leq \kappa _{m}\kappa _{p}/\left( \kappa
_{m}+\kappa _{p}\right)$. In this review, we will limit
ourselves to two elementary cases: i) the two-port, two-mode,
non-degenerate parametric amplifier with Hamiltonian:%
\begin{equation}
\label{hamndpa}
\frac{H^{\rm{NDPA}}}{\hbar }=\omega _{a}a^{\dag }a+\omega _{b}b^{\dag
}b+ig_{ab}\left( abe^{i\left( \Omega _{ab}t+\theta \right)
}-\rm{H.c.}\right) 
\end{equation}%
and port coupling constants $\kappa _{a}$ and $\kappa _{b}$, and ii) the
one-port, one-mode, degenerate parametric amplifier with Hamiltonian:%
\begin{equation}
\label{hamdpa1}
\frac{H^{\rm{DPA}}}{\hbar }=\omega _{a}a^{\dag }a+ig_{\rm{aa}}\left(
a^{2}e^{i(\Omega _{\rm{aa}}t+\theta)}-\rm{H.c.}\right). 
\end{equation}%
In this last case, there is a single port coupling constant $\kappa _{a}$. The
frequency landscapes corresponding to the two cases are represented
schematically on Fig. \ref{Fig_5}. 

We deal with these model systems using the Quantum Langevin Equations (QLE-s), under rotating wave and Markov approximations (for derivation, see Appendix \ref{appQLE}). For the modes $a,b$ in the non-degenerate case, the QLE-s are given by:
\begin{align}
\label{qle_ndpa}
\frac{da}{dt} &= -\frac{i}{\hbar}[a,H^{\rm{NDPA}}]-\frac{\kappa_a}{2}a+\sqrt{\kappa_a}a^{\rm{in}},\nonumber\\
\frac{db}{dt} &= -\frac{i}{\hbar}[b,H^{\rm{NDPA}}]-\frac{\kappa_b}{2}b+\sqrt{\kappa_b}b^{\rm{in}}
\end{align}
where the input and output fields satisfy the boundary conditions: 
\begin{align}
\label{inoutNDPA}
a^{\rm{out}}=-a^{\rm{in}}+\sqrt{\kappa_a}a, \quad b^{\rm{out}}=-b^{\rm{in}}+\sqrt{\kappa_b}b. 
\end{align}
Similarly, in the degenerate case, the QLE takes the form:
\begin{align}
\label{qle_dpa}
\frac{da}{dt} &= -\frac{i}{\hbar}[a,H^{\rm{DPA}}]-\frac{\kappa_a}{2}a+\sqrt{\kappa_a}a^{\rm{in}}
\end{align}
with the boundary condition: 
\begin{align}
\label{inoutDPA}
a^{\rm{out}}=-a^{\rm{in}}+\sqrt{\kappa_a}a.
\end{align}
From Eqs. \eqref{qle_ndpa}, \eqref{inoutNDPA}, we obtain for the non-degenerate case,
\begin{align}
{}&F(\omega_a, \kappa_a)a^{\rm{out}}(t) +g_{ab}\sqrt{\frac{\kappa_a}{\kappa_b}}e^{-i(\Omega _{ab}t+\theta)}{b^{\rm{out}}}(t)^{\dag }\nonumber\\ &\qquad=-F(\omega_a, -\kappa_a)a^{\rm{in}}(t)-g_{ab}\sqrt{\frac{\kappa_a}{\kappa_b}}e^{-i(\Omega _{ab}t + \theta)}{b^{\rm{in}}}(t)^{\dag },\nonumber \\\label{ab_jpc_eqs}
{}&F(\omega _{b}, \kappa _{b}) b^{\rm{out}}(t) +g_{ab}\sqrt{\frac{\kappa_b}{\kappa_a}}e^{-i(\Omega _{ab}t+\theta)}{a^{\rm{out}}}(t)^{\dag }\nonumber\\&\qquad=-F(\omega _{b}, -\kappa _{b}) b^{\rm{in}}(t) -g_{ab}\sqrt{\frac{\kappa_b}{\kappa_a}}e^{-i(\Omega _{ab}t+\theta)}{a^{\rm{in}}}(t)^{\dag }, 
\end{align}
where $F(\omega, \kappa) = d/dt + i\omega + \kappa/2$. For the degenerate paramp, there is only one equation:
\begin{align}
{}&F(\omega _{a}, \kappa _{a}) a^{\rm{out}}(t) +2
g_{\rm{aa}}e^{-i(\Omega _{\rm{aa}}t+\theta)}{a^{\rm{out}}}(t)^{\dag }\nonumber\\&\qquad=-F(\omega _{a}, -\kappa _{a})a^{\rm{in}}(t) -2g_{\rm{aa}}e^{-i(\Omega _{\rm{aa}}t+\theta)}{a^{\rm{in}}}(t)^{\dag }.
\end{align}%
Going to the Fourier domain and solving for outgoing waves as a function of
the incoming waves we find for the non-degenerate case:%
\begin{equation}
\label{jpc_scatmat}
\left[ 
\begin{array}{c}
a^{\rm{out}}\left[ +\omega _{S}\right]  \\ 
a^{\rm{out}}\left[ -\omega _{S}\right]  \\ 
b^{\rm{out}}\left[ +\omega _{I}\right]  \\ 
b^{\rm{out}}\left[ -\omega _{I}\right] 
\end{array}%
\right] =\left[ 
\begin{array}{cccc}
r_{SS} & 0 & 0 & s_{SI} \\ 
0 & r_{SS}^{\ast } & s_{SI}^{\ast } & 0 \\ 
0 & s_{IS}^{\ast } & r_{II}^{\ast } & 0 \\ 
s_{IS} & 0 & 0 & r_{II}%
\end{array}%
\right] \left[ 
\begin{array}{c}
a^{\rm{in}}\left[ +\omega _{S}\right]  \\ 
a^{\rm{in}}\left[ -\omega _{S}\right]  \\ 
b^{\rm{in}}\left[ +\omega _{I}\right]  \\ 
b^{\rm{in}}\left[ -\omega _{I}\right] 
\end{array}%
\right], 
\end{equation}%
where $\omega _{S}, \omega _{I}$ are the two signal and image (or
idler) frequencies, respectively, linked precisely by $\omega _{S}+\omega
_{I}=\Omega _{ab}$. The operators at the negative frequencies should be interpreted as Hermitian conjugates of the same operators at the positive frequencies: $a^{\rm{in/out}}[-\omega] = a^{\rm{in/out}}[\omega]^\dagger, b^{\rm{in/out}}[-\omega] = b^{\rm{in/out}}[\omega]^\dagger$ (see Appendix \ref{appQuantumSignals} for details).
Unlike in the case of simple harmonic circuits, an
input signal at one frequency can here be processed into an output signal at
another frequency. There is also a change in sign of the frequency in this
process, which is called phase conjugation, and this is why we need to represent the
scattering by a $4\times 4$ matrix. This matrix can be separated into two
blocks related by complex conjugation relating Fourier
coefficients with opposite frequencies, a mathematical operation independent
of the physical phenomenon of phase conjugation. Phase conjugation manifests
itself practically in the following manner: if one advances the phase of the
input signal by a given quantity, the phase of the conjugated output signal
becomes retarded by the same quantity. The elements of the scattering matrix
are given by 
\begin{align}
\label{rss}
r_{SS} &=\frac{\chi _{a}^{-1}\left( \omega _{S}\right) ^{\ast }\chi
_{b}^{-1}\left( \omega _{I}\right) ^{\ast }+\ \rho _{ab}^{2}}{\chi
_{a}^{-1}\left( \omega _{S}\right) \chi _{b}^{-1}\left( \omega _{I}\right)
^{\ast }-\rho _{ab}^{2}}, \\
r_{II} &=\frac{\chi _{a}^{-1}\left( \omega _{S}\right) \chi _{b}^{-1}\left(
\omega _{I}\right) +\rho _{ab}^{2}}{\chi _{a}^{-1}\left( \omega _{S}\right)
\chi _{b}^{-1}\left( \omega _{I}\right) ^{\ast }-\rho _{ab}^{2}}, \\
s_{SI} &=\frac{-2\rho _{ab}e^{-i\theta }}{\chi _{a}^{-1}\left( \omega
_{S}\right) \chi _{b}^{-1}\left( \omega _{I}\right) ^{\ast }-\rho _{ab}^{2}},
\\\label{sis}
s_{IS} &=\frac{-2\rho _{ab}e^{i\theta }}{\chi _{a}^{-1}\left( \omega
_{S}\right) \chi _{b}^{-1}\left( \omega _{I}\right) ^{\ast }-\rho _{ab}^{2}}.
\end{align}
These expressions contain two ingredients: the single mode bare
susceptibilities $\chi $, which are given by%
\begin{equation}
\chi _{m}\left( \omega \right) =\frac{1}{1-2i\left( \omega -\omega
_{m}\right) /\kappa _{m}}
\end{equation}%
and the reduced effective mode coupling given by%
\begin{equation}
\rho _{ab}=\frac{2g_{ab}}{\sqrt{\kappa _{a}\kappa _{b}}}.
\end{equation}%
Its modulus squared is often called the mode cooperativity.
When the drive tone of the amplifier is optimally tuned at $\Omega _{ab}=\omega _{a}+\omega _{b}$ 
and the monochromatic input signals are on resonance with their
corresponding mode $\omega _{S} =\omega _{a}, \omega _{I} =\omega _{b}$, 
the scattering matrix takes the simpler form, with  $r_{SS}=r_{II}=\sqrt{G_0}$ and $s_{SI}=s_{IS}^*=-\sqrt{G_{0}-1}e^{-i\theta }$. 
Here, the zero-detuning, optimal amplifier power gain $G_{0}$ is%
\begin{equation}
G_{0}=\left( \frac{1+\rho _{ab}^{2}}{1-\rho _{ab}^{2}}\right) ^{2}.
\end{equation}%
It can be shown that the stability of the amplifier requires that $\rho
_{ab}<1$, i.e. there is a ceiling to the effective coupling between modes of
the circuit, beyond which amplification turns into spontaneous parametric oscillation. We will see later in Sec. \ref{sec_5} how this raw notion of stability ceiling is modified when the pump is treated more realistically.

Note that the determinant of the scattering matrix is unity even in the
fully general case. Also, it is important to realize that, quite generally,
the scattering is not reciprocal. A wave going from port \textit{b} to port 
\textit{a} acquires a phase factor $e^{-i\theta }$ from the drive which is
conjugate to the phase factor $e^{i\theta }$ accompanying the scattering
from port a to port b.

We now turn to the degenerate case in which the scattering relations, involving
only one port, still are expressed as a $4\times 4$ matrix:%
\begin{equation}
\left[ 
\begin{array}{c}
a^{\rm{out}}\left[ +\omega _{S}\right]  \\ 
a^{\rm{out}}\left[ -\omega _{S}\right]  \\ 
a^{\rm{out}}\left[ +\omega _{I}\right]  \\ 
a^{\rm{out}}\left[ -\omega _{I}\right] 
\end{array}%
\right] =\left[ 
\begin{array}{cccc}
r_{SS} & 0 & 0 & s_{SI} \\ 
0 & r_{SS}^{\ast } & s_{SI}^{\ast } & 0 \\ 
0 & s_{IS}^{\ast } & r_{II}^{\ast } & 0 \\ 
s_{IS} & 0 & 0 & r_{II}%
\end{array}%
\right] \left[ 
\begin{array}{c}
a^{\rm{in}}\left[ +\omega _{S}\right]  \\ 
a^{\rm{in}}\left[ -\omega _{S}\right]  \\ 
a^{\rm{in}}\left[ +\omega _{I}\right]  \\ 
a^{\rm{in}}\left[ -\omega _{I}\right] 
\end{array}%
\right].\nonumber
\end{equation}%
Now, the different frequencies are carried on the same port and are all in
the vicinity of the single resonance of the unique mode. Nevertheless, the
scattering coefficients are similar to the previous
expressions, with the substitution $\chi _{b}\left( \omega \right) =\chi
_{a}\left( \omega \right) $. A simplification occurs if the drive frequency
is precisely tuned to twice the effective resonant frequency, i.e. $\Omega
_{\rm{aa}}=2\omega _{a}$, in which case $\chi _{a}\left( \omega _{I}\right)
=$ $\chi _{a}\left( \omega _{S}\right) ^{\ast }$ and the subblock of the
scattering matrix takes the form%
\begin{align}
\label{jba_scatmat}
\left[ 
\begin{array}{c}
a^{\rm{out}}\left[ +\omega _{S}\right]  \\ 
a^{\rm{out}}\left[ -\omega _{I}\right] 
\end{array}%
\right] &=\frac{1}{D}\left[ 
\begin{array}{cc}
M_1 & M_2 \\ 
M_2^* & M_1%
\end{array}%
\right] \left[ 
\begin{array}{c}
a^{\rm{in}}\left[ +\omega _{S}\right]  \\ 
a^{\rm{in}}\left[ -\omega _{I}\right] 
\end{array}%
\right],
\end{align}%
with $M_1=|\chi _{a}^{-1}\left( \omega _{S}\right)| ^{2}+\rho
_{\rm{aa}}^{2}$, $M_2=-2\rho _{\rm{aa}}e^{-i\theta }$ and $D=\chi _{a}^{-2}\left( \omega _{S}\right)-\rho
_{\rm{aa}}^{2}$. Here, $\rho _{\rm{aa}}=4g_{\rm{aa}}/\kappa _{a}$. We also introduce the in-phase and quadrature components of the
incoming and outgoing waves:%
\begin{eqnarray}
a_{\Vert, \bot }^{\rm{in, out}}[ \delta \omega] &=&a^{\rm{in, out}}\left[ \omega
_{S}\right] \pm e^{-i\theta }a^{\rm{in, out}}\left[ -\omega _{I}\right], 
\end{eqnarray}%
where $\delta \omega =\omega _{S}-\omega _{a}=\omega _{a}-\omega
_{I}$. The meaning of this transformation can be illustrated by the
following consideration, which supposes $\theta =0$ for simplicity.
Classically, if a signal is such that 
\begin{equation}
y\left( t\right) =f\left( t\right) \cos \left( \omega _{a}t\right)
+g\left( t\right) \sin \left( \omega _{a}t\right),
\end{equation}%
with in-phase and quadrature modulation components $f\left( t\right) $ and $%
g\left( t\right) $ slow compared to $\left( \omega _{a}\right) ^{-1}$,
then 
\begin{eqnarray}
y_{\Vert }[\delta \omega]  =f[ \delta \omega],\hspace{1cm}y_{\bot }[ \delta \omega ] =g[\delta \omega]
\end{eqnarray}%
One can easily check that the effect of the angle $\theta $ associated with
the time dependence of the effective Hamiltonian is just to rotate the
component signals in the Fresnel plane. In the representation where the
in-phase and quadrature components form the basis signals, we find that the
scattering matrix is diagonal 
\begin{eqnarray}
a_{\Vert }^{\rm{out}}[\delta \omega]  &=&\frac{\left\vert \chi
_{a}^{-1}\left( \omega _{S}\right) \right\vert ^{2}+2\rho _{\rm{aa}}+\ \rho
_{\rm{aa}}^{2}}{D_-}a_{\Vert }^{\rm{in}}[\delta \omega]\nonumber\\ &=&\Lambda _{\Vert
}\left( \delta \omega \right) a_{\Vert }^{\rm{in}}[\delta \omega], \\
a_{\bot }^{\rm{out}}[\delta \omega]  &=&\frac{\left\vert \chi
_{a}^{-1}\left( \omega _{S}\right) \right\vert ^{2}-2\rho _{\rm{aa}}+\ \rho
_{\rm{aa}}^{2}}{D_-}a_{\bot }^{\rm{in}}[\delta \omega]\nonumber\\ &=&\Lambda _{\bot
}\left( \delta \omega \right) a_{\bot }^{\rm{in}}[\delta \omega].
\end{eqnarray}%
The property of the scattering matrix to have unity determinant imposes%
\begin{equation}
G_{\Vert }G_{\bot }=1.
\end{equation}%
where $G_{\Vert ,\bot }=\left\vert \Lambda _{\Vert ,\bot }\left( \delta
\omega \right) \right\vert ^{2}$. Thus, in this mode of operation of the
degenerate parametric amplifier, one quadrature of the signal is amplified
while the other is de-amplified. If the input signal consists only of vacuum
fluctuations, the amplifier squeezes these fluctuations for one quadrature,
making it less uncertain than the so-called standard quantum limit, which is
associated to a standard deviation corresponding to the square root of a
quarter of a photon (the half photon of the zero-point motion is split
evenly between the two quadratures, and only one is squeezed). \cite{Wall_Milburn_2008, Drummond_Ficek_2013}

In the non-degenerate case (i) a more complex form of squeezing -- two-mode
squeezing -- occurs in the four dimensional phase space of the quadratures
of the two propagating signals incident on the circuit. \cite{Wall_Milburn_2008, Drummond_Ficek_2013}
The non-degenerate parametric amplifier is usually employed as a sort of RF
op-amp: the idler port is connected to a cold matched load emulating an
infinite transmission line at zero-temperature and the device viewed from
the signal port functions as a reflection amplifier operating in the phase
preserving mode: for signals having a bandwidth small compared to that of
the amplifier, we have 
\begin{equation}
a^{\rm{out}}=\sqrt{G}\left(a^{\rm{in}}+\sqrt{1-\frac{1}{G}}b^{\rm{in}\dag }\right).
\label{simple_gain_input-output_expression}
\end{equation}%
The second term on the right of this last expression shows that quantum
noise entering through the \textit{b} port must necessarily be added to the
amplified signal. \cite{Caves_1982} This added noise contribution amounts, in the large gain
limit $G\gg 1$ and for an idler port at zero temperature, to a half-photon
at the signal frequency, referred to the input. It can be seen as an evil necessary to preserve the commutation relation%
\begin{equation}
\left[ a^{\rm{out}},a^{\rm{out}\dag }\right] =\left[ a^{\rm{in}},a^{\rm{in}\dag }\right].
\end{equation}%
More practically, the extra half-photon of noise can also be seen as a
consequence of the Heisenberg Uncertainty Principle. A phase preserving
amplifier processes equally both quadratures, which in quantum mechanics are
non-commuting observables. Since the process of amplification is equivalent
to measurement, the extra noise forbids that both quadratures are known
precisely simultaneously, in accordance with the central principle of
quantum mechanics. An amplifier functioning in this Heisenberg regime where
the efficiency of the amplification process is only limited by irreducible
quantum fluctuations is said to be quantum-limited.

\section{Dynamic range of amplifiers}
\label{sec_5}
In previous section, we have described the linear scattering theory of  ideal amplifiers using effective, time-dependent, quadratic Hamiltonians. With such models, one can, in principle, achieve arbitrary high gains in amplifiers. However, in realistic situations, there are two major effects that reduce of the gain of the amplifiers, as described below. 

First, the nonlinearity giving rise to the desired mode-mixing rarely ever yields only the desired quadratic Hamiltonian. In practice, higher order terms such as quartic (Kerr) terms cannot be neglected even if their amplitude is smaller than the main quadratic terms. In the presence of pump and incident signals, these relatively small quartic terms are sufficient to shift the resonator frequencies appreciably (this effect is also known as the AC-Stark shift). This results in the incident pump of Eqs. (\ref{hamndpa}, \ref{hamdpa1}) being off-resonant, and thus, the gain of these devices is lowered [see Eqs. (\ref{rss}-\ref{sis})]. Recent results (see Ref. \onlinecite{Liu2017}) indicate that this is the dominant source of gain saturation in current parametric amplifiers. More details on how this arises is given in the discussion of implementation of parametric amplifiers [see Sec. \ref{sec_jba}, in particular Eqs. (\ref{alpha0}-\ref{eq_b})]. We note that recently, a method to implement three-wave mixing while suppressing the detrimental effects of the fourth order nonlinearity has been proposed. \cite{Zorin2016, Frattini2017}

Second, pump depletion, which occurs due to the large magnitude of the signal being amplified, also limits the gains of amplifiers. When the amplitude of the incident signal being amplified is infinitesimal compared to that of the pump, the number of photons required from the pump for amplification is negligible compared to the total number of photons present in it. Thus, the pump tone can be considered to be stiff and its amplitude and phase can be treated as parameters without dynamics. This is the case in Sec. \ref{sec_4}. However, as the amplitude of the signal grows, the number of photons required for amplification also grows. As a result, the pump tone gets depleted, which leads to a reduction of the gain of the device. Note that even in absence of any incident signal, vacuum fluctuations are always incident on all the ports, which forbid, in a precise analysis, to consider the pump as perfectly stiff. This is because amplification of these vacuum fluctuations also leads to a reduction of the dynamic range of the device. 

In this section, we calculate the gain saturation of our model amplifiers. In our analysis, we only consider pump depletion due to the finite size of the incident signal. We model the pump as a classical drive incident on a low-Q mode coupled to the modes being amplified. We calculate the resultant gain and the output signal power in a self-consistent manner. In effect, we perform a mean-field calculation for the incident pump tone, neglecting its quantum fluctuations. For simplicity, we consider the situation when the pump tones meet the resonance condition for amplification and neglect incident thermal photons on all the ports \footnote{For similar analysis in a related topic, see Refs. \onlinecite{Lahteenmaki2013, Lahteenmaki2016}.}.   

Thus, we start with the following Hamiltonian for the non-degenerate parametric amplifier: 
\begin{equation}
\label{nondegham}
\tilde{H}^{\rm{NDPA}} = \sum_{\alpha = a,b,c}\omega_\alpha \alpha^\dagger \alpha+i\hbar g_3(abc^\dagger -a^\dagger b^\dagger c),
\end{equation}
where where $a,b, c$ respectively denote the signal, idler and pump modes for the device \footnote{Such a trilinear Hamiltonian arises also in treatment of Hawking radiation from a quantized source \cite{Nation2010}.} (see Sec. \ref{sec_7d} for a practical implementation). The QLE-s for the modes $a,b,c$ are given by
\begin{align}\label{a_qle_jpc}
\frac{da}{dt}&= -i\omega_a a-\frac{\kappa_a}{2}a-g_3b^\dagger c+\sqrt{\kappa_a}a^{\rm{in}}\\\label{b_qle_jpc}
\frac{db}{dt}&= -i\omega_b b-\frac{\kappa_b}{2}b-g_3a^\dagger c+\sqrt{\kappa_b}b^{\rm{in}},\\\label{c_qle_jpc}
\frac{dc}{dt}&= -i\omega_c c-\frac{\kappa_c}{2}c+g_3ab+\sqrt{\kappa_c}c^{\rm{in}}.
\end{align}
Here $\omega_a$, $\omega_b$, $\omega_c$ denote the frequencies and $\kappa_a$, $\kappa_b$, $\kappa_c$ the decay-rates for the modes $a,b,c$. Furthermore, $g_3$ denote the nonlinear coupling between them. We require the Q-factor of mode $c$ to be much lower than that of $a,b$: $\kappa_c\gg\kappa_a, \kappa_b$. The annihilation operators for the signal, idler and the pump are given by $a^{\rm{in}}, b^{\rm{in}}$ and $c^{\rm{in}}$. The incident pump tone is in a coherent state with amplitude (phase) $|\langle c^{\rm{in}}\rangle|\ (\theta_c)$, on resonance with the cavity frequency $\omega_c$. In practice, the pump tone is often delivered through a rather high-Q, but off-resonance resonator which ends up being populated with very few photons.  This type of impedance mismatch while closer to reality, involves more elements to be described correctly, and we have simplified it without losing the end results. In any case, $\sqrt{\kappa_c}|\langle c^{\rm{in}}\rangle|\gg g_2n_a$, where $n_a$ is the number of photons in steady-state in the $a$-mode.
Under the stiff pump approximation, the term $g_3ab$ in Eq. \eqref{c_qle} is neglected. Thus, the mode $c$ is in a coherent state, with its amplitude, in steady state, given by
\begin{equation}
\langle c\rangle_0 = \frac{2}{\sqrt{\kappa_c}}|\langle c^{\rm{in}}\rangle| e^{-i(\omega_c t+\theta_c)}.
\end{equation}
Replacing $c\rightarrow \langle c\rangle_0$ in Eqs. (\ref{a_qle_jpc}, \ref{b_qle_jpc}) results in the linearized dynamical equations for $a,b$, given earlier in Eq. \eqref{ab_jpc_eqs}, with 
\begin{equation}
g_{\rm{ab}}= \frac{2g_3|\langle c^{\rm{in}}\rangle|}{\sqrt{\kappa_c}},\quad \Omega_{\rm{ab}} = \omega_c, \theta = \theta_c.
\end{equation}
For the degenerate case, the corresponding Hamiltonian is given by:
\begin{equation}
\label{degham}
\tilde{H}^{\rm{DPA}} = \sum_{\alpha = a,c}\omega_\alpha \alpha^\dagger \alpha+i\hbar g_2(a^2c^\dagger -a^\dagger{}^2c),
\end{equation}
where $a,c$ respectively denote the signal and pump modes for the device (for its practical implementation, see Sec. \ref{sec_7c}). The QLE for the modes $a,c$ are given by
\begin{align}\label{a_qle}
\frac{da}{dt}&= -i\omega_a a-\frac{\kappa_a}{2}a-2g_2a^\dagger c+\sqrt{\kappa_a}a^{\rm{in}}\\\label{c_qle}
\frac{dc}{dt}&= -i\omega_b b-\frac{\kappa_b}{2}b+g_2a^2+\sqrt{\kappa_c}c^{\rm{in}},
\end{align}
The stiff pump approximation, in this case, leads to a linear QLE for $a$, giving rise to the scattering matrix in Eq. \eqref{jba_scatmat} in Sec. \ref{sec_4}. The parameters of the scattering matrix relate to the pump tone parameters as follows: 
\begin{equation}
g_{\rm{aa}} = \frac{2g_2|\langle c^{\rm{in}}\rangle|}{\sqrt{\kappa_c}},\quad \Omega_{\rm{aa}} = \omega_c, \theta_c = \theta.
\end{equation}
In order to incorporate the effect of pump depletion for the non-degenerate (degenerate) case, one needs to include the previously neglected term $g_3ab\ (g_2a^2)$ in the analysis and solve for $a,b,c\ (a,c)$ self-consistently. Since we are interested in parameter regimes where the pump depletion is small, yet non-negligible, we will treat the pump mode as a  coherent state and perform a perturbation analysis around the un-depleted value for $\langle c\rangle$: 
\begin{equation}
\langle c\rangle = \langle c\rangle_0 e^{-i\omega_c t}+ \langle\delta c(t)\rangle
\end{equation}
and solve for $\langle \delta c(t)\rangle$ self-consistently. The equation of motion for $\langle\delta c(t)\rangle$, for the non-degenerate and degenerate cases, are respectively given by
\begin{equation}
\label{eq_deltac_jpc}
\frac{d\langle \delta c(t)\rangle}{dt}=\Big(-i\omega_c-\frac{\kappa_c}{2}\Big) \langle \delta c\rangle+g_3\langle a b\rangle
\end{equation}
and 
\begin{equation}
\frac{d\langle \delta c(t)\rangle}{dt}=\Big(-i\omega_c-\frac{\kappa_c}{2}\Big) \langle \delta c\rangle+g_2\langle a^2\rangle.
\end{equation}
Consider the non-degenerate case. In the Fourier domain, Eq. \eqref{eq_deltac_jpc} can be written as
\begin{align}
\langle \delta c[\omega]\rangle&=\frac{1}{\frac{\kappa_c}{2}-i(\omega-\omega_c)}\frac{g_3}{\sqrt{2\pi
\kappa_a\kappa_b}}\int_0^\infty d\omega' \langle(a^{\rm{in}}[\omega']\nonumber\\&\qquad+a^{\rm{out}}[\omega'])(b^{\rm{in}}[\omega-\omega']+b^{\rm{out}}[\omega-\omega'])\rangle,\nonumber
\end{align}
where in the last line, we have used the input-output relations Eq. \eqref{inoutNDPA}. We treat the case when there is an incident tone on the $a$-mode with a photon flux per unit time $P_{a, \rm{coh}}^{\rm{in}}$ (see Eq. \eqref{photflux1} for definition). Our results can be generalized easily to the more complicated case of both $a$ and $b$ modes being driven. Substituting the expression of $a^{\rm{out}}[\omega-\omega'])$, $a^{\rm{out}}[\omega'])$ from Eq. \eqref{jpc_scatmat}, after several lines of algebra, we arrive at
\begin{align}
\langle \delta c[\omega]\rangle&=\frac{1}{\frac{\kappa_c}{2}-i(\omega-\omega_c)}\frac{g_3}{\sqrt{2\pi\kappa_a\kappa_b}}\delta(\omega-\omega_c)\nonumber\\&\quad\Bigg[-\frac{8\pi\rho_{\rm{ab}}e^{-i\theta}P_{a, \rm{coh}}^{\rm{in}}}{(1-\rho_{\rm{ab}}^2)^2}-4\rho_{\rm{ab}}e^{-i\theta}\nonumber\\&\quad\int_0^{\omega_c} d\omega'\frac{\chi_b^{-1}(\omega_c-\omega')^*}{\big|{\chi_a}^{-1}(\omega'){\chi_b}^{-1}(\omega_c-\omega')^*-\rho_{\rm{ab}}^2\big|^2}\Bigg],\nonumber
\end{align}
where we have neglected the number of thermal photons present in the transmission lines. Fourier transforming back, we get
\begin{equation}
\langle \delta c(t)\rangle = \frac{8g_3e^{-i(\omega_c t+\theta)}}{\sqrt{\kappa_a\kappa_b}\kappa_c}\Bigg[\frac{-\rho_{\rm{ab}}}{(1-\rho_{\rm{ab}}^2)^2}P_{a, \rm{coh}}^{\rm{in}}-\frac{\sqrt{\kappa_a\kappa_b}}{8}\frac{\rho_{\rm{ab}}}{1-\rho_{\rm{ab}}^2}\Bigg].\nonumber
\end{equation}
Using the above expression for $\langle \delta c(t)\rangle$, we arrive at the self-consistency relation for $\rho_{\rm{ab}}$:
\begin{align}
\rho_{\rm{ab}} = \rho_{\rm{ab}}^0\Bigg|1-\rho_{\rm{ab}}^0\frac{\rho_{\rm{ab}}}{(1-\rho_{\rm{ab}}^2)^2}\frac{P_{a, \rm{coh}}^{\rm{in}}}{P_c^{\rm{in}}}-\frac{g_3}{2\sqrt{\kappa_c}|\langle c^{\rm{in}}\rangle|}\frac{\rho_{\rm{ab}}}{1-\rho_{\rm{ab}}^2}\Bigg|,\nonumber
\end{align}
where the second and third terms on the right denote the pump depletion due to incident signal and vacuum fluctuations on the signal port. Here, $\rho_{\rm{ab}}^0 = 4g_3|\langle c^{\rm{in}}\rangle|/\sqrt{\kappa_a\kappa_b\kappa_c}$. Using the above equation, one can compute the gain and the output signal power (see Appendix \ref{output_power_calc}) given by
\begin{align}
\label{eqGPout1}
G &= \Big(\frac{1+\rho_{\rm{ab}}^2}{1-\rho_{\rm{ab}}^2}\Big)^2, \\\label{eqGPout2}
P_{\rm{a, tot}}^{\rm{out}} &= GP_{\rm{a, coh}}^{\rm{in}} + \frac{\kappa_a}{\sqrt{G}}(G-1)\frac{1+\rho_{\rm{ab}}^2}{8}.
\end{align}

\begin{figure}
\centering
\includegraphics[width=0.5\textwidth]{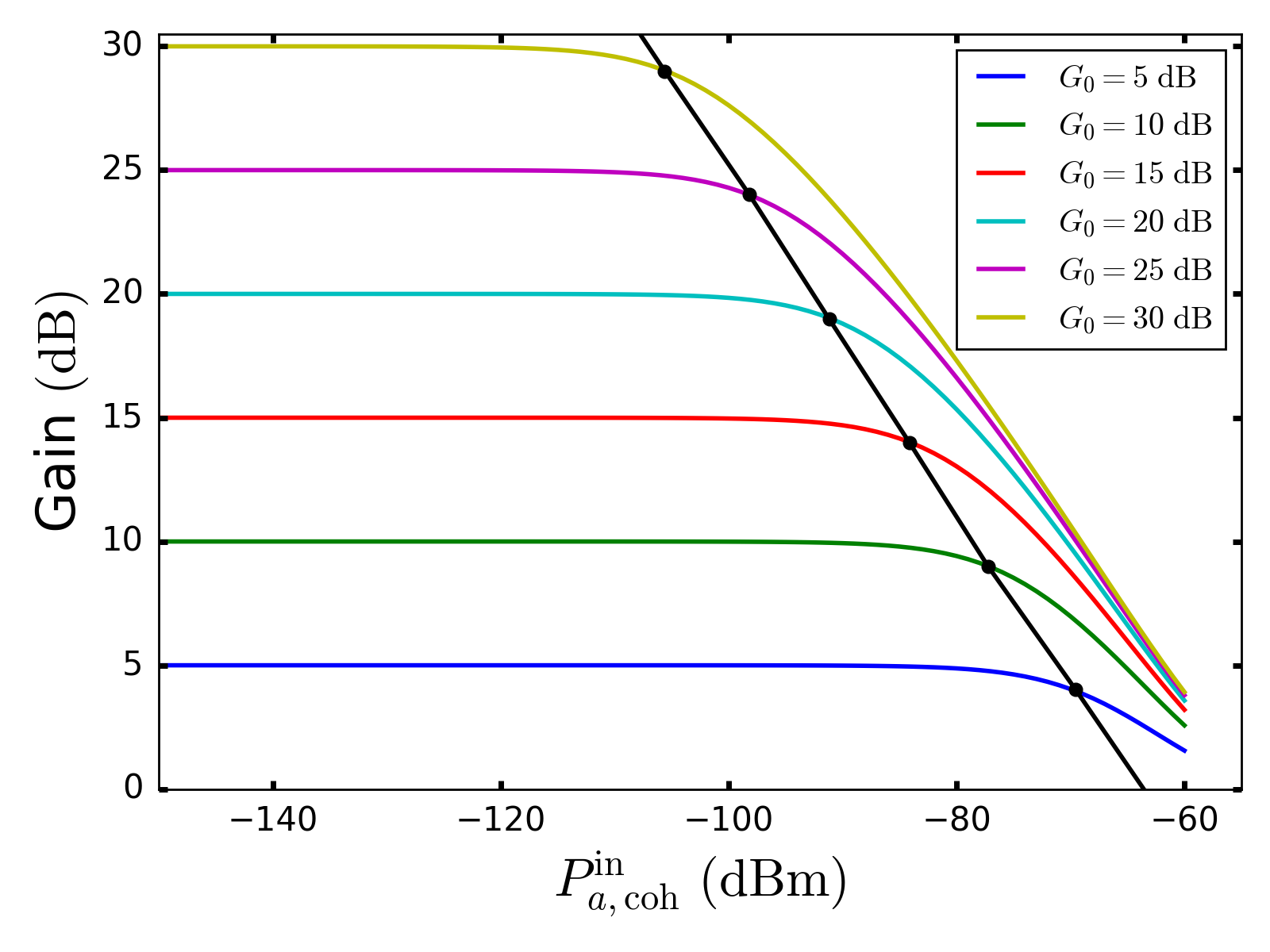}
\caption{\label{jpc_gain} Theoretical gain of a non-degenerate parametric amplifier as a function coherent incident signal power, taking into account the effect of pump depletion. The different solid lines correspond to un-depleted gain of $5$ to $30$ dB, in steps of $5$ dB. For definite, realistic system parameters, we have chosen $\omega_a/2\pi = 10$ GHz, $\omega_b/2\pi = 7$ GHz, $\omega_c/2\pi=17$ GHz, $\kappa_a/2\pi=\kappa_b/2\pi = 100$ MHz, $\kappa_c/2\pi = 600$ MHz and $g_3/2\pi=0.1$ MHz. The black dots on each curve correspond to the 1 dB compression point, where the gain of the amplifier drops by 1 dB. These dots lie on a straight line (the black line in the figure), whose slope in the given plot is $\sim -0.7$. The asymptotic value of the slope in the limit of high gain is $-2/3$ (see Ref. \onlinecite{Abdo_Devoret_2013_b}). }
\end{figure}

Fig. \ref{jpc_gain} shows the resulting gain as a function of coherent incident signal power ($P_{\rm{a, coh}}^{\rm{in}}$) for the non-degenerate case. The different curves correspond to un-depleted gain of $5$ to $30$ dB, in steps of $5$ dB. The black dots on each curve correspond to the point where the gain of the amplifier is lowered by 1 dB. These points lie on a straight line, with a slope of $\sim -0.7$. This is in reasonable agreement with the asymptotic value of the slope in the limit of high gain, predicted to be $-2/3$ [see Eq. (103) in Ref. \onlinecite{Abdo_Devoret_2013_b}].

\begin{figure}
\centering
\includegraphics[width=0.5\textwidth]{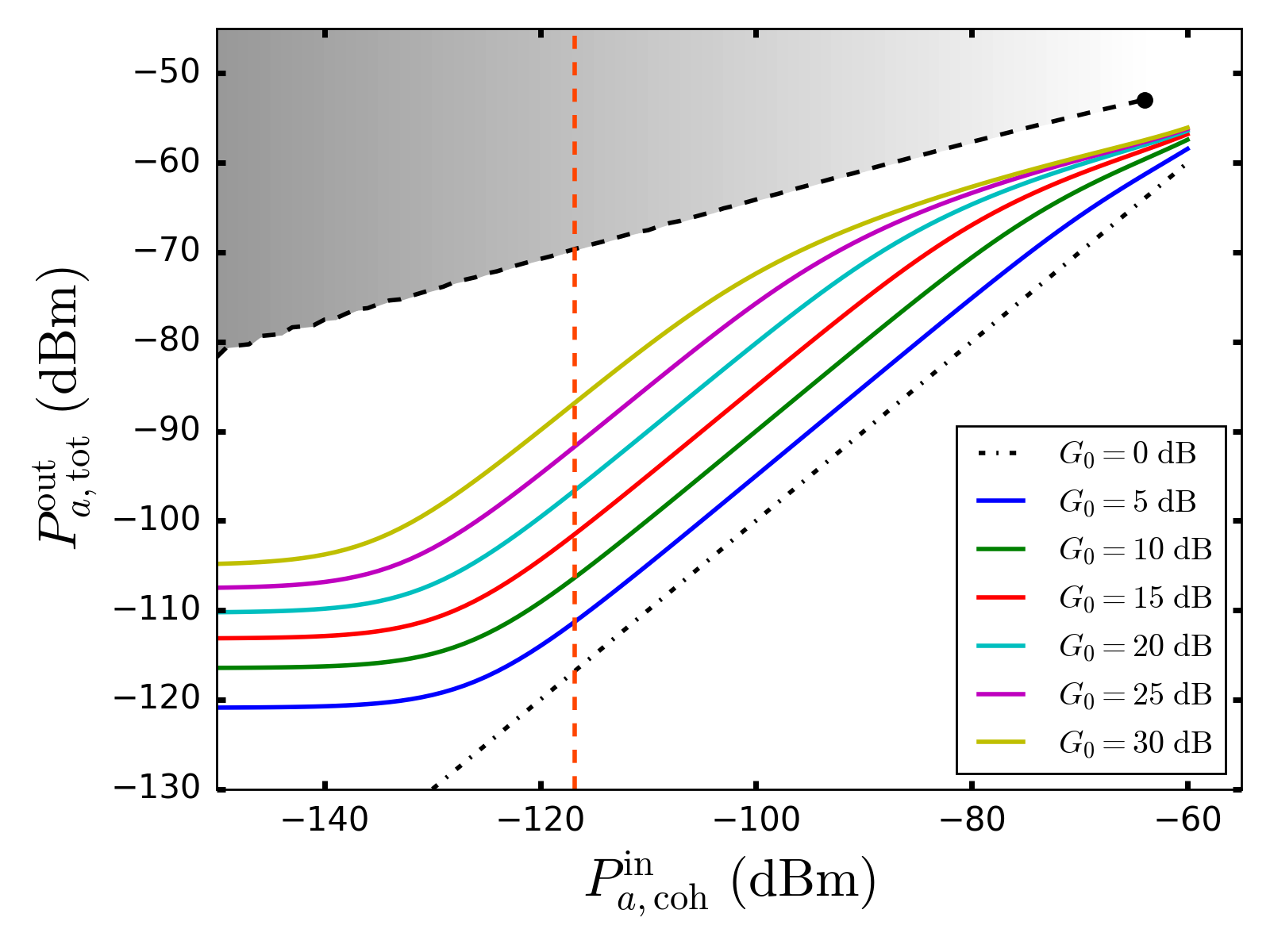}
\caption{\label{jpc_pout} Total output signal power as a function of coherent incident signal power for the non-degenerate paramp.  The different solid lines correspond to un-depleted gain of $5$ to $30$ dB, in steps of $5$ dB. The system parameters are chosen as in Fig. \ref{jpc_gain}. As incident coherent signal power goes to zero, the output power tends to a constant value corresponding to amplified vacuum fluctuations incident on the signal port. 
The black dot-dashed line corresponds to the pump tone being switched off. The black dashed line corresponds to the maximum output signal power before the onset of spontaneous oscillation. With sufficient increase of the incident signal power, the system ceases to exhibit spontaneous oscillation, hence the black dot at the end of the line. In the entire shaded region, the system shows parametric oscillation. The gray color gradient schematically indicates the difference between the two possible classical amplitudes of the output signal in the region of parametric oscillation. This difference goes to zero when the incident power is large enough for the system to stop showing parametric oscillation. The vertical orange line corresponds to the power of half a photon of noise incident on the signal port. Note that in addition to the constraint imposed on the magnitude of incident pump power by spontaneous oscillation threshold, additional restrictions arise due to the Josephson nonlinearity. This is because higher order nonlinear effects become relevant in addition to those presented in the Hamiltonian of Eq. \eqref{nondegham}. The exact value of the maximum output power depends on the exact implementation of this device using Josephson circuits and will scale with the junction energy $E_J$. For some typical values for the Josephson Parametric Amplifier, see Secs. III, IV of Ref. \onlinecite{Abdo_Devoret_2013_b}. } 
\end{figure}

\begin{figure}
\centering
\includegraphics[width=0.5\textwidth]{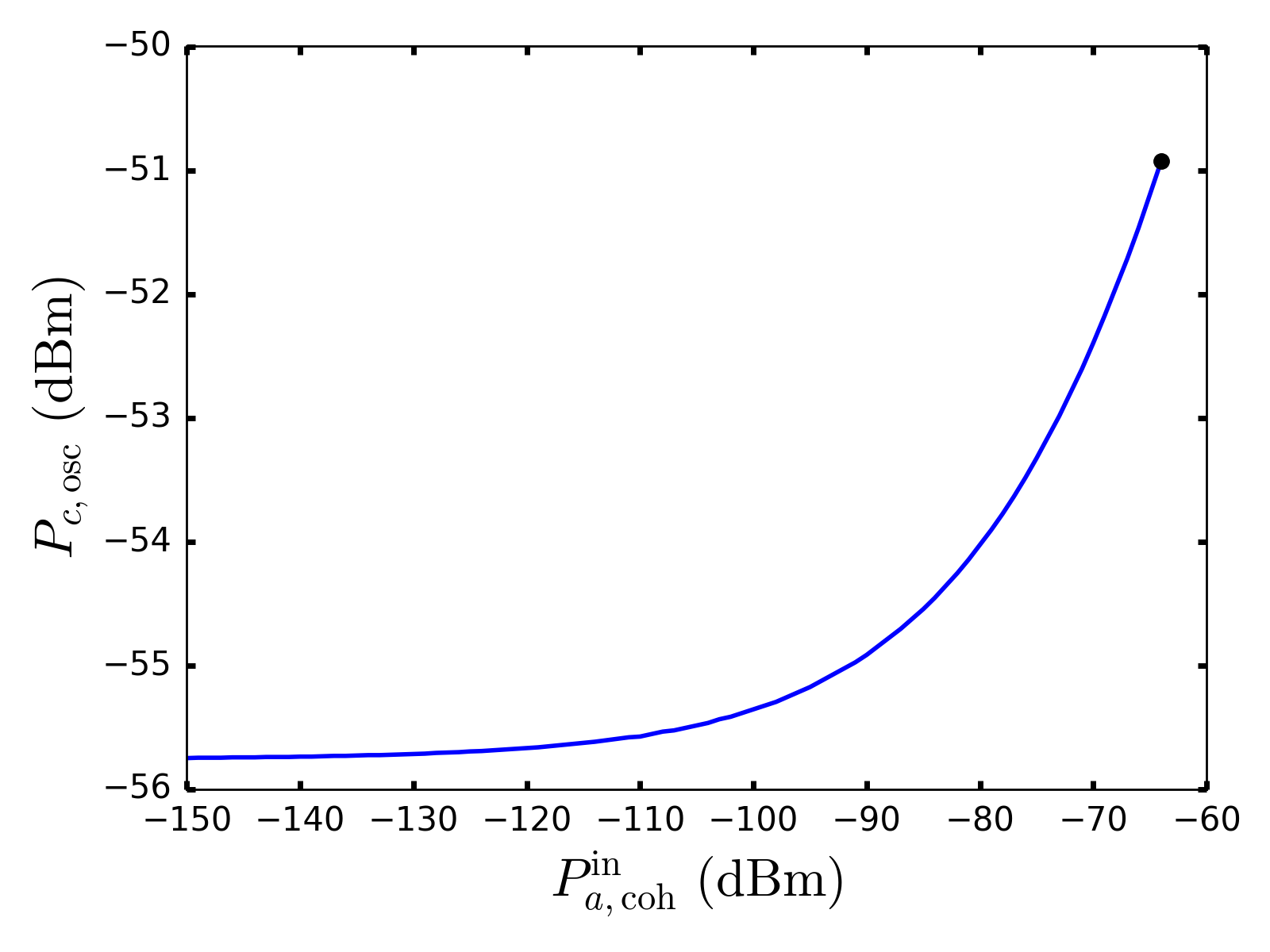}
\caption{\label{jpc_thresh} Shift of threshold of spontaneous oscillation upon increase of coherent signal power. The system parameters are chosen as in Fig. \ref{jpc_gain}. }
\end{figure}

In Fig. \ref{jpc_pout}, the total output signal power is plotted as a function of $P_{\rm{a, coh}}^{\rm{in}}$. The solid curves denote the output power for un-depleted gain of $5$ to $30$ dB, in steps of $5$ dB. As the incident coherent signal power goes to zero, the output power tends to a constant value which corresponds to amplified vacuum fluctuations incident on the signal port. The black dot-dashed line correspond to unity un-depleted gain when the pump tone is switched off. Finally, the dashed black line correspond the maximum output power that the device can produce before the onset of spontaneous parametric oscillation\footnote{This curve can be obtained  as follows. Solving the coupled nonlinear set of equations [Eqs. (\ref{a_qle_jpc}, \ref{b_qle_jpc}, \ref{c_qle_jpc})] in steady state, one obtains the smallest pump power which gives rise to multiple, stable, steady-state solutions for the modes $a,b,c$. The pump power immediately below this value is the maximum allowed pump power. In practice, the equations are solved numerically for different pump powers. }. For low enough incident power, the dashed line eventually approaches a constant value like the solid curves. This regime falls outside the range of incident powers shown in the plot. The shaded region above the dashed line indicates the region where the system shows spontaneous oscillation. In this region, the noise spectrum of the amplifier output develops a peak at the self-oscillation frequency, which rides on top of a continuous background. Note that as the incident signal power is increased, the pump power needed for onset of oscillation, $i.e.$ the threshold pump power, increases (see Fig. \ref{jpc_thresh}). For sufficiently high incident signal power, the system ceases to exhibit parametric oscillation (denoted by the black circle in Figs. (\ref{jpc_pout}, \ref{jpc_thresh}). We note that the solid curves asymptotically do not approach the dashed line of the onset of parametric oscillation (there is a gap of about 5 dB). The reason for that is as follows. The self-consistent theory for the solid curves [Eqs. (\ref{eqGPout1}, \ref{eqGPout2}] is derived using the linear scattering relations of the parametric amplifier [see Eq. \eqref{jpc_scatmat}]. Thus, onset of parametric oscillation, which is inherently a nonlinear effect, is beyond the scope of this theory. We believe a more complete theory of pump depletion will be able to address this discrepancy. The details of such a theory is left for a future work. 

The aforementioned parametric oscillation in these parametric amplifiers is analogous to the second order phase-transition occurring in ferromagnets. For the non-degenerate case, consider a two-dimensional ferromagnet (which has a complex scalar magnetization order parameter) in four or higher dimensions. Here, the dimension is chosen to make the analogy with our mean-field calculations more appropriate. For simplicity, we choose the pump phase appropriately so that Eqs. (\ref{a_qle_jpc}-\ref{c_qle_jpc}) allow for solutions for $a_0e^{-i\omega_a t}$,$b_0e^{-i\omega_b t}$ where $a_0,b_0$ are real. In this analogy, the role of the complex magnetization order parameter is played by $a_0+ib_0$. The role of the inverse temperature is played by the incident pump tone amplitude $c^{\rm{in}}$. The incident signal $a^{\rm{in}}$ acts like an external applied magnetic field. In this analogy, the black dashed line corresponds to a second-order phase-transition, terminating at a critical point. However, note the qualitative difference to conventional phase-transitions in this analogy. In ferromagnets, in presence of magnetic field, the second order phase-transition disappears in favor of a smooth crossover. In the non-degenerate amplifier, the parametric oscillation persists for a range of the incident signal. 

The calculations for the degenerate case proceeds analogously and leads to
\begin{equation}
\langle \delta c(t)\rangle = \frac{8g_2e^{-i(\omega_c t+\theta)}}{\kappa_a\kappa_c}\Bigg[\frac{-2\rho_{\rm{aa}}}{(1-\rho_{\rm{aa}}^2)^2}P_{a, \rm{coh}}^{\rm{in}}-\frac{\kappa_a}{8}\frac{\rho_{\rm{aa}}}{1-\rho_{\rm{aa}}^2}\Bigg].\nonumber
\end{equation}
Here, we have neglected the generated second order harmonic. This harmonic does not contribute to amplification within the RWA. Using the above expression for $\langle \delta c(t)\rangle$, we arrive at the self-consistency relation for $\rho_{\rm{aa}}$:
\begin{align}
\rho_{\rm{aa}} = \rho_{\rm{aa}}^0\Bigg|1-\rho_{\rm{aa}}^0\frac{\rho_{\rm{aa}}}{(1-\rho_{\rm{aa}}^2)^2}\frac{P_{a, \rm{coh}}^{\rm{in}}}{P_c^{\rm{in}}}-\frac{g_2}{2\sqrt{\kappa_c}|\langle c^{\rm{in}}\rangle|}\frac{\rho_{\rm{aa}}}{1-\rho_{\rm{aa}}^2}\Bigg|,\nonumber
\end{align}
where $\rho_{\rm{aa}}^0=8g_2|\langle c^{\rm{in}}\rangle/(\kappa_a\sqrt{\kappa_c})$.  
The gain and the output power on the signal port are respectively given by
\begin{align}
G &= \Big(\frac{1+\rho_{\rm{aa}}^2}{1-\rho_{\rm{aa}}^2}\Big)^2, \\
P_{\rm{a, tot}}^{\rm{out}} &= (2G-1)P_{\rm{a, coh}}^{\rm{in}} + \frac{\kappa_a}{\sqrt{G}}(G-1)\frac{1+\rho_{\rm{aa}}^2}{8}.
\end{align}
The plots for the degenerate case are shown in Appendix \ref{dyn_range_deg_paramp}. Analogies with phase-transitions in ferromagnets can also be constructed  for the degenerate case and is omitted for brevity.  

\section{Phase-space distribution of parametric amplifiers: below, at and above threshold}
\label{sec_6}
Whereas in the previous sections, we have analyzed degenerate and non-degenerate parametric amplifiers in the Heisenberg picture, here in this section, we turn to the Schr\"odinger picture and the associated Lindblad master equation to analyze the phase-space distributions of the modes. This approach has the merit of directly yielding the variance of the fluctuations of the signal and idler waves. Starting from the Lindblad equation, we derive an effective Fokker-Planck equation for the Wigner distribution for the relevant modes. These are subsequently solved analytically to get the results. Our calculations extend those given in Chap. 10 of Ref. \onlinecite{Carmichael_2007}. Here, we begin with the case of a degenerate paramp since it is computationally easier. 

\subsection{Degenerate parametric amplifier}
\label{sec_6a}
The Lindblad equation for the density matrix for the modes of a degenerate paramp is given by
\begin{align}
\frac{d\rho}{dt}=-\frac{i}{\hbar}[\tilde{H}^{\rm{DPA}}+H_{\rm{drive}}, \rho] + \big(\kappa_a{\cal D}[a]+ \kappa_c{\cal D}[c]\big)\rho,
\end{align}
where 
$\tilde{H}^{\rm{DPA}}$ is defined in Eq. \eqref{degham}, 
\begin{align}
H_{\rm{drive}}=\hbar(\epsilon_a a^\dagger e^{-i\omega_a t} + \epsilon_c c^\dagger e^{-i\omega_c t} + \rm{H.c.})
\end{align}
and ${\cal D}[a]\cdot = a\cdot a^\dagger -(a^\dagger a\cdot+\cdot a^\dagger a)/2$ is the Lindblad superoperator. Here $\epsilon_{a(c)}$ denote the  incident signal (pump) tone. Using the definition of the Wigner distribution $(W)$ in terms of the symmetrized characteristic function (see Chap. 10 of Ref. \onlinecite{Carmichael_2007} and Chap. 4 of Ref. \onlinecite{Carmichael_1999}), one can write down the equation of motion for $W(\alpha,\alpha^*, \gamma, \gamma^*)$, where $\alpha, \gamma$ denote the complex numbers corresponding to amplitudes of the modes $a,c$ respectively. The resulting equation is given by
\begin{align}
\frac{\partial W}{\partial t} &= \Big[\frac{\partial }{\partial \alpha}\bigg\{\bigg(\frac{\kappa_a}{2}+i\omega_a\bigg)\alpha + 2g_2 \alpha^*\gamma + i\epsilon_ae^{-i\omega_a t}\bigg\}\nonumber\\&\quad +\frac{\partial }{\partial \gamma}\bigg\{\bigg(\frac{\kappa_c}{2}+i\omega_c\bigg)\gamma - g_2 \alpha^2 + i\epsilon_ce^{-i\omega_c t}\bigg\} + {\rm{c.c.}}\nonumber\\&\quad+\frac{\kappa_a}{2}\frac{\partial^2}{\partial\alpha\partial\alpha^*}+\frac{\kappa_c}{2}\frac{\partial^2}{\partial\gamma\partial\gamma^*}\nonumber\\&\quad-\frac{g_2}{4}\bigg(\frac{\partial^3}{\partial\alpha^2\partial\gamma^*}+{\rm{c.c.}}\bigg)\Big]W.
\end{align}
While the above equation describes the full dynamics of the Wigner distribution, it is not amenable to analytical solution. To achieve the latter, we expand the mode amplitudes around the semi-classical solutions and look at small fluctuations around these solutions. For the degenerate case, this approach is valid for all values of pump and signal power (see below for the non-degenerate case). For nonzero incident signal power ($\epsilon_a\neq0$), the fluctuations of the modes are small compared to their semiclassical values by a factor of  $1/(n_p^{\rm{thr}})^{1/2}$ (the proof is identical to that presented in Chap. 10 of Ref. \onlinecite{Carmichael_2007} and Chap. 8 of Ref. \onlinecite{Carmichael_1999}). Here $n_p^{\rm{thr}}$ is the number of pump ($c$-mode) photons required to give rise to spontaneous oscillation of the mode $a$ for $\epsilon_a=0$. The formula for $n_p^{\rm{thr}}$ is
\begin{equation}
n_p^{\rm{thr}} = \frac{\kappa_a^2}{16g_2^2},
\end{equation}
which is obtained by setting $\rho_{\rm{aa}}^0$ of Sec. \ref{sec_5} to 1. This scaling of the fluctuations is a manifestation of the fluctuations of the modes remaining Gaussian for all values of the pump power (see Fig. \ref{jba_wigner} below, bottom panels). However, the situation is different for zero incident signal power ($\epsilon_a=0$) at the threshold of spontaneous oscillation. For $\epsilon_a=0$, at the threshold, the fluctuations are damped by a factor of $1/(n_p^{\rm{thr}})^{1/4}$ (see Chap. 10 of Ref. \onlinecite{Carmichael_2007} for proof). This is a manifestation of the non-Gaussian behavior of the fluctuations of the signal mode (see Fig. \ref{jba_wigner} below, top center panel). In the following, we perform the calculations for $\epsilon_a\neq0$. Since the details for $\epsilon_a=0$ are given in Chap. 10 of Ref. \onlinecite{Carmichael_2007}, we do not repeat that calculation here. We define
\begin{align}
\tilde{\alpha} = \sqrt{n_p^{\rm{thr}}}\langle \tilde{a}\rangle + \tilde{z}, \quad\tilde{\gamma} = \sqrt{n_p^{\rm{thr}}}\langle \tilde{c}\rangle + \tilde{u},
\end{align}
where $\tilde{\alpha}=e^{i\omega_a t}\alpha$, $\tilde{\gamma}=e^{i\omega_c t}\gamma$, $\tilde{z}= e^{i\omega_a t}z$, $\tilde{u}= e^{i\omega_c t}u$ and $\tilde{a}=e^{i\omega_a t}a$, $\tilde{c}=e^{i\omega_c t}c$. Here,  $z,u$ denote the fluctuations around a semi-classical solutions $\langle a\rangle, \langle c\rangle$. 
Performing the analysis self-consistently and omitting some algebra, we arrive at the equations of motion for the semi-classical solutions:
\begin{align}
\label{semi-class-1}
\frac{d\langle \tilde{a}\rangle}{dt}&=-\frac{\kappa_a}{2}\langle\tilde{a}\rangle - \frac{\kappa_a}{2}\langle \tilde{a}^\dagger\rangle\langle\tilde{c}\rangle+\lambda_a\\\label{semi-class-2}\frac{d\langle \tilde{c}\rangle}{dt}&=-\frac{\kappa_c}{2}\langle\tilde{c}\rangle + \frac{\kappa_a}{4}\langle \tilde{a}^\dagger\rangle^2+\lambda_c,
\end{align}
where $\lambda_{a,c}=-4ig_2\epsilon_{a,c}/\kappa_{a,c}$. Note that these equations are merely scaled, semi-classical versions of the QLE-s derived in Sec. \ref{sec_5}. The fluctuation around these semi-classical solutions is given by the Wigner distribution $\bar{W}(\tilde{z}, \tilde{z}^*,\tilde{u}, \tilde{u}^*)$, which obeys
\begin{align}\label{Wdeg_eqn}
\frac{\partial\bar{W}}{\partial t}&=\Big[\frac{\partial}{\partial \tilde{z}}\big\{\frac{\kappa_a}{2}\tilde{z}+\frac{\kappa_a}{2}(\langle \tilde{a}^\dagger \rangle \tilde{u} + \langle \tilde{c}\rangle \tilde{z}^*)\big\}\nonumber\\&\quad+\frac{\partial}{\partial\tilde{u}}\big\{\frac{\kappa_b}{2}\tilde{u}-\frac{\kappa_a}{2}\langle \tilde{a} \rangle \tilde{z}\big\}+{\rm{c.c.}}\nonumber\\&\quad+\frac{\kappa_a}{2}\frac{\partial^2}{\partial \tilde{z}\partial \tilde{z}^*}+\frac{\kappa_c}{2}\frac{\partial^2}{\partial \tilde{u}\partial \tilde{u}^*}\Big]\bar{W}.
\end{align}
This equation for $\bar{W}$ is a Fokker-Planck equation and thus, can be exactly solved as a function of time. Therefore, any correlation function of the modes can also be analytically computed using this equation. Next, we give the steady state properties of $\bar{W}$ when $\lambda_{a,c}\in\Re$. Computations for other choices of $\lambda_{a,b}$ can be performed analogously. In steady state, 
\begin{equation}
\label{wig}
\bar{W}(\tilde{z}, \tilde{z}^*, \tilde{u}, \tilde{u}^*)=F_1(\tilde{z}_1,\tilde{u}_1)F_2(\tilde{z}_2,\tilde{u}_2),
\end{equation}
where $\tilde{z} = \tilde{z}_1+i\tilde{z}_2$, $\tilde{u} = \tilde{u}_1+i\tilde{u}_2$ and $F_1, F_2$ individually obey the following Fokker-Planck equations:
\begin{align}\label{wdeg_eqn1}
{}&\bigg[\frac{\partial}{\partial\tilde{z}_j}\bigg\{\frac{\kappa_a}{2}(1-(-1)^jc_0)\tilde{z}_j+\frac{\kappa_a}{2}a_0\tilde{u}_j\bigg\}\nonumber\\&\quad+\frac{\partial}{\partial\tilde{u}_j}\bigg\{-\frac{\kappa_a}{2}a_0\tilde{z}_j+\frac{\kappa_c}{2}\tilde{u}_j\bigg\}\nonumber\\&\quad+\frac{\kappa_a}{8}\frac{\partial^2}{\partial \tilde{z}_j^2}+\frac{\kappa_c}{8}\frac{\partial^2}{\partial \tilde{u}_j^2}\bigg]F_j(\tilde{z}_j,\tilde{u}_j)=0,\quad j=1,2.
\end{align}
Here, $a_0, c_0$ denote the semi-classical solutions obtained by solving Eqs. (\ref{semi-class-1}, \ref{semi-class-2}) in steady state. For nonzero $\epsilon_a$, the fluctuations around the semi-classical solutions remain Gaussian for all values of the pump $\epsilon_c$, even though the system shows parametric oscillation (see Sec. \ref{sec_5}) for a range of $\epsilon_a$. This should be compared to the case when $\epsilon_a=0$. In that case, below and above the threshold for spontaneous oscillation, the fluctuations around the semi-classical solution are Gaussian, while the system exhibits non-Gaussian behavior at threshold. The results of the calculations are shown in Fig. \ref{jba_wigner}. The top panels show the case when $\epsilon_a=0$. The top left  and top right panel shows the fluctuations of the $a$-mode around semi-classical solution below and above the spontaneous oscillation threshold. The top center panel shows the non-Gaussian behavior of the system at the threshold in absence of incident signal, as indicated by the contour lines which are no longer ellipses. The bottom panels show the Wigner function of the signal mode when $\epsilon_a\neq0$ for the corresponding pump powers. 
\begin{figure}
\centering
\includegraphics[width=0.5\textwidth]{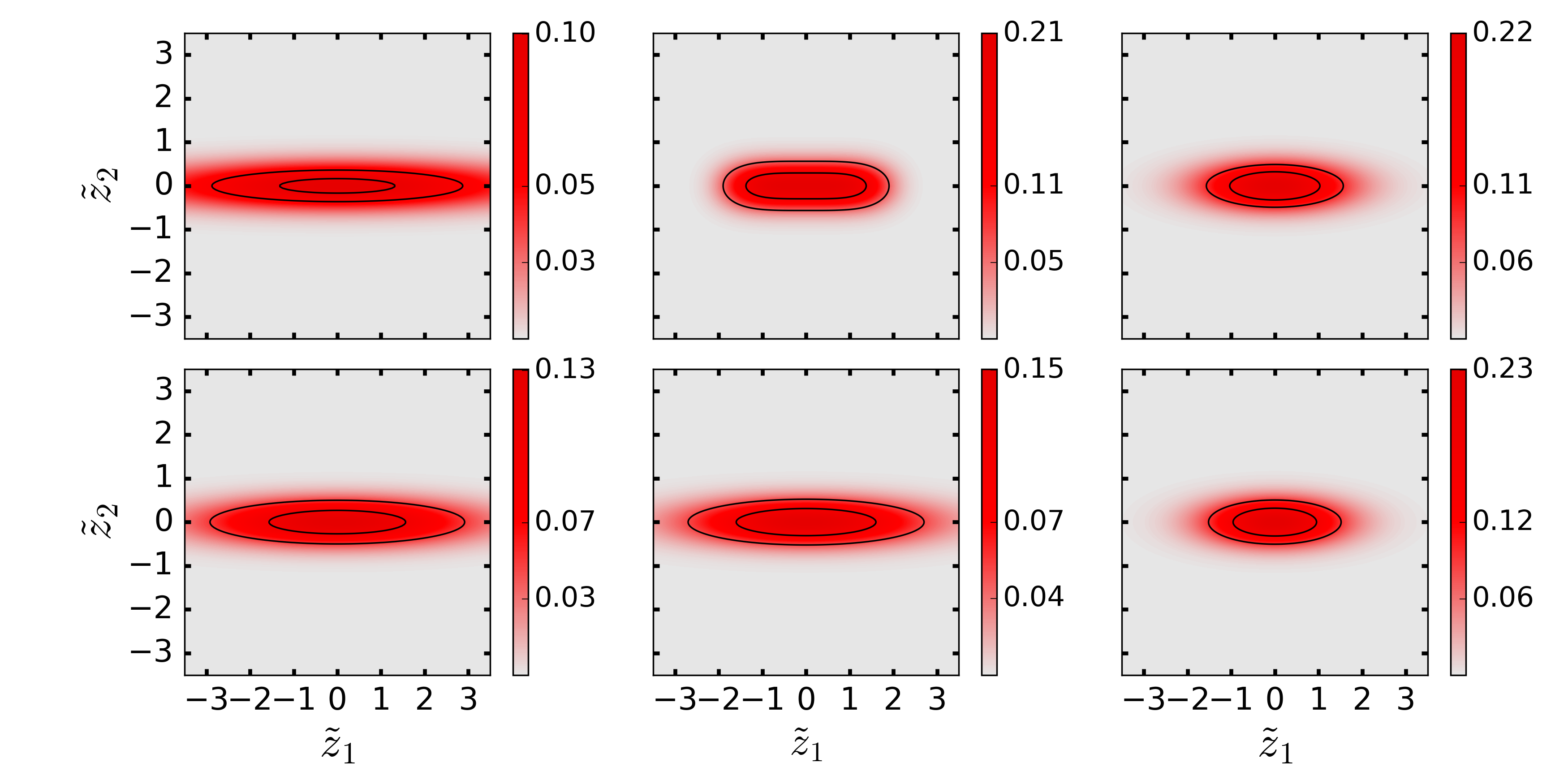}
\caption{\label{jba_wigner} Fluctuations of the signal mode for a degenerate parametric amplifier. For realistic system parameters, we have chosen $\omega_a/2\pi = 10$ GHz, $\omega_c/2\pi = 20$ GHz, $\kappa_a/2\pi = 100$ MHz, $\kappa_c/2\pi = 600$ MHz and $g_2/2\pi=0.1$ MHz. The top (bottom) panels shows the phase space fluctuations in absence (presence) of incident signal. The top left (right) panel show the fluctuations when the system is below (above) the oscillation threshold. The top center panel shows the non-Gaussian fluctuations of the system at threshold, as indicated by the non-elliptical contour lines. In presence of an incident signal (here, we have chosen an incident signal power of $-110$ dBm), upon increase of pump power, the fluctuations remain Gaussian (bottom panels). }
\end{figure}
Note that for $\epsilon_a\neq0$, the output amplified signal exhibits correlations with the pump tone. This is evident from the factorization of the Wigner function [see Eq. \eqref{wig}], where one quadrature of the signal is correlated with one of the pump. As the amplitude of the incoming signal becomes smaller, this correlation decreases and vanishes completely only in the limit $\epsilon_a=0$. 

\subsection{Non-degenerate parametric amplifier}
\label{sec_6b}
\begin{figure}
\centering
\includegraphics[width=0.5\textwidth]{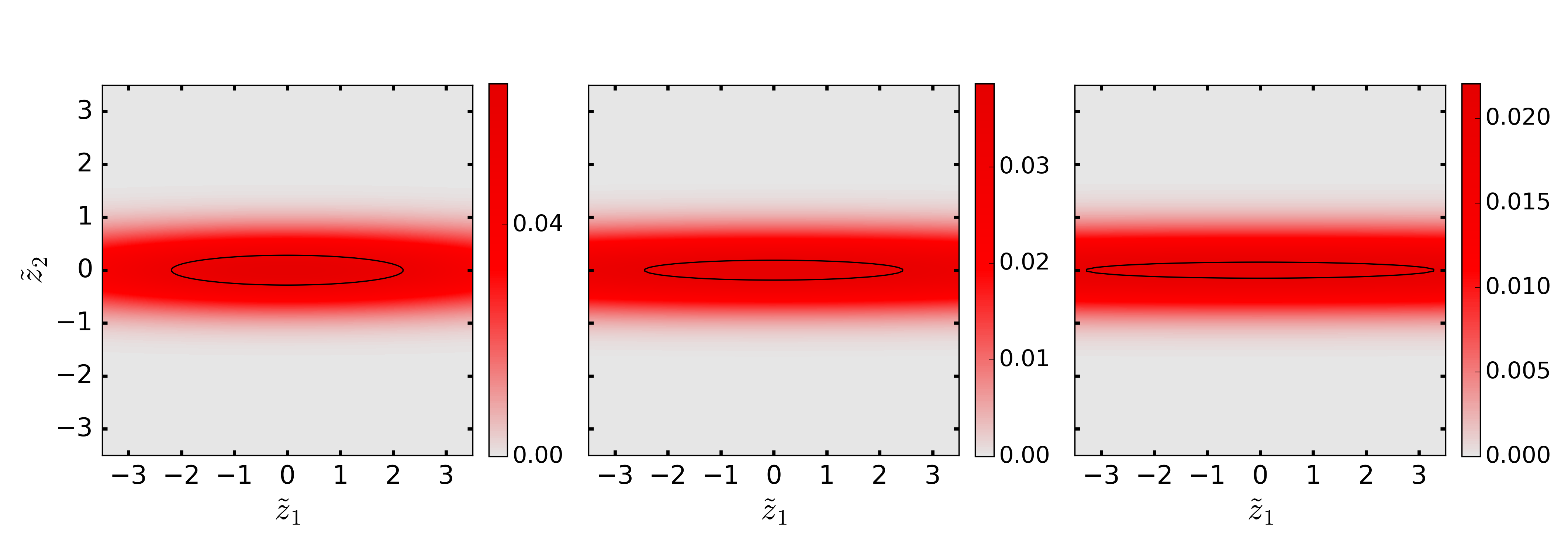}
\caption{\label{jpc_wigner} Phase-space fluctuations of the $a$-mode of a non-degenerate parametric amplifier in presence of incident signal. For definite, realistic system parameters, we have chosen $\omega_a/2\pi = 10$ GHz, $\omega_b/2\pi = 7$ GHz, $\omega_c/2\pi=17$ GHz, $\kappa_a/2\pi=\kappa_b/2\pi = 100$ MHz, $\kappa_c/2\pi = 600$ MHz and $g_3/2\pi=0.1$ MHz. The incident signal power is chosen to be $-110$ dBm. The left, center and right panels respectively corresponds to below, at and above the threshold for spontaneous oscillation. The fluctuations remain Gaussian for all pump powers. Note that here we show only the case of nonzero incident signal. A treatment of the case of zero incident signal below and at threshold gives identical behavior to the degenerate case (see Fig. \ref{jba_wigner}, top left and center panels). For zero incident signal power, above threshold, the system shows phase-diffusion. The behavior of the system for this case is treated using the positive-P distribution\cite{Reid1989} and is beyond the scope of the article. }
\end{figure}

Next, we describe the phase-space distribution of the modes of a non-degenerate parametric amplifier when there is an incident signal on the $a$-mode. The case when both $a$ and $b$ modes are driven can be treated analogously. The equation of motion for the system is given by
\begin{align}
\frac{d\rho}{dt}&=-\frac{i}{\hbar}[\tilde{H}^{\rm{NDPA}}+H_{\rm{drive}}, \rho]\nonumber\\&\quad + \big(\kappa_a{\cal D}[a]+ \kappa_b{\cal D}[b]+\kappa_c{\cal D}[c]\big)\rho
\end{align}
where $\tilde{H}^{\rm{NDPA}}$ is defined in Eq. \eqref{nondegham} and 
\begin{equation}
H_{\rm{drive}}=\hbar(\epsilon_a a^\dagger e^{-i\omega_a t} + \epsilon_c c^\dagger e^{-i\omega_c t} + \rm{H.c.}).
\end{equation}
As in Sec. \ref{sec_5}, we consider the case when only the signal mode is driven, with $\epsilon_{a(c)}$ denoting the incident signal (pump) tone.  Then, the equation of motion for the Wigner distribution $W(\alpha, \alpha^*,\beta, \beta^*,\gamma, \gamma^*)$ is 
\begin{align}
\frac{\partial W}{\partial t} &= \Big[\frac{\partial }{\partial \alpha}\bigg\{\bigg(\frac{\kappa_a}{2}+i\omega_a\bigg)\alpha + g_3 \beta^*\gamma + i\epsilon_ae^{-i\omega_a t}\bigg\}\nonumber\\&\quad +\frac{\partial }{\partial \beta}\bigg\{\bigg(\frac{\kappa_b}{2}+i\omega_b\bigg)\beta + g_3 \alpha^*\gamma \bigg\}\nonumber\\&\quad+\frac{\partial }{\partial \gamma}\bigg\{\bigg(\frac{\kappa_c}{2}+i\omega_c\bigg)\gamma - g_3 \alpha\beta + i\epsilon_ce^{-i\omega_c t}\bigg\}+{\rm{c.c.}}\nonumber\\&\quad+\frac{\kappa_a}{2}\frac{\partial^2}{\partial\alpha\partial\alpha^*}+\frac{\kappa_b}{2}\frac{\partial^2}{\partial\beta\partial\beta^*}+\frac{\kappa_c}{2}\frac{\partial^2}{\partial\gamma\partial\gamma^*}\nonumber\\&\quad-\frac{g_3}{4}\bigg(\frac{\partial^3}{\partial\alpha^2\partial\beta^*}+{\rm{c.c.}}\bigg)\Big]W.
\end{align}
For non-zero incident signal power ($\epsilon_a\neq 0$), the analysis is analogous to the degenerate case. The fluctuations around the semiclassical amplitudes remain Gaussian for all values of pump powers. They are smaller than the classical amplitudes by the factor of  $1/(n_p^{\rm{thr}})^{1/2}$, where now
\begin{equation}
n_p^{\rm{thr}}=\frac{\kappa_a\kappa_b}{4g_3^2}.
\end{equation}
This is obtained by setting $\rho_{\rm{ab}}^0$ of Sec. \ref{sec_5} to 1. In absence of incident signal, below and at the threshold of oscillation, the fluctuations of the modes $a,b$ are respectively Gaussian and non-Gaussian and the scaling is identical as in the degenerate case. However, above threshold there is a qualitative difference between the degenerate and non-degenerate cases. Unlike the degenerate case, the phases of self-oscillating amplitude of the modes $a,b$ are undetermined above threshold in absence of an incident signal. This, together with the decay of the modes $a,b$, gives rise to phase-diffusion. A full treatment of this effect needs the use of the positive-P distribution (see Ref. \onlinecite{Reid1989} for more details) and goes beyond the scope of this article. Here, we will only treat the case when there is an incident signal which is relevant for amplifiers. Then, the Wigner distribution is sufficient to analyze the system. We define 
\begin{align}
\tilde{\alpha} &= \sqrt{n_p^{\rm{thr}}}\langle \tilde{a}\rangle + \tilde{z}, \quad\tilde{\beta} = \sqrt{n_p^{\rm{thr}}}\langle \tilde{b}\rangle + \tilde{w},\nonumber\\\tilde{\gamma} &= \sqrt{n_p^{\rm{thr}}}\langle \tilde{c}\rangle + \tilde{u}.
\end{align}
Here, $\tilde{\alpha}=e^{i\omega_a t}\alpha$, $\tilde{\beta}=e^{i\omega_b t}\beta$, $\tilde{\gamma}=e^{i\omega_c t}\gamma$, $\tilde{z}= e^{i\omega_a t}z$, $\tilde{w}= e^{i\omega_b t}w$, $\tilde{u}= e^{i\omega_c t}u$ and $\tilde{a}=e^{i\omega_a t}a$, $\tilde{b}=e^{i\omega_b t}b$, $\tilde{c}=e^{i\omega_c t}c$. The variables $z,w, u$ denote the fluctuations around the semi-classical solutions $\langle a\rangle, \langle b\rangle, \langle c\rangle$. Expanding self-consistently, one arrives at the semi-classical equations of motion:
\begin{align}
\label{semi-class-3}
\frac{d\langle \tilde{a}\rangle}{dt}&=-\frac{\kappa_a}{2}\langle\tilde{a}\rangle - \frac{\sqrt{\kappa_a\kappa_b}}{2}\langle \tilde{b}^\dagger\rangle\langle\tilde{c}\rangle+\lambda_a\\\label{semi-class-4}\frac{d\langle \tilde{b}\rangle}{dt}&=-\frac{\kappa_b}{2}\langle\tilde{b}\rangle - \frac{\sqrt{\kappa_a\kappa_b}}{2}\langle \tilde{a}^\dagger\rangle\langle\tilde{c}\rangle\\\label{semi-class-5}\frac{d\langle \tilde{c}\rangle}{dt}&=-\frac{\kappa_c}{2}\langle\tilde{c}\rangle + \frac{\sqrt{\kappa_a\kappa_b}}{2}\langle \tilde{a}\rangle\langle \tilde{b}\rangle+\lambda_c,
\end{align}
where $\lambda_{a,c} = -2ig_3\epsilon_{a,c}/\sqrt{\kappa_a\kappa_b}$. Once again, these are merely the scaled, semi-classical versions of the QLE-s derived in Sec. \ref{sec_5}. The fluctuations are determined by the Wigner distribution $\bar{W}(\tilde{z},\tilde{z}^*,\tilde{w},\tilde{w}^*,\tilde{u},\tilde{u}^*)$, which obeys the following:
\begin{align}
\label{Wdeg_eqn}
\frac{\partial\bar{W}}{\partial t}&=\bigg[\frac{\partial}{\partial \tilde{z}}\bigg\{\frac{\kappa_a}{2}\tilde{z}+\frac{\sqrt{\kappa_a\kappa_b}}{2}(\langle \tilde{b}^\dagger \rangle \tilde{u} + \langle \tilde{c}\rangle \tilde{w}^*)\bigg\}\nonumber\\&\quad+\frac{\partial}{\partial \tilde{w}}\bigg\{\frac{\kappa_b}{2}\tilde{w}+\frac{\sqrt{\kappa_a\kappa_b}}{2}(\langle \tilde{a}^\dagger \rangle \tilde{u} + \langle \tilde{c}\rangle \tilde{z}^*)\bigg\}\nonumber\\&\quad+\frac{\partial}{\partial\tilde{u}}\bigg\{\frac{\kappa_c}{2}\tilde{u}-\frac{\sqrt{\kappa_a\kappa_b}}{2}(\langle \tilde{a} \rangle \tilde{w}+\langle \tilde{b} \rangle \tilde{z})\bigg\}+{\rm{c.c.}}\nonumber\\&\quad+\frac{\kappa_a}{2}\frac{\partial^2}{\partial \tilde{z}\partial \tilde{z}^*}+\frac{\kappa_b}{2}\frac{\partial^2}{\partial \tilde{w}\partial \tilde{w}^*}+\frac{\kappa_c}{2}\frac{\partial^2}{\partial \tilde{u}\partial \tilde{u}^*}\bigg]\bar{W}.
\end{align}
For simplicity, we choose $\kappa_a=\kappa_b=\kappa$ and the phases of $\epsilon_a, \epsilon_c$ so that the semi-classical solutions are real. In steady state, the above Fokker-Planck equation factorizes and $\bar{W}=\prod_{j=1,2}F_j(\tilde{z}_j,\tilde{w}_j, \tilde{u}_j)$, where $F_j$ obeys 
\begin{align}
\label{wnondeg_1}
{}&\bigg[\frac{\kappa}{2}\frac{\partial}{\partial \tilde{z}_j}\bigg\{\tilde{z}_j-(-1)^jc_0\tilde{w}_j+b_0\tilde{u}_j\bigg\}+\frac{\kappa}{2}\frac{\partial}{\partial \tilde{w}_j}\bigg\{\tilde{w}_j\nonumber\\&\quad-(-1)^jc_0\tilde{z}_j+a_0\tilde{u}_j\bigg\}+\frac{\kappa_c}{2}\frac{\partial}{\partial \tilde{u}_j}\bigg\{\tilde{u}_j-\frac{\kappa}{\kappa_c}a_0\tilde{w}_j\nonumber\\&\quad-\frac{\kappa}{\kappa_c}b_0\tilde{z}_j\bigg\}+\frac{\kappa}{8}\bigg(\frac{\partial^2}{\partial \tilde{z}_j^2}+\frac{\partial^2}{\partial \tilde{w}_j^2}\bigg)+\frac{\kappa_c}{8}\frac{\partial^2}{\partial \tilde{u}_j^2}\bigg]F_j=0, 
\end{align}
$j = 1,2$ and $a_0, b_0$ and  $c_0$ denote the steady state solutions of Eqs. (\ref{semi-class-3}-\ref{semi-class-5}). The above equations can be solved analytically. The cut of the Wigner distribution showing the phase-space fluctuations of the signal mode is given in Fig. \ref{jpc_wigner}. Due to the presence of the incident signal, the fluctuations are always Gaussian (as in the degenerate case). The three panels correspond to pump powers that, in absence of incident signal, correspond to below, at and above threshold of spontaneous oscillation. Note that the incident signal also removes the phase-diffusion present in the non-degenerate parametric oscillator above threshold \cite{Reid1989} (see right panel). The modes  $a,b$ are correlated with the pump mode for finite $\epsilon_a$ [this can be also seen from the factorization of the Wigner function in Eq. \eqref{wnondeg_1}]. This correlation is exactly zero only for $\epsilon_a=0$. Of course, in this case of two-mode squeezing, the modes $a,b$ stay correlated for all values of $\epsilon_a$. 

\section{Practical amplifier circuits based on Josephson junction circuits}
\label{sec_7}
In this section, we provide several circuit realizations of the degenerate amplifiers (Secs. \ref{sec_jba}, \ref{sec_7b}, \ref{sec_7c}) and a non-degenerate amplifier (Sec. \ref{sec_7d}). The circuit construction, together with pros and cons of each approach, is discussed. A concise summary of the different circuit constructions is provided in Fig. \ref{Fig_9}. 

\subsection{Driven microwave oscillator whose inductance is a single
Josephson element: Duffing-like dynamics}
\label{sec_jba}
\begin{figure}
\centering
\includegraphics[width=0.5\textwidth]{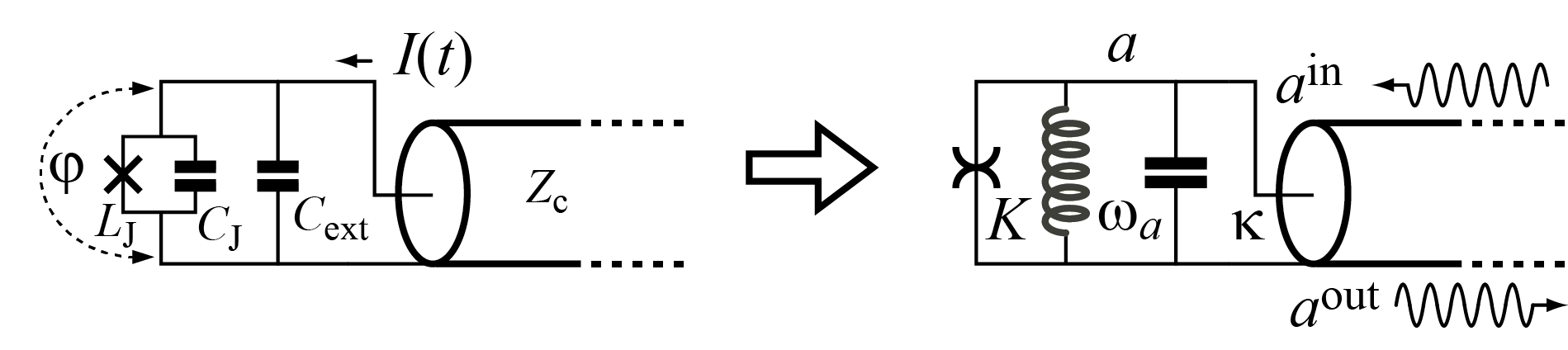}
\caption{\label{Fig_6}1-mode, 1-port Josephson amplifier involving only the Josephson
inductance. Left panel is schematic of Josephson tunnel junction, itself
consisting of a Josephson tunnel element playing the role of non-linear
inductance (cross, $L_{J}$) and a junction capacitance $C_{J}$, in parallel
with an external capacitance $C_{\rm{ext}}$ and a transmission line bringing in
the current $I(t)$. The variable $\protect\varphi $ is the phase across the
junction. On the right panel, simplified schematic based on an RWA treatment
where the oscillator is reduced to its frequency $\protect\omega _{a}$ and
damping rate $\protect\kappa $, with the Josephson non-linearity manifesting
itself as a simple Kerr component (opposing arcs symbols) characterized by the parameter K, the shift in frequency of the oscillator corresponding to 1 photon. The degree of
freedom is described by the standing photon ladder operator $a.$}
\end{figure}

We first examine the simplest case of a one-mode, one-port circuit in which the
inductance is just the Josephson element of superconducting tunnel junction
(see Fig. \ref{Fig_6}) \cite{Siddiqi_Devoret_2004, Vijay_2008}. 
The Hamiltonian of such systems is given by%
\begin{eqnarray}
H &=&H_{\rm{circ}}-\frac{\hbar }{2e}\varphi\cdot I+\text{ 
}H_{\rm{env}}, \\
H_{\rm{circ}} &=&-E_{J}\cos \varphi+\frac{Q^{2}}{2C_{\Sigma }},
\end{eqnarray}%
where $E_{J}=\left( \frac{\hbar }{2e}\right) ^{2}/L_{J}$ is the Josephson
energy, $C_{\Sigma }=C_J+C_{\rm{ext}}$ is the total capacitance in parallel
with the Josephson element,  $\varphi$ the gauge-invariant phase
difference across the junction, $Q$ the charge conjugate to the phase $%
\left[ \varphi,Q\right] =2ei$, $H_{\rm{env}}$ the Hamiltonian
of the transmission line, including the pump arriving through this channel
and $I$ the current operator belonging to the degrees of freedom of
the line. The amplifier functions with $\langle\varphi\rangle$ having excursions
much less than $\pi /2$ and the cosine function in the Hamiltonian can be
expanded to $4^{th}$ order only, with the $\varphi^{4}$ term treated as a
perturbation \cite{Manucharyan_Devoret_2007, Nigg_Girvin_2012}. Introducing the ladder operators of the
single mode of the circuit%
\begin{equation}
\varphi=\varphi ^{\rm{ZPF}}\left( a+a^{\dag }\right) 
\end{equation}%
and working in the framework of both an expansion in $\varphi ^{\rm{ZPF}}=\left(
2e^2/\hbar\right) ^{1/2}\left( L_{J}/C_{\Sigma }\right) ^{1/4}$ and RWA, the
hamiltonian of the circuit simplifies to%
\begin{equation}
\frac{H_{\rm{circ}}}{\hbar }=\tilde{\omega} _{a}a^{\dag }a+\frac{K}{2}a^{\dag
}a\left( a^{\dag }a-1\right), 
\end{equation}%
where, in the regime $\varphi ^{\rm{ZPF}}\ll 1$, $K=-e^{2}/(2\hbar C_{\Sigma })$ and $%
\tilde{\omega} _{a}=1/\sqrt{L_{J}C_{\Sigma }} + K$. The QLE
applied to this system yields%
\begin{equation}\label{qle_jba}
\frac{d}{dt}a=-i\left( \tilde{\omega} _{a}+Ka^{\dag }a\right) a-\frac{\kappa }{2}a+%
\sqrt{\kappa }a^{\rm{in}}\left( t\right),
\end{equation}%
where $\tilde{\omega}_{a}\gg \kappa _{a}\gg K$. This last equation is the quantum
version, in the RWA approximation, of the equation describing systems
modeled by the Duffing equation. The classical Duffing oscillator obeys the
equation 
\begin{equation}
m\ddot{x}+\eta \dot{x}+m\omega _{0}^{2}x(1+\mu x^{2})=f_{D}\cos \left(
\omega _{D}t\right) +f_{P}\left( t\right) 
\end{equation}%
for the position variable $x$ having mass $m$, small amplitude spring
constant $m\omega _{0}^{2}$, friction coefficient $\eta $ and driven at
frequency $\omega _{D}$, which is close to the small amplitude resonant
frequency $\omega _{0}$. A small probe force $f_{P}\left( t\right) $ allows
to study the displacement response of the system. Non-linearity of the
oscillator corresponds here to the spring constant being dependent
quadratically on position.

Here, for our amplifier, $\kappa $ plays the role of the damping rate $\eta
/m$, and $K$ plays the role of $\mu $. Let us now suppose that, in addition
to the signal to be processed, the $a$ port also receives an intense drive
tone described by a propagating coherent state with amplitude $\alpha ^{\rm{in}}$
and frequency $\Omega $. We treat this drive by the change of variable%
\begin{eqnarray}
a^{\rm{in}}\left( t\right)  &=&\alpha ^{\rm{in}}e^{-i\Omega t}+ \delta a^{\rm{in}}\left(
t\right), \\\label{jbaeq1}
a\left( t\right)  &=&\alpha e^{-i\Omega t}+\delta a\left( t\right) .
\end{eqnarray}%
We aim to solve for the semi-classical amplitude $\alpha$ from Eq. \eqref{qle_jba}:
\begin{equation}
\frac{d\alpha}{dt} - i\Omega\alpha=-i\tilde{\omega} _{a}\alpha-iK|\alpha|^2\alpha-\frac{\kappa _{a}}{2}\alpha+\sqrt{%
\kappa _{a}}\alpha ^{\rm{in}}.
\end{equation}%
By treating the non-linear term as a perturbation, we obtain the
self-consistent algebraic equation in steady state:%
\begin{equation}
\alpha =\frac{i\sqrt{\kappa _{a}}\alpha ^{\rm{in}}}{\left( \Omega -\tilde{\omega}
_{a}\right) +\frac{i\kappa _{a}}{2}-K\left\vert \alpha \right\vert ^{2}},
\end{equation}%
which in general yields for the c-number $\alpha $ a complex value $\left\vert \alpha -\alpha _{0}\right\vert\ll1$. Here,  
\begin{eqnarray}
\label{alpha0}
\alpha _{0} =\frac{i\sqrt{\kappa _{a}}\alpha ^{\rm{in}}}{\left( \Omega
-\tilde{\omega} _{a}\right) +\frac{i\kappa _{a}}{2}-4K\left\vert \alpha
^{\rm{in}}\right\vert ^{2}/\kappa _{a}}.
\end{eqnarray}%
Expanding $a$ around this value [using Eq. \eqref{jbaeq1}] and keeping up to second order terms, we arrive at the effective Hamiltonian for the degenerate parametric
amplifier arising from the pumping of the Josephson junction%
\begin{equation}
\label{jbaham11}
\frac{H}{\hbar }=\omega _{a}\delta a^{\dag }\delta a+\left[
g_{\rm{aa}}e^{i\left( \Omega _{\rm{aa}}t+\theta \right) }\left( \delta a\right)
^{2}+\rm{H.c.}\right] 
\end{equation}%
with%
\begin{eqnarray}
\label{eq_b}
\omega _{a} =\tilde\omega _{a}+2K\left\vert \alpha \right\vert
^{2}, 
g_{\rm{aa}}e^{i\theta } =\frac{K\alpha ^{\ast 2}}{2},
\Omega _{\rm{aa}} =2\Omega.
\end{eqnarray}%
The analysis of the amplifier based on the Hamiltonian given by Eq. \eqref{jbaham11} is justified only for $d\alpha/dt$ small compared to other spurious couplings/measurement times. Note that Eq. \eqref{eq_b} puts into light two drawbacks of this type of amplifier: the center frequency of the band of the amplifier shifts as the pump amplitude is increased and the pump tone needs to be at the center of the band for optimal amplification.
The use of two pumps frequencies $\Omega _{1}$ and $\Omega _{2}$ such
that $\Omega _{\rm{aa}}=\Omega _{1}+$ $\Omega _{2}$ facilitates the use of this
parametric amplifier. \cite{Kamal_Devoret_2009}

The device has noticeable gain when $K\left\vert \alpha ^{\rm{in}}\right\vert
^{2}/\kappa _{a}^{2}$ is of order unity, implying that the number of pump
photons in the oscillator is of order $\kappa _{a}/K$, a large number by
hypothesis. This justifies our treatment of the pump drive as a c-number.
Neglected terms such as the non-RWA term $\left( \delta a\right)
^{3}e^{i\Omega t}$ have smaller factors and can themselves be treated as
perturbations on top of the standard degenerate parametric amplifier
formalism. It is worth noting that for this device, the pump tone and the
signal tone must enter the circuit on the same port, which is inconvenient given the widely different amplitude levels of these two waves. 

Amplifiers based on the same Duffing type of non-linearity can also be
fabricated with two-port circuits containing arrays of Josephson junctions. \cite{Castellanos-Beltran_Lehnert_2008, Lehnert2009, Macklin_Siddiqi_2015}

\subsection{A parametrically driven oscillator: the DC-SQUID driven by RF
flux variation}
\label{sec_7b}
\begin{figure}
\centering
\includegraphics[width=0.5\textwidth]{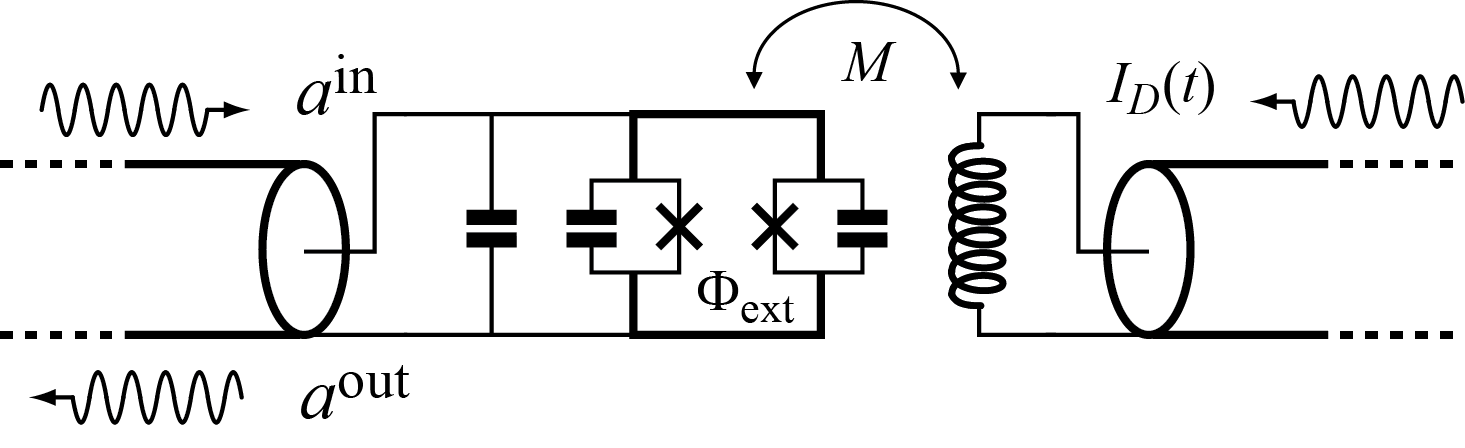}
\caption{\label{Fig_7}Parametrically driven oscillator
based on the property of the Josephson inductance of a DC-SQUID (two
Josephson junctions in parallel forming a loop, here represented by a
thicker line) to be modulated by the variation of an external flux $\Phi
_{ext}$. The modulation arises from an RF drive current $I_{D}\left(
t\right) =I_{D}^{RF}\cos \Omega t$ in the primary of a transformer that
creates though its mutual inductance $M$ a sinusoidal flux variation in the
loop of the DC-SQUID.}
\end{figure}

Another class of Josephson circuit implementing parametric amplifiers at
microwave frequencies is the RF-Flux-driven DC-SQUID (see Fig. \ref{Fig_7}).

It turns out that
this parametric drive can be implemented in Josephson circuits by taking a
DC-SQUID, which is formed by two nominally identical Josephson junctions in
parallel and modulating at the RF\ pump frequency the flux $\Phi _{ext}$
threading the superconducting loop between them (here the term DC refers to
the circulating current in the loop due to an external bias flux). One
exploits the functional form of the Josephson inductance of the DC-SQUID%
\begin{equation}
L_{J}^{SQUID}=\frac{L_{J}}{\cos \left\vert \pi \frac{\Phi _{ext}}{\Phi _{0}}%
\right\vert },
\end{equation}%
where $\Phi _{0}=h/2e$ is the flux quantum and $L_{J}/2$ is the Josephson
inductance of each individual junction. When%
\begin{equation}
\Phi _{ext}=\frac{\Phi _{0}}{4}\left[ 1+\varepsilon \cos \left( \Omega
t\right) \right],
\end{equation}%
with $\Omega $ close to the resonant frequency of the SQUID $1/\sqrt{%
C_{\Sigma }L_{J}^{\rm{SQUID}}}$ and $\varepsilon \ll 1$, one implements the
parametrically driven harmonic oscillator with relative frequency modulation
parameter $\mu _{r}=\pi \varepsilon /4$. This modulation is produced by a
drive current $I_{D}\left( t\right) =I_{D}^{RF}\cos \left( \Omega t\right) $
at the primary of the transformer coupling the transmission line of an RF
pump to the flux of the SQUID (see Fig. \ref{Fig_7}).  Classically, the parametrically
driven harmonic oscillator obeys the equation%
\begin{equation}
m\ddot{x}+\eta \dot{x}+m\omega _{0}^{2}x\left[ 1+\mu _{r}\cos \left( \Omega
t\right) \right] =f_{P}\left( t\right).
\end{equation}%
In contrast with the Duffing oscillator above, this system is described by a
fully linear, albeit time-dependent, equation. The drive, instead of
appearing as a force coupled directly to position, now modulates the spring
constant with a relative amplitude $\mu _{r}$. The system behaves as an
amplifier when the argument of the cosine modulation term is such that the
drive frequency $\Omega $ is close to the resonant frequency $\omega _{0}$.
In the weak damping limit $\eta \ll m\omega _{0}$, the quantum version of
this oscillator is directly a one-port, one-mode system described by our
degenerate amplifier Hamiltonian%
\begin{equation}
\frac{H}{\hbar }=\omega _{a}a^{\dag }a+\left[ g_{\rm{aa}}e^{i\Omega
_{\rm{aa}}t}a^{2}+\rm{H.c.}\right],
\end{equation}%
where%
\begin{eqnarray}
\Omega _{\rm{aa}} &=&\Omega,  \\
g_{\rm{aa}} &=&\mu _{r}\omega _{0}/4.
\end{eqnarray}

Note that now the drive frequency needs to be near twice the resonance
frequency of the amplified mode, unlike in the Duffing case, and it is thus
easier to decouple the pump tone from the weak signal to be amplified. This
type of amplifier has been implemented in several labs. \cite{Yamamoto_Tsai_2008, Johansson_Nori_2009, Zhong2013, Zhou_Esteve_2014, Mutus_Martinis_2014}

\subsection{Double-pumped degenerate parametric amplifier}
\label{sec_7c}
\begin{figure}
\centering
\includegraphics[width=0.5\textwidth]{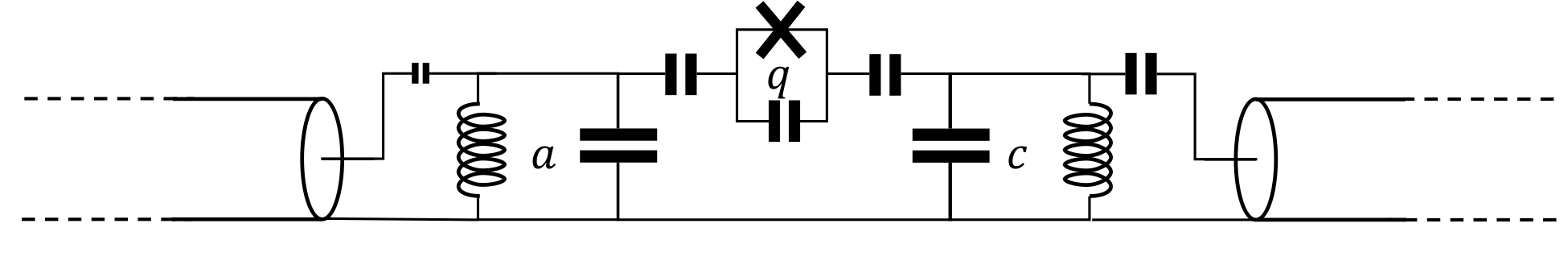}
\caption{\label{Fig_7_1} Implementation of a degenerate parametric amplifier using two cavity modes ($a, c$) coupled by a Josephson junction (mode $q$). A stiff pump and a weak resonant drive on mode $c$, together with the Josephson nonlinearity, gives rise the desired Hamiltonian given in Eq. \eqref{degham}. }
\end{figure}
Consider two cavity modes $a,c$ coupled by a single Josephson junction, treated as a qubit mode $q$ (see Fig. \ref{Fig_7_1}). The resonant frequencies (decay rates) of the three modes are denoted by $\omega_x(\kappa_x)$, $x= a,c,q$. By applying a stiff pump and a weak resonant drive on the mode $c$, for $\kappa_c\gg\kappa_a$, this system was used for stabilization of Schr\"odinger cat states. \cite{Leghtas_Devoret_2015} We now show that the same system, for $\kappa_a\sim\kappa_c$, acts as a degenerate parametric amplifier described in Secs. \ref{sec_4}, \ref{sec_5}. We will follow derivation given in the Supplementary Material of Ref. \onlinecite{Leghtas_Devoret_2015}. 

The total Hamiltonian of the system is given by
\begin{align}
H &= \omega_q q^\dagger q + \omega_a a^\dagger a + \omega_c c^\dagger c -E_J \big(\cos\varphi + \varphi^2/2)\nonumber\\&\quad +2\Re(\epsilon_p e^{-i\omega_p t} + \epsilon_c e^{-i\omega_dt})(c+c^\dagger),\\
\varphi &= {\varphi_q}^{\rm{ZPF}} (q+q^\dagger) + {\varphi_a}^{\rm{ZPF}} (a+a^\dagger) + \varphi^{\rm{ZPF}}_c (c+c^\dagger).\nonumber 
\end{align}
Here, $\epsilon_{p,c}$ are the drive strengths. The frequencies of the incident drives $\omega_p, \omega_d$ will be determined from the following calculation. Under rotating wave approximation and eliminating the fast dynamics (for details, see Supplementary Material of Ref. \onlinecite{Leghtas_Devoret_2015}), we arrive at the resultant Hamiltonian for the modes $a,c$:
\begin{align}
H &=\tilde\omega_a a^\dagger a+\tilde\omega_c c^\dagger c +(ig_2 a^\dagger c + \epsilon_c c^\dagger + {\rm{H.c.}})\nonumber\\&\quad-\frac{\chi_{\rm{aa}}}{2}{a^\dagger}^2a^2-\frac{\chi_{\rm{cc}}}{2}{c^\dagger}^2c^2-\chi_{\rm{ac}}a^\dagger ac^\dagger c.
\end{align}
Here, the self-Kerr, cross-Kerr and nonlinear couplings of the modes $a,c$ are respectively given by
\begin{align}
\chi_{\rm{mm}}&=\frac{E_J\varphi_m^4}{2\hbar},\ \chi_{\rm{mm'}}=\frac{E_J \varphi_m^2\varphi_{m'}^2}{\hbar},\ g_2 = \frac{\chi_{\rm{ac}}\xi_p^*}{2}
\end{align}
where $\xi_p = -i\epsilon_p/[\kappa_c/2+i(\omega_c-\omega_p)]$. We have omitted the Hamiltonian terms involving the qubit mode since it does not participate in the dynamics and merely provides the nonlinearity for the interaction between $a$ and $c$. Here, 
\begin{align}
\label{eqomegaa}
\tilde\omega_a &= \omega_a-\frac{\omega_d+\omega_p}{2}-\chi_{\rm{aa}}-\chi_{\rm{ac}}|\xi_p|^2, \\\label{eqomegab}\tilde\omega_c &= \omega_c-\omega_d-\chi_{\rm{cc}}-\chi_{\rm{cc}}|\xi_p|^2.
\end{align}
Here, the terms $\chi_{\rm{ac}}|\xi_p|^2$, $\chi_{\rm{cc}}|\xi_p|^2$ denote the AC-stark shift due to the presence of the incident pump. To arrive at the degenerate parametric amplifier Hamiltonian [Eq. \eqref{degham}], we solve for $\omega_p, \omega_d$ by setting $\tilde\omega_a=\tilde\omega_c=0$. The resultant Hamiltonian for $\tilde\omega_a=\tilde\omega_c=0$ corresponds to that of the degenerate paramp when the modes $a,c$ are considered in their respective rotating frames. The physical process that underlies the amplification mechanism can be understood easily for negligible cross-Kerr and self-Kerr couplings, when the frequency constraint becomes
\begin{equation}
\omega_c = \omega_d, \ 2\omega_a = \omega_p+\omega_d. 
\end{equation}
Thus, the amplification of the $a$-mode occurs when one photon of the pump (at $\omega_p$) and one photon of the drive (at $\omega_d$) are converted to two photons of the signal (at $\omega_a$).  

\subsection{Three-mode circuit employing the purely dispersive Josephson 3-wave
mixer}
\label{sec_7d}
\begin{figure}
\centering
\includegraphics[width=0.5\textwidth]{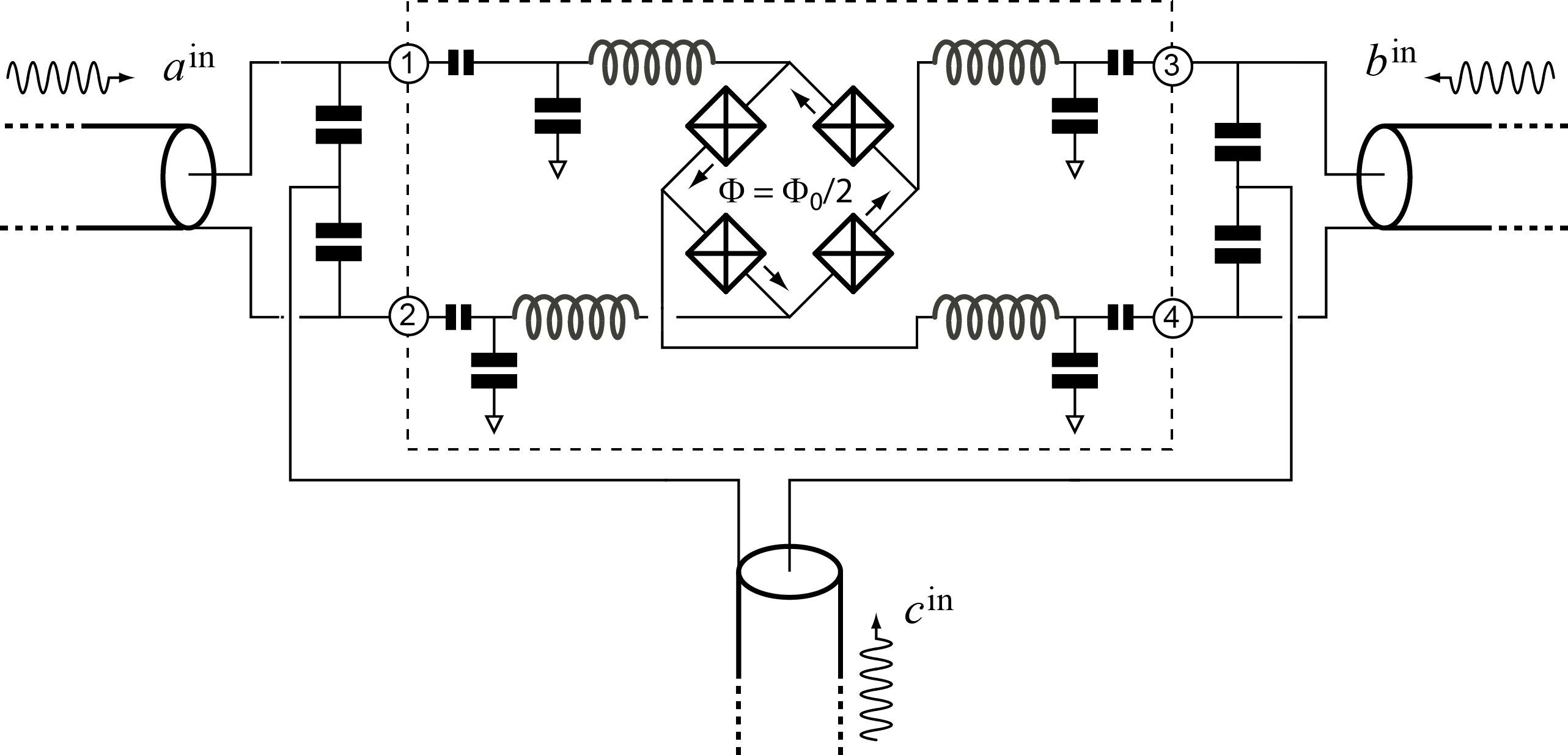}
\caption{\label{Fig_8}Schematic circuit of the
purely dispersive 3-wave mixer (dashed line) involving three microwave modes,
themselves coupled to three ports. The system functions as a non-degenerate
parametric amplifier with mode $a$ and $b$ playing the role of the signal 
and idler, while mode $c$ is used to couple in the pump tone. While there
are in principle four modes coupled by the junctions (cross inside a square,
denoting both the Josephson element and associated capacitance), the symmetry
of the circuit when the junctions are
identical, imposes that only three modes participate in the nonlinear interaction. A flux threading the ring of junctions induces a current (arrows) that replaces one of the four waves coupled by the junction.}
\end{figure}

We have just described several ways in which a circuit involving 
Josephson junctions can implement the degenerate parametric amplifier. The non-degenerate parametric amplifier can be implemented by a 3-mode, 3-port circuit employing four
junctions forming the so-called Josephson ring modulator, a purely
dispersive 3-wave mixer (see Fig. \ref{Fig_8} and further details in Ref. \onlinecite{Bergeal_Devoret_2010, Bergeal_Devoret_2010_a, Abdo_Devoret_2013_b}). 

Three microwave standing wave resonators are coupled by this
last element and are described by the Hamiltonian%
\begin{align}
\frac{H}{\hbar }&=\omega _{a}a^{\dag }a+\omega _{b}b^{\dag }b+\omega
_{c}c^{\dag }c\nonumber\\&\quad+g_{3}\left( a+a^{\dag }\right) \left( b+b^{\dag }\right)
\left( c+c^{\dag }\right),  \label{JPC_ideal_Hamiltonian}
\end{align}%
together with their port coupling $\kappa _{a}$, $\kappa _{b}$ and $\kappa
_{c}$. The frequency scales are such that%
\begin{equation}
\omega _{c}\gg \omega _{b}>\omega _{a}>\kappa _{c}\gg \kappa _{a}\simeq
\kappa _{b}\gg g_{3}.
\end{equation}%
The trilinear coupling term, treated as a perturbation, possesses the
precious property that it does not, at the lowest order, offset the
frequency of the quadratic terms when the modes are occupied by coherent
signals, unlike the Kerr term above $\frac{K}{2}a^{\dag }a\left( a^{\dag
}a-1\right) $. Other terms of higher order have been neglected in the
Hamiltonian (\ref{JPC_ideal_Hamiltonian}). They ensure that the system
remains stable when the amplitudes become large, as the trilinear coupling
renders by itself the Hamiltonian unstable. In the regime where the $c$ mode
is driven by a large coherent field $\alpha _{c}^{\rm{in}}e^{-i\Omega t}$, we can
neglect the fluctuating part of the corresponding operator. The previous
Hamiltonian can be treated as%
\begin{align}
\frac{H^{\rm{eff}}}{\hbar }&=\omega _{a}a^{\dag }a+\omega _{b}b^{\dag
}b\nonumber\\&\quad+2g_{ab}\left( a+a^{\dag }\right) \left( b+b^{\dag }\right) \cos
\left( \Omega _{ab}t+\theta \right),
\end{align}%
where%
\begin{eqnarray}
g_{ab}\cos \left( \Omega _{ab}t+\theta \right)  &=&g_{3}\Re\left[
\frac{\sqrt{\kappa _{c}}\alpha _{c}^{\rm{in}}e^{-i\Omega t}}{-i\left( \omega
_{ab}^{D}-\omega _{c}\right) +\kappa _{c}}\right],  \\
\Omega _{ab} &=&\Omega .
\end{eqnarray}%
When one works within the framework of the Rotating Wave Approximation and $%
\Omega _{ab}\simeq \omega _{a}+\omega _{b}$, the fast rotating terms can be
neglected and one recovers the Hamiltonian of the generic non-degenerate
parametric amplifier%
\begin{equation}
\frac{H^{\rm{NDPA}}}{\hbar }=\omega _{a}a^{\dag }a+\omega _{b}b^{\dag }b+\left[
g_{ab}abe^{i\left( \Omega _{ab}t+\theta \right) }+\rm{H.c.}\right].
\end{equation}
We note that the presence of higher order Kerr nonlinear terms in the Hamiltonian \cite{Bergeal_Devoret_2010, Bergeal_Devoret_2010_a} not described here lowers the gain of the device due to AC-stark shift (see discussion at the beginning of Sec. \ref{sec_5}). A remedy for this drawback has been proposed recently in Refs. \onlinecite{Zorin2016, Frattini2017}.

\section{Related other topics and future directions}
\label{7.5}
Before concluding, we point out several recent amplifier developments which are subject of current research, but were not discussed in this work.

First, a DC and AC flux-driven SQUID performing as an amplifier can be viewed as a nonlinear circuit element, referred to as the ``pumpistor", with a phase-sensitive impedance which can turn negative.\cite{Sundqvist2013} This approach was not presented in this paper since we focussed on the interaction between the signal and the pump, an effect that cannot be addressed by the ``pumpistor" idea.

Second, Ref. \onlinecite{Roy_Vijay_2015} proposes an impedance-engineering approach to evade the gain-bandwidth product limitation of amplifiers. In this approach, the nonlinear circuit element based on a Josephson junction is connected to a multi-pole environment. The resulting impedance seen by this element is modified by tuning this environment, thereby providing an additional control to enhance the bandwidth of the amplifier. This impedance-engineering approach is qualitatively different from the one presented in this paper. In contrast to starting with a given Hamiltonian and analyzing its scattering properties, Ref. \onlinecite{Roy_Vijay_2015} starts with a susceptibility function and subsequently, constructs the Hamiltonian and the circuit subjecting the susceptibility function to certain requirements. Thus, while our paper provides the `direct' scattering analysis of a given Hamiltonian, Ref. \onlinecite{Roy_Vijay_2015} provides the `inverse' scattering approach of Hamiltonian construction. Furthermore, the amplifier constructed in this way differs from those presented in this paper in another aspect. Due to the specific form demanded of the susceptibility function of Ref. \onlinecite{Roy_Vijay_2015}, the pumped nonlinear part of the circuit sees a fundamentally non-Markovian bath, unlike in the amplifiers described here. Of course, the input-output framework used in our work is powerful enough to model such a non-Markovian environment by considering additional degrees of freedom in the system intercalated between the non-linear element and the transmission line (see Part IV of Ref. \onlinecite{Breuer2002}). 

Third, as pointed out in the introduction, a key desired property of an amplifier is its unidirectionality [point (v) of the desiderata]. For the amplifiers described in the body of the paper, additional circuit elements such as circulators and isolators play this crucial function. However, they suffer from the drawback of being lossy and bulky components. Amplifiers which are intrinsically directional and do not suffer from the aforementioned drawbacks have been proposed and for some, realized \cite{Kamal_Devoret_2011, Abdo_Devoret_2014, Macklin_Siddiqi_2015, Ranzani_Aumentado_2015}. This is an actively researched topic and goes beyond the scope of this article. 

Fourth,  amplification using dissipation engineering evading both the gain-bandwidth compromise and the instability at the onset of parametric oscillation in amplifiers \cite{Metelmann_Clerk_2014, Ranzani_Aumentado_2015} were also not covered in this article.

\section{Summary}
\label{sec_8}
\begin{figure*}[p]
\centering
\includegraphics[width=\textwidth]{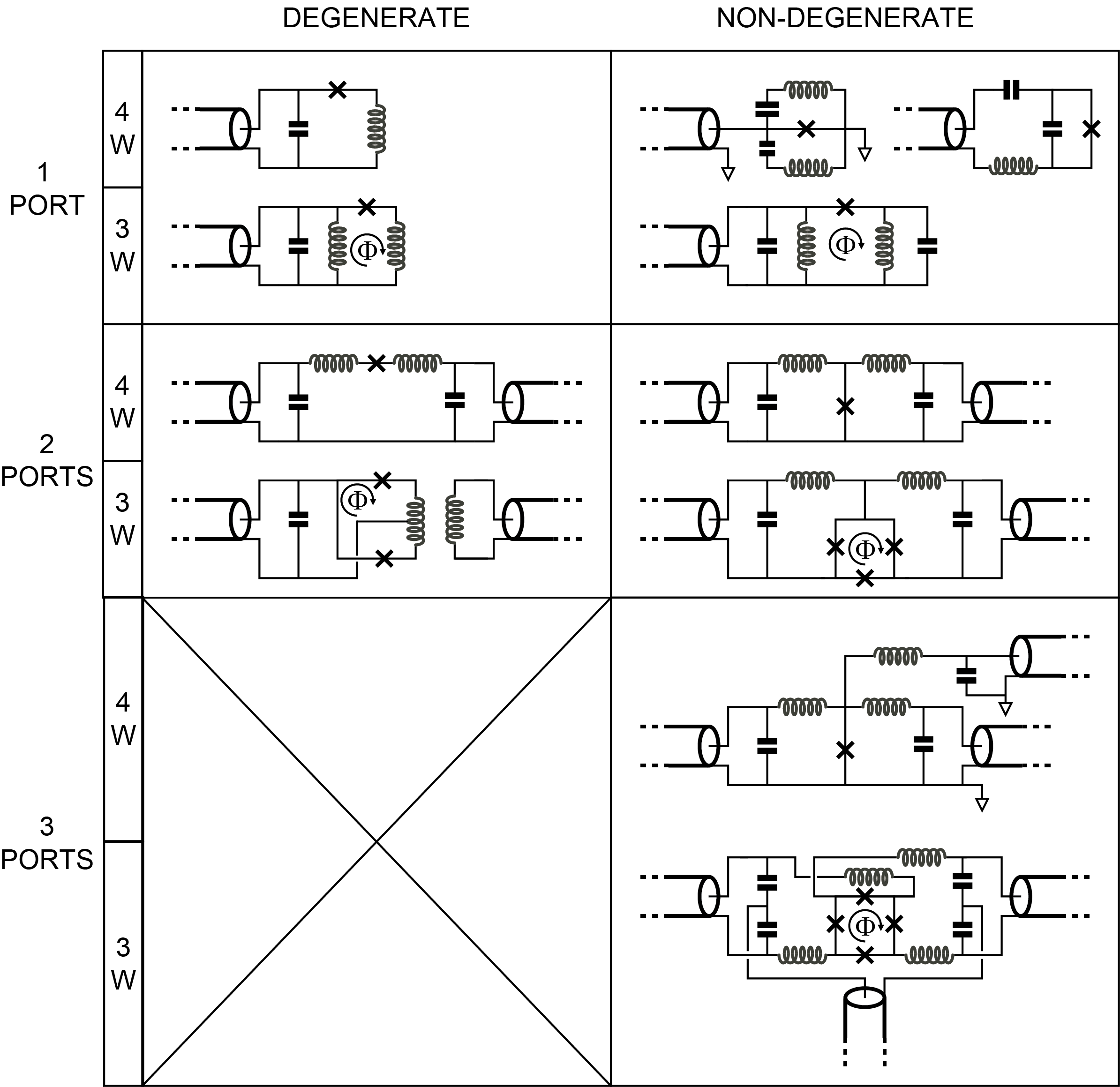}
\caption{\label{Fig_9} Minimalistic versions of the various Josephson circuits implementing parametric amplification of quantum signals. The circuits are classified according to the degenerate/non-degenerate character of the amplifying process (one or two standing modes). The four-wave (4W) or the three-wave (3W) labels characterize the mixing process  for the signal and idler waves taking place in the Josephson junctions. In the three-wave process, the place of one of the four-waves incident on the junction is replaced by a DC current generated by the externally applied flux $\Phi$. We also distinguish circuits by the number of ports through which the signal, idler and the pump waves are delivered. In the upper left corner (minimal complexity), the three waves are approximately at the same frequency and arrive through the same port, whereas in the lower right corner (maximal complexity), the three waves are both spatially and spectrally separated. In the upper right corner, we have represented two implementations of the 1-port, 4-wave, non-degenerate parametric amplifier. In the circuit on the left-handside, the two modes share a common junction but are symmetrically coupled to the port, whereas on the right-handside, the two modes are gauge-coupled and are asymmetrically coupled to the port. An important direction not fully explored by this table is the use of extra modes to do broadband parametric amplification with impedance engineering. \cite{Roy_Vijay_2015} In the example shown in the upper right corner, a one-port, non-degenerate four-wave device constructed from two modes can be engineered, by proper choice of its parameters,  to evade the constraint on the gain-bandwidth product discussed, for instance, in Ref. \onlinecite{Abdo_Devoret_2013_b}. }
\end{figure*}

To summarize, we have described in this article degenerate and non-degenerate parametric amplifiers based on Josephson junction circuits. The key organizing concept is the effective quadratic time-dependent Hamiltonian which comes in two forms: degenerate and non-degenerate, depending on whether the signal and idler waves occupy the same physical degree of freedom or two separate ones. In Fig. \ref{Fig_9}, we summarize the different circuit configurations leading, on one hand, to the degenerate case and, on the other hand, to the non-degenerate case (left and right columns, respectively). The figure also classifies circuits depending on the number of access ports, the simpler case being that of 1-port carrying the signal, idler and pump waves (upper left panels), while the case in which the signal, idler and the pump waves are separated in both temporally and spatially is shown in the bottom right panel. The circuit complexity increases when going from the upper left corner of Fig. \ref{Fig_9} to the lower right one. Moreover, we have described the linear scattering properties of these amplifiers in the stiff-pump approximation. Subsequently, we computed the effects of pump depletion and the phase-space properties of signal modes for both degenerate and non-degenerate cases. Finally, we reviewed some practical implementations of such amplifiers. 

\section*{Acknowledgments}
The authors are grateful to Benjamin Huard, Michael Hatridge, Archana Kamal and Katrina Sliwa for valuable discussions in the preparation of this review. AR acknowledges the support through the
ERC Consolidator Grant No. 682726 and the support of the Alexander von Humboldt Foundation. AR and MD acknowledge the support of by ARO under Grant No.W911NF-14-1-0011. 

\appendix

\section{Quantum signals propagating along a transmission line}
\label{appQuantumSignals}
Crudely speaking, quantum signals are electromagnetic excitations of a
transmission line that involve only a few photons. The  state of these
excitations must display some degree of quantum purity for the signals to
carry quantum information, which is the subject of interest in Josephson
circuits. In this section, we provide the basic mathematical background for
the concept of quantized electromagnetic excitations in the microwave domain. \cite{Gardiner_Zoller_2004, Wall_Milburn_2008}
Starting from the microscopic Hamiltonian of a transmission line, we arrive at the concept of left and right moving propagating photon flux operators (see Fig. \ref{disp}). As will be shown below, they obey the same commutation relations as those of elementary bosonic modes in vacuum. These operators are subsequently used to define the concept of a photonic excitation in a propagating wavepacket.  
\begin{figure}[H]
\centering
\includegraphics[width=0.3\textwidth]{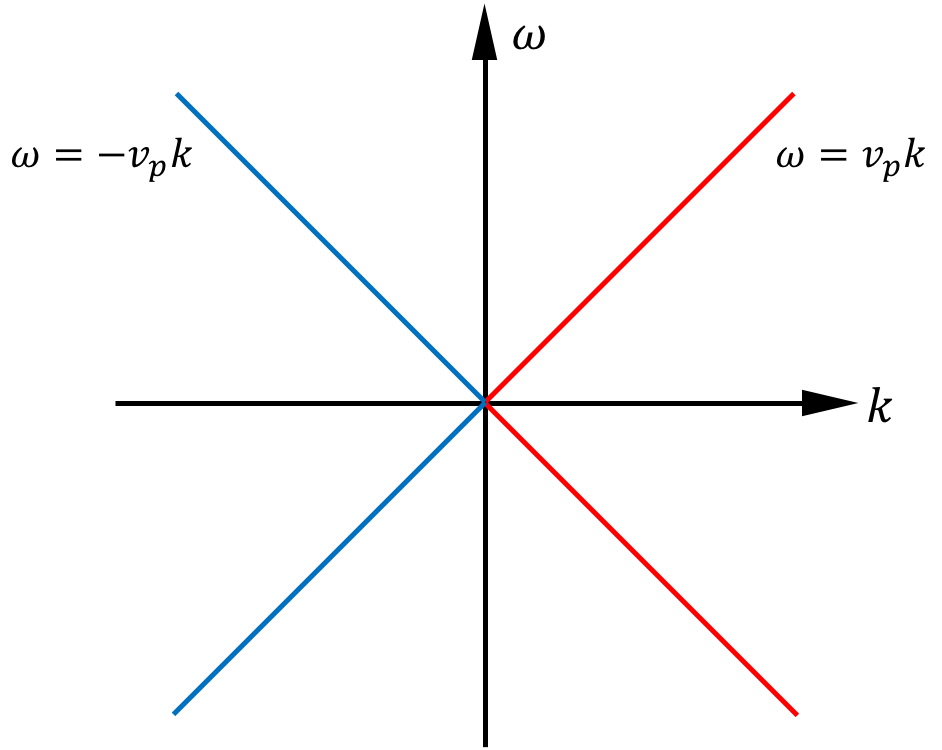}
\caption{\label{disp} Schematic of dispersion of quantized left-moving (shown in blue) and right-moving (shown in red) waves in a dispersionless one-dimensional medium (transmission line) with propagation velocity $v_p$. Reflection about the vertical axis ($k=0$) axis corresponds to transformation of a left-moving wave into a right-moving wave and vice versa, while reflection about the horizontal axis ($\omega =0$) corresponds to Hermitian conjugation (see below for more details). }
\end{figure}

\subsection{Hamiltonian description of a quantum transmission line}
\begin{figure}[H]
\centering
\includegraphics[width = 0.5\textwidth]{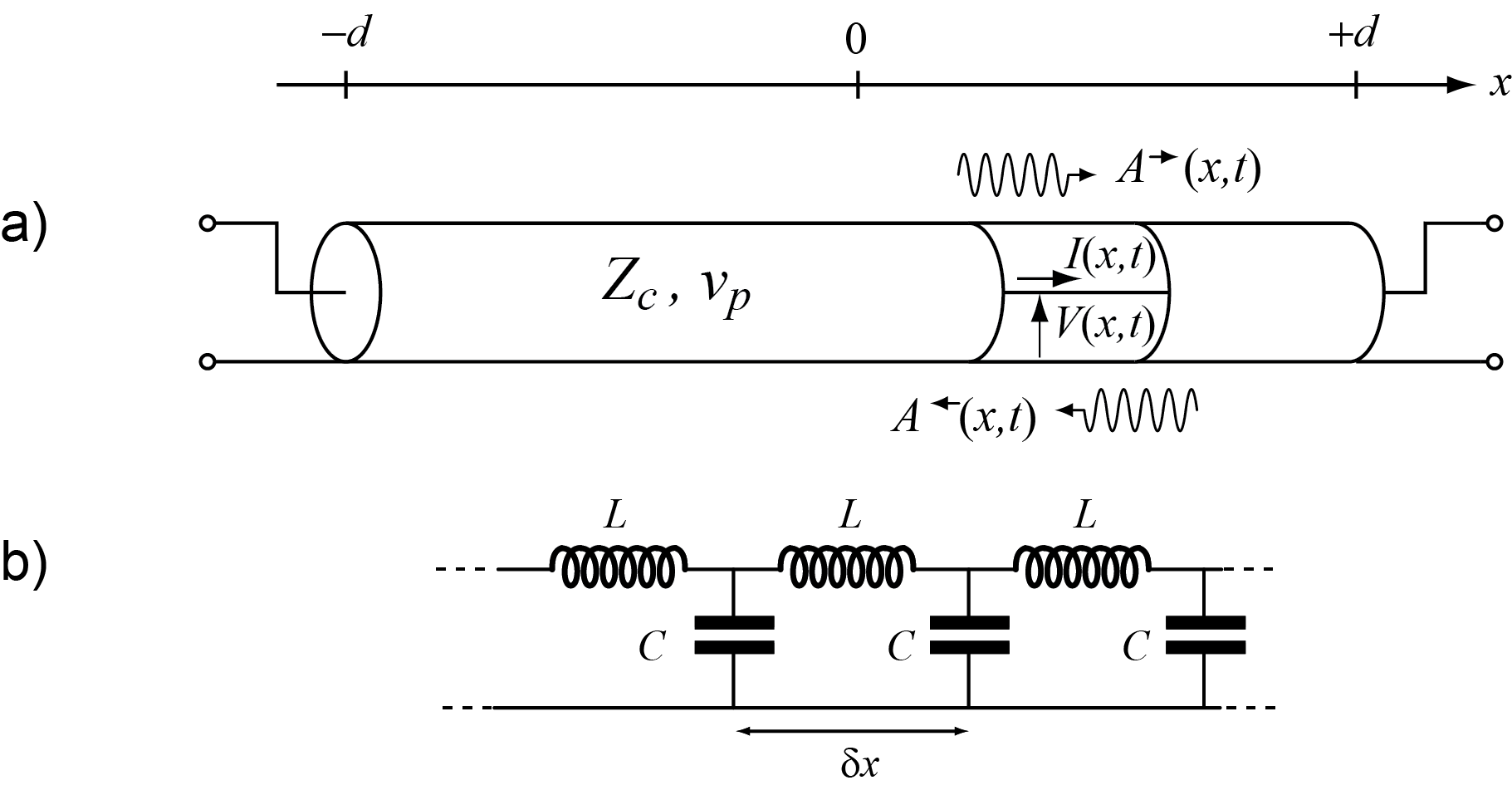}
\caption{\label{Fig_1} (a) Electromagnetic transmission line implemented as a
coaxial cable. The parameter $x$ denotes the position along the line, $I$
denotes the current along the line in the positive direction and $V$ the
voltage between the inner and outer conductors. The characteristic impedance
and the propagation velocity are denoted by $Z_{c}$ and $v_{p}$,
respectively. The line has a continuous density of modes in the limit where
its length $2d\rightarrow \infty $. In (b), a ladder circuit model with cell
dimension $\protect\delta x$ models the infinite transmission line. Its
capacitance and inductance per unit length are given by $L_{\ell }=L/\protect%
\delta x$ and $C_{\ell }=C/\protect\delta x$, respectively. In the limit
where the signal frequency $\protect\omega $ is small compared to $1/\protect%
\sqrt{LC}$, $Z_{c}=\protect\sqrt{L/C}$ and $v_{p}=1/\protect\sqrt{L_{\ell
}C_{\ell }}$. }
\end{figure}

Here we follow a route found in previous works (see Chap. 3 of Ref. \onlinecite{Drummond_Ficek_2013}, Ref. \onlinecite{Gardiner_Zoller_2004}, and Ref. \onlinecite{Clerk_Schoelkopf_2010}) with some clarifications particularly necessary for electrical circuits that we hope the reader will find useful. Consider an infinite transmission line, a one-dimensional
electromagnetic medium characterized by a propagation velocity $v_{p}$ and a
characteristic impedance $Z_{c}$. A microwave coaxial line serves as the
canonical example of such a medium (see Fig. \ref{Fig_1}). 

Position along the line is indexed by the real number $x\in ( -\infty ,+\infty )$. We suppose that the line is ideal, with both $v_{p}$ and $Z_{c}$ independent of frequency $\omega $. The TEM modes propagating on this transmission line can be equivalently described by propagating modes sustained by the infinite LC ladder shown in Fig. \ref{Fig_1} (b), in the limit when the wavelength of the propagating modes are much larger than the size of the unit cell. The inductance and capacitance per unit length are given by: $L_\ell = L/{\delta x}$ and $C_{\ell} = C/{\delta x}$ respectively \footnote{This model of a coaxial transmission line is due to Nyquist and corresponds to an actual microscopic description of the line, where $L_\ell, C_\ell$ correspond to the inductance and capacitance per unit length of the line. This should be differentiated from the Caldeira-Leggett model of resistance and impedance in terms of an infinite number of LC oscillators. The Caldeira-Leggett model describes the effective behavior of the resistor in terms of the LC oscillators even though underlying microscopic description of the resistor has nothing to do with LC oscillators.}. In terms of $L_\ell, C_\ell$, the propagating velocity and the characteristic impedance are given by: 
\begin{equation}
\label{trans_line_char}
v_p= \frac{1}{L_\ell C_\ell}, \ Z_c = \sqrt{\frac{L_\ell}{C_\ell}}. 
\end{equation}

Following Ref. [\onlinecite{Devoret_1997}], we define a flux operator: 
\begin{equation}
\label{trans_line_flux_defn}
\Phi(x,t) = \int_{-\infty}^t dt' V(x,t'),
\end{equation}
where $V(x,t)=\partial_t\Phi(x,t)$ is the local voltage operator at position $x$ on the transmission line at time $t$. The average voltage drop across a segment of length $\delta x$ with inductance $L_\ell\delta x$ is $-\delta x\partial_x\partial_t\langle\Phi(x,t)\rangle$. The average flux through the inductance is given by $-\delta x\partial_x \langle\Phi(x,t)\rangle$ and the operator for the current flowing through the inductance is given by the usual relation:
\begin{equation}
\label{trans_line_current_defn}
I(x,t) = -\partial_x\Phi(x,t)/L_\ell.
\end{equation}
The Lagrangian that describes the system is given by: 
\begin{eqnarray}
{\cal L}_{\rm{line}} &=& \int_{-\infty}^\infty\ dx \bigg\{\frac{C_\ell}{2}\Big(\frac{\partial \Phi}{\partial t}\Big)^2- \frac{1}{2L_\ell}\Big(\frac{\partial \Phi}{\partial x}\Big)^2\bigg\}.
\end{eqnarray}
which, through the Euler-Lagrange's equation of motion, results in the dispersionless wave propagation equation: 
\begin{equation}
\frac{\partial^2\Phi}{\partial x^2}-\frac{1}{v_p^2}\frac{\partial^2\Phi}{\partial t^2}=0.
\end{equation}
The canonical conjugate momentum is the charge density $\Pi(x,t)$: 
\begin{equation}
\Pi(x,t)\equiv \frac{\partial{ \cal L}_{\rm{line}}}{\partial(\partial_t\Phi)} = C_\ell\frac{\partial \Phi}{\partial t} = C_\ell V(x,t)
\end{equation}
and thus, the Hamiltonian describing the transmission is given as: 
\begin{equation}
H_{\rm{line}} = \int_{-\infty}^\infty dx \bigg\{\frac{1}{2C_\ell}\Pi(x,t)^2+ \frac{1}{2L_\ell}\Big(\frac{\partial \Phi}{\partial x}\Big)^2\bigg\}.
\end{equation}
We define the following Fourier transforms: 
\begin{eqnarray}
\label{trans_line_ft_defn}
 q(\omega,t)&\equiv& \frac{1}{\sqrt{2\pi v_p}}\int_{-\infty}^\infty dx\ \Phi(x,t) e^{-i\omega x/v_p},\\
 p(\omega,t)&\equiv& \frac{1}{\sqrt{2\pi v_p}}\int_{-\infty}^\infty dx\ \Pi(x,t)e^{-i\omega x/v_p}.
\end{eqnarray}
Here, the variable $\omega/v_p$ denotes the wave-vector of the spatial Fourier component and is {\it not} a frequency (the reason for this notation choice will become clear later). The positive and negative values of $\omega$ indicate wave-momentum in $+x$ and $-x$ direction, respectively (the red and blue lines of Fig. \ref{disp}). For a mode propagating with wave-vector $\omega/v_p$, the energy is given by $\hbar|\omega|$ (see below). Also, 
note that $p(\omega, t)$, and \textit{not} $q(\omega,t)$, is the Fourier transform of the charge density. We adopt this notation because the flux and charge operators, $\Phi(x,t)$ and $\Pi(x,t)$, respectively correspond to the position and momentum operators of the equivalent mechanical system. This choice is appropriate because there are nonlinear, non-dissipative inductances, but no nonlinear, non-dissipative capacitances. Since $\Phi(x,t), \Pi(x,t)$ are Hermitian operators, it follows trivially that: 
\begin{equation}
q(\omega,t)^\dagger = q(-\omega,t), \ p(\omega,t)^\dagger = p(-\omega,t).
\end{equation}
In terms of the Fourier transformed operators, the Hamiltonian is given by: 
\begin{eqnarray}
{\cal H }_{\rm{line}} &=& \int_{-\infty}^\infty d\omega \bigg\{\frac{1}{2C_\ell}p(\omega,t)p(-\omega,t)\nonumber\\&& + \frac{\omega^2C_\ell}{2}q(\omega,t)q(-\omega,t)\bigg\},
\end{eqnarray}
where we have used Eq. \eqref{trans_line_char}. Next, we define $\textgoth{a}(\omega,t)$, which plays the role of annihilation operator for different modes of the transmission line: 
\begin{equation}
\textgoth{a}(\omega,t) = \sqrt{\frac{|\omega|C_\ell}{2\hbar}}q(\omega,t) + \frac{i}{\sqrt{2\hbar|\omega| C_\ell}}p(\omega,t).
\end{equation}
These $\textgoth{a}(\omega,t)$ operators will be used in turn to define the propagating field operators. Note that with this definition, $\textgoth{a}(\omega,t)^\dagger \neq \textgoth{a}(-\omega,t)$. Thus, the Hamiltonian can be rewritten in terms of these operators as 
\begin{equation}
H_{\rm{line}} = \int_{-\infty}^\infty d\omega \frac{\hbar|\omega|}{2}\Big\{\textgoth{a}(\omega,t)^\dagger \textgoth{a}(\omega,t) + \textgoth{a}(\omega,t)\textgoth{a}(\omega,t)^\dagger\Big\}.
\end{equation}
Next, we use the canonical quantization relation for the continuous field operators: 
\begin{equation}
\big[\Phi(x,t),\Pi(x',t)\big] = i\hbar\delta(x-x'), 
\end{equation}
which in the Fourier domain, becomes
\begin{equation}
\big[q(\omega,t),p(\omega',t)\big] = i\hbar\delta(\omega + \omega').
\end{equation}
This leads to the following standard commutation relation for the annihilation field operator $\textgoth{a}(\omega,t)$: 
\begin{eqnarray}
\big[\textgoth{a}(\omega,t),\textgoth{a}(\omega',t)^\dagger\big]&=&\delta(\omega-\omega'), \nonumber\\ \big[\textgoth{a}(\omega,t),\textgoth{a}(\omega',t)\big]&=&0.
\end{eqnarray}
The Heisenberg equation of motion for the operator $\textgoth{a}(\omega,t)$ is given by: 
\begin{eqnarray}
\frac{d\textgoth{a}(\omega,t)}{dt} &=& -\frac{i}{\hbar}\big[\textgoth{a}(\omega,t),H_{\rm{line}}\big]\nonumber\\&=& -i|\omega|\textgoth{a}(\omega,t),
\end{eqnarray}
which can be solved to give: 
\begin{equation}
\textgoth{a}(\omega,t) = e^{-i|\omega| (t-t_0)}\textgoth{a}(\omega,t_0),
\end{equation}
where $t_0$ is some initial time (eventually, we will take $t_0$ to be $-\infty$). This, together with Eq. \eqref{trans_line_ft_defn}, leads to: 
\begin{eqnarray}
q(\omega,t) &=& \sqrt{\frac{\hbar}{2|\omega|C_\ell}}\Big\{\textgoth{a}(\omega,t_0)e^{-i|\omega|(t-t_0)}\nonumber\\&& + \textgoth{a}(-\omega,t_0)^\dagger e^{i|\omega|(t-t_0)}\Big\}\\p(\omega,t) &=& -i\sqrt{\frac{\hbar|\omega|C_\ell}{2}}\Big\{\textgoth{a}(\omega,t_0)e^{-i|\omega|(t-t_0)}\nonumber\\&& - \textgoth{a}(-\omega,t_0)^\dagger e^{i|\omega|(t-t_0)}\Big\}.
\end{eqnarray}
Now, we can solve for the field operators $\Phi(x,t), \Pi(x,t)$ arriving at: 
\begin{align}
\Phi(x,t) &= \sqrt{\frac{Z_c}{2\pi}}\int_{-\infty}^\infty d\omega \sqrt{\frac{\hbar}{2|\omega|}}e^{i\omega x/v_p}\Big\{\textgoth{a}(\omega,t_0)e^{-i|\omega|(t-t_0)} \nonumber\\&\qquad+ {\rm{H.c.}}\Big\}\nonumber\\&=\sqrt{\frac{Z_c}{2\pi}}\int_{0}^\infty d\omega \sqrt{\frac{\hbar}{2\omega}}\Big\{\textgoth{a}(\omega,t_0)e^{-i\omega(t-x/v_p)}e^{i\omega t_0} \nonumber\\&\quad+ \textgoth{a}(-\omega,t_0) e^{-i\omega(t+x/v_p)}e^{i\omega t_0} + \rm{H.c.}\Big\},
\end{align}
\begin{align}
\Pi(x,t) &= -\frac{i}{v_p\sqrt{2\pi Z_c}}\int_{-\infty}^\infty d\omega \sqrt{\frac{\hbar|\omega|}{2}}e^{i\omega x/v_p}\nonumber\\&\qquad\Big\{\textgoth{a}(\omega,t_0)e^{-i|\omega|(t-t_0)} -{\rm{H.c.}}\Big\}\nonumber\\&=-\frac{i}{v_p\sqrt{2\pi Z_c}}\int_{0}^\infty d\omega \sqrt{\frac{\hbar\omega}{2}}\Big\{\textgoth{a}(\omega,t_0)e^{-i\omega(t-x/v_p)}\nonumber\\&\quad e^{i\omega t_0}+ \textgoth{a}(-\omega,t_0) e^{-i\omega(t+x/v_p)}e^{i\omega t_0} - \rm{H.c.}\Big\}.
\end{align}
Note that, as expected, the operators $\Phi(x,t),\Pi(x,t)$ are Hermitian and have two traveling wave components corresponding to the two opposite  traveling directions (see Fig. \ref{disp}). From these quantities, it is easy to calculate the voltage and current operators using Eqs. \eqref{trans_line_flux_defn}, \eqref{trans_line_current_defn}, leading to: 
\begin{eqnarray}
V(x,t) &=& V^\rightarrow(x,t) + V^\leftarrow(x,t),\nonumber\\ I(x,t) &=& I^\rightarrow(x,t) - I^\leftarrow(x,t),\\I^\rightleftarrows(x,t) &=& \frac{1}{Z_c}V^\rightleftarrows(x,t),
\end{eqnarray}
where
\begin{align}
V^\rightleftarrows(x,t) &= -i\sqrt{\frac{Z_c}{2\pi}}\int_0^\infty d\omega \sqrt{\frac{\hbar\omega}{2}}\nonumber\\&\quad\Big\{\textgoth{a}(\pm\omega, t_0)e^{-i\omega(t\mp x/v_p)}e^{i\omega t_0}-\rm{H.c.}\Big\}.
\end{align}
Here, we have expressed the current and voltage operators as superpositions of those operators propagating in opposite directions. As expected, the right(left)-propagating waves involve the operators $\textgoth{a}(\omega,t_0)$ with positive (negative) wave-vectors. Next, we define the propagating wave amplitude operators in terms of these propagating current and voltage operators: 
\begin{eqnarray}
\label{trans_line_trav_wav_ampl_defn}
A^\rightleftarrows(x,t) &=& A^\rightleftarrows(t\mp x/v_p)=\frac{1}{2}\Big(\frac{V}{\sqrt{Z_c}}(x,t) \pm \sqrt{Z_c}I(x,t)\Big),\nonumber\\A^\rightleftarrows (x,t) &=& \frac{-i}{\sqrt{2\pi}} \int_0^\infty d\omega \sqrt{\frac{\hbar\omega}{2}}\Big\{\textgoth{a}(\pm\omega, t_0)e^{-i\omega(t\mp x/v_p)}e^{i\omega t_0}\nonumber\\&&-\rm{H.c.}\Big\},
\end{eqnarray}
Eq. \eqref{trans_line_trav_wav_ampl_defn} shows that the spatial dependence can be obtained trivially from $A^\rightleftarrows(t)\equiv A^\rightleftarrows(x=0,t)$ by setting $t\rightarrow t\mp x/v_p$: 
\begin{eqnarray}
\label{trans_line_A_x_0_defn}
A^\rightleftarrows (t) &=& \frac{-i}{\sqrt{2\pi}} \int_0^\infty d\omega \sqrt{\frac{\hbar\omega}{2}}\Big\{\textgoth{a}(\pm\omega, t_0)e^{-i\omega (t-t_0)}\nonumber]\\&&-\rm{H.c.}\Big\}.
\end{eqnarray}
These simple and physical propagating wave-amplitudes satisfy the following commutation relation: 
\begin{align}
\label{trans_line_A_t_comm_reln}
\big[A^{l_1}(t_1),A^{l_2}(t_2)\big] &= \frac{i\hbar}{2}\frac{d}{d(t_1-t_2)}\delta(t_1-t_2)\delta_{l_1l_2},
\end{align}
where $l_1,l_2=0,1$ according as the denote the direction of propagation is $\rightarrow, \leftarrow$. These traveling wave amplitudes have also been introduced in earlier treatments of input-output theory. \cite{Yurke_Denker_1984, Gardiner_Collett_1985, Gardiner_Zoller_2004, Drummond_Ficek_2013} The wave amplitude, whose dimension is [watt]$^{1/2}$, is such that its square is the energy flux of waves traveling in the direction indicated by the arrow. These propagating wave amplitude operators describe the power-flow in the transmission line. The net power flowing in the $+x$ direction is given by: 
\begin{equation}
P = \langle A^\rightarrow(x,t)\rangle^2-\langle A^\leftarrow(x,t)\rangle^2,
\end{equation}
and is equivalent to the usual Poynting vector of electrodynamics. In terms of these propagating wave-amplitudes, the Hamiltonian can be written as: 
\begin{equation}
H_{\rm{line}} = \frac{1}{v_p}\int_{-\infty}^\infty dx \big\{A^\rightarrow(x,t)^2 + A^\leftarrow(x,t)^2\big\},
\end{equation}
which expresses the fact that the total energy is the sum of the energies of these propagating waves. 

In earlier treatments of dispersionless quantum transmission lines (Chap. 3 of Ref. \onlinecite{Drummond_Ficek_2013}, Chap. 3 of Ref. \onlinecite{Gardiner_Zoller_2004}, and Ref. \onlinecite{Clerk_Schoelkopf_2010}), it is at this point where the rotating wave approximation (RWA) is made before discussing traveling photon wavepackets. These treatments are sufficient when the spectral width of the traveling pulse is much smaller than the center frequency, which is usually the case in usual quantum optical systems operating with center frequencies in the THz range. However, for microwave circuit-QED systems operating with center frequencies in the GHz range, it is easy to conceive of a temporal wave packet that is not well-described by these approximations. Therefore, we go a step further and introduce the concept of a traveling-wave field-ladder operator without making RWA. This is described below. 

Define the Fourier transforms of the propagating wave amplitudes as 
\begin{equation}
\label{trans_line_ft_A_defn}
A^\rightleftarrows [\omega] \equiv \frac{1}{\sqrt{2\pi }}\int_{-\infty}^\infty dt A^\rightleftarrows (t)e^{i\omega t},
\end{equation}
Here, we have used the squared parenthesis to distinguish the case when the Fourier transform is taken with respect to time from the earlier case when the Fourier transform was with respect to space. Since $A^\rightleftarrows(t)$ in a Hermitian operator, it follows that $A^\rightleftarrows [\omega]^\dagger = A^\rightleftarrows [-\omega]$. Using Eqs. \eqref{trans_line_A_x_0_defn}, \eqref{trans_line_ft_A_defn}, one readily obtains: 
\begin{eqnarray}
A^\rightleftarrows[\omega] &=&  -i\sqrt{\frac{\hbar|\omega|}{2}}\textgoth{a}(\pm|\omega|,t_0)e^{+i|\omega| t_0},\ \omega>0,\nonumber\\&=&i\sqrt{\frac{\hbar|\omega|}{2}}\textgoth{a}(\pm|\omega|,t_0)^\dagger e^{-i|\omega| t_0},\ \omega<0,
\end{eqnarray}
where the upper (lower) sign in front of $|\omega|$ correspond to right (left) moving waves. The two signs correspond to the two branches of the blue (leftarrow) and red (rightarrow) lines representing the propagating modes in Fig. \ref{disp}, while the argument of $A$ denotes the position along the branch. 
These result in the following commutator relation:
\begin{equation}
\Big[A^{l_1}[\omega_1],A^{l_2}[\omega_2]\Big] = \frac{\hbar(\omega_1-\omega_2)}{4}\delta(\omega_1 + \omega_2)\delta_{l_1l_2},
\end{equation}
which is the frequency-domain counterpart of Eq. \eqref{trans_line_A_t_comm_reln} and $l_1,l_2$ stand for the sense of propagation. 

Now, we are ready to define the traveling-wave field ladder operators:
\begin{equation}
a^{l}[\omega] \equiv\frac{1}{\sqrt{\hbar \left\vert \omega \right\vert /2}}A^{l}[\omega].
\end{equation}%
These frequency-domain, traveling-wave, field ladder operators $a^{l} \left[ \omega \right] $ have commutation relations bearing a marked
resemblance to the ladder operators of a set of standing wave harmonic oscillators (in the continuum):
\begin{widetext}
\begin{equation}
\boxed{\big[ a^{l_{1}}[\omega _{1}] ,a^{l_{2}}[\omega _{2}]\big] =\mathrm{sgn}\left( \omega _{1}-\omega _{2}\right) \delta \left( \omega _{1}+\omega _{2}\right) \delta _{l_{1}l_{2}}.}
\end{equation}
\end{widetext}
This formula represents the central result of this subsection. Note that
\begin{equation}
a^{l}[\omega] ^{\dag }=a^{l}[ -\omega].
\end{equation}%
In terms of the reciprocal space field operators $\textgoth{a}(\omega,t)$, these traveling-wave field ladder operators are given by: 
\begin{eqnarray}
a^\rightleftarrows[\omega] &=& -i\textgoth{a}(\pm|\omega|,t_0)e^{+i|\omega| t_0}, \ \omega>0,\\&=&i\textgoth{a}(\pm|\omega|,t_0)^\dagger e^{-i|\omega| t_0}, \ \omega<0.
\end{eqnarray}
As expected, we see that the right(left)-propagating waves involve the field operators $\textgoth{a}(\omega,t)$ with positive (negative) wavevectors. The above equations establish the connection between the traveling field operators $a^{l}[\omega]$ defined on the upper and lower half of Fig. \ref{disp}. 

Going back to the time domain, one can evaluate the propagating traveling-wave field ladder operators  at $x=0$ to be: 
\begin{eqnarray}
\label{trans_line_a_t_defn}
a^\rightleftarrows(t) &=& \frac{1}{\sqrt{2\pi }}\int_{-\infty}^\infty d\omega a^\rightleftarrows[\omega]e^{-i\omega t}\nonumber\\&=& \frac{-i}{\sqrt{2\pi}}\int_0^\infty d\omega \big\{\textgoth{a}(\pm\omega, t_0) e^{-i\omega (t-t_0)}\nonumber\\&& -\textgoth{a}(\pm\omega, t_0)^\dagger e^{i\omega (t-t_0)}\big\} .
\end{eqnarray}
It is important to note that $a^\rightleftarrows(t)$ is a Hermitian operator, and it satisfies the commutation relation: 
\begin{equation}
\left[ a^{l_{1}}\left( t_{1}\right) ,a^{l_{2}}\left( t_{2}\right) \right]
=\frac{i}{\pi} {\rm{p.p}}\frac{1}{t_2-t_1}\delta _{l_{1}l_{2}}.
\end{equation}
In Appendix \ref{appQLE}, we will make the rotating wave approximation on this propagating field-operator $a^\rightleftarrows(t)$, which amounts to dropping the second term involving in $\textgoth{a}(\omega,t_0)^\dagger$ in Eq. \eqref{trans_line_a_t_defn}. 

With the results obtained in this subsection, we can proceed to define propagating photon excitations of the transmission line. This is done below. We emphasize that for defining the photon excitations of the line, we will not need the RWA.   

\subsection{Definition of a traveling photon wavepacket}
In order to properly define the photons of the line, one needs to introduce
an orthonormal signal basis consisting of \textquotedblleft
first-quantization\textquotedblright\ wavelets\cite{Mallat_1999} $w_{mp}^{l}\left( t\right) $
such that%
\begin{eqnarray}
\int\nolimits_{-\infty }^{+\infty }dt\;w_{m_{1}p_{1}}^{l_{1}}\left( t\right)
w_{m_{2}p_{2}}^{l_{2}}\left( t\right) ^{\ast } &=&\delta
_{m_{1},m_{2}}\delta _{p_{1},p_{2}}\delta _{l_{1},l_{2}}, \nonumber\\
w_{mp}^{l}\left( t\right) ^{\ast } &=&w_{-mp}^{l}\left( t\right) ,\nonumber \\
\sum_{m=-\infty }^{+\infty }\sum_{p=-\infty }^{+\infty }w_{mp}^{l}\left(
t_{1}\right) w_{-mp}^{l}\left( t_{2}\right) &=&\delta \left(
t_{1}-t_{2}\right).
\end{eqnarray}%
The pair of indices $\left( \left\vert
m\right\vert ,p\right) \in \mathbb{N}^{+}\times \mathbb{Z}$ defines a
propagating temporal mode of the line, and the combined amplitudes of the
two corresponding wavelets can be seen as an elementary degree of freedom of
the field. There are two conjugate wavelets per mode since the phase space
of each mode is bi-dimensional.

It is necessary to request that the support of $%
w_{mp}^{l}\left[ \omega \right] $, the Fourier transform of $%
w_{mp}^{l}\left( t\right) $, is entirely contained in the positive frequency
sector if $m>0$ and in the negative frequency sector if $m<0$.%
\begin{eqnarray}
w_{mp}^{l}\left[ \omega \right] &=&w_{mp}^{l}\left[ \omega \right] \Theta
\left( \omega \right) \qquad \textrm{if}\qquad m>0, \\
w_{mp}^{l}\left[ \omega \right] &=&w_{mp}^{l}\left[ \omega \right] \Theta
\left( -\omega \right) \qquad \textrm{if}\qquad m<0.
\end{eqnarray}%
In these last expressions, $\Theta \left( \omega \right) $ is the Heaviside
function \footnote{The index value $m=0$ corresponds to special wavelets that have to be treated separately.}.

This complete wavelet basis is a purely classical signal processing concept
and its existence solely results from the property of the signals to be
square-integrable functions. Any continuous signal $f\left( t\right) $ such
that $\int_{-\infty }^{+\infty }\left\vert f\left( t\right) \right\vert
^{2}dt<\infty$
can indeed be decomposed into a countable infinite number of elementary
signals 
\begin{eqnarray}
f\left( t\right) &=&\sum_{m=-\infty }^{+\infty }\sum_{p=-\infty }^{+\infty
}f_{-mp}w_{mp}\left( t\right), \\
f_{mp} &=&\int\nolimits_{-\infty }^{+\infty }dt\;w_{mp}\left( t\right)
f\left( t\right).\label{fmus}
\end{eqnarray}%
A common example of such a wavelet is the Shannon wavelet

\begin{equation}
\mathfrak{w}_{mp}\left( t\right) =2\sqrt{\frac{\tau }{2\pi }}\frac{\sin
\left( \frac{\pi }{\tau }(t-p\tau )\right)}{t}e^{i2\pi mt/\tau }
\end{equation}%
whose Fourier transform is 
\begin{equation}
\mathfrak{w}_{mp}\left[ \omega \right] =\sqrt{\frac{\tau }{2\pi }}1_{\frac{%
2\pi }{\tau }(m-1/2),\frac{2\pi }{\tau }(m+1/2)}\left( \omega \right)
e^{ip\omega \tau },
\end{equation}%
where $1_{x_{1},x_{2}}\left( x\right) $ is the indicator function which is $%
0 $ everywhere except in the interval $\left[ x_{1},x_{2}\right] $, where
its value is unity. Many other useful bases, involving more continuous
wavelets, exist. \cite{Mallat_1999} In the above example, the center frequency
and time location of the wavelet is $2\pi m/\tau $ and $p\tau $, respectively
(in order to form a complete basis, the pitch in frequency $\Delta \omega $
and pitch in time $\Delta t$ of the wavelet basis has to satisfy $\Delta
\omega .\Delta t\leq 2\pi $) . 

The discreteness of the signal component indices is the justification for
the term \textquotedblleft first-quantization\textquotedblright\ and no
quantum mechanics is involved here since all functions are at this stage
c-number valued. Second-quantization intervenes when we define the discrete ladder
field operators, with indices $m>0$ and $p$
\begin{eqnarray}
\bm{\psi}_{mp}^{l} &=& \int_{-\infty }^{+\infty }d\omega
w_{mp}^{l}\left( \omega \right) a^{l}[\omega ]\\&=&\int\nolimits_{-\infty }^{+\infty }dt
w_{mp}^{l}(-t) a^{l}(t)\\
\bm{\psi}_{-mp}^{l} &=&\bm{\psi}_{mp}^{l\dag }.
\end{eqnarray}
We introduce the short-hand $\mu = (l, |m|, p)$ as the index of the spatio-temporal mode, also called the flying oscillator. The photon-number operator is given by:  
\begin{equation}
\mathbf{n}_{\mu }=\bm{\psi}_{\mu }^{\dag }\bm{\psi}_{\mu }
\end{equation}
and the discrete ladder operators $\bm{\psi}_{\mu}$ satisfy the same commutation relation as
standing mode ladder operators:
\begin{eqnarray}
\left[ \bm{\psi}_{\mu_1},\bm{\psi}_{\mu_2}^\dag %
\right] &=&\int\nolimits_{-\infty }^{+\infty }\int\nolimits_{-\infty
}^{+\infty }d\omega_{1}d\omega_{2}w_{m_{1}p_{1}}^{l_{1}}\left( \omega_{1}\right)
w_{m_{2}p_{2}}^{l_{2}}\left( \omega_{2}\right) ^{* }\nonumber\\&&[a^{l_{1}}[
\omega_{1}], {a^{l_{2}}[\omega_{2}]}\nonumber] \\
&=&\delta _{\mu_1, \mu_2}.
\end{eqnarray}%
An important remark can be made: if the photon amplitude operator $\bm{\psi}_{\mu }$ is
non-hermitian, this is only because its first quantization component $%
w_{mp}^{l}\left( t \right) $ is complex. Its second quantization
component $a^{l}\left( t \right) $ is an hermitian operator. It is also
important to note that, in general, the frequency of a photon is
ill-defined, in contrast with what could be inferred from elementary
introductions to quantum mechanics. This feature happens as soon as the
duration of the wavelet corresponding to that particular photon is not very
long compared with the inverse of the wavelet center frequency. Thus the
concept of photon for a propagating signal has to be clearly distinguished
from an energy quantum. A propagating photon is an elementary excitation of
the field carrying a quantum of action, and corresponds to a field
wavefunction orthogonal to the vacuum.%
\begin{eqnarray}
\left\vert \Psi _{1\mu }\right\rangle &=&\bm{\psi}_{\mu }^{\dag }\left\vert 
\mathrm{vac}\right\rangle, \\
\left\langle \mathrm{vac}|\Psi _{1\mu }\right\rangle &=&0.
\end{eqnarray}%
A wavelet can contain several photons in mode $\mu $ 
\begin{equation}
\left\vert \Psi _{n\mu }\right\rangle =\frac{1}{\sqrt{n!}}\left( \bm{\psi}%
_{\mu }^{\dag }\right) ^{n}\left\vert \mathrm{vac}\right\rangle
\end{equation}%
and each multi-photon state (Fock state) is orthogonal to the others\newline
\begin{equation}
\left\langle \Psi _{n_{2}\mu }|\Psi _{n_{1}\mu }\right\rangle =\delta
_{n_{1}n_{2}}.
\end{equation}%
Several modes can simultaneously be excited%
\begin{equation}
\left\vert \Psi _\sigma\right\rangle =\frac{1}{%
\sqrt{n_{1}!}}\left( \bm{\psi}_{\mu _{1}}^{\dag }\right) ^{n_{1}}\frac{1}{%
\sqrt{n_{2}!}}\left( \bm{\psi}_{\mu _{2}}^{\dag }\right) ^{n_{2}}\ldots\left\vert \mathrm{vac}\right\rangle,
\end{equation}%
where the sequence of indices $\sigma =\left( n_{1},\mu _{1};n_{2},\mu _{2};n_{3},\mu
_{3};\dots\right) $ is a mode photon occupancy configuration. Finally, the
most general wavefunction of the field of the transmission line is a
superposition of all field photon configurations in all the spatio-temporal
modes of the line:%
\begin{equation}
\left\vert \Psi \right\rangle =\sum_{\sigma }C_{\sigma }\left\vert \Psi
_{\sigma }\right\rangle.
\end{equation}%
There are exponentially many more quantum coefficients $C_{\sigma }$ than
the classical coefficients $f_{\mu }$ in Eq.~(\ref{fmus})! And it is also important to
understand that a state with a well defined number of photons in a certain
wavelet basis can be fully entangled in another basis.

A wavelet can also support a so-called coherent state instead of a well
defined number of photons:

\begin{eqnarray}
\left\vert \alpha _{\mu }\right\rangle &=&e^{-\left\vert \alpha _{\mu
}\right\vert ^{2}/2}\sum_{n}\frac{\alpha _{\mu }^{n/2}}{\sqrt{n!}}\left\vert
\Psi _{n\mu }\right\rangle, \\
&=&e^{-\left\vert \alpha _{\mu }\right\vert ^{2}/2}e^{\alpha _{\mu }\bm{\psi}
_{\mu }^{\dag }}\left\vert \mathrm{vac}\right\rangle,
\end{eqnarray}%
and if all wavelets are in a coherent state, we obtain a coherent field state%
\begin{eqnarray}
\left\vert \Psi _{\left\{ \alpha \right\} }\right\rangle
&=&\prod\limits_{\mu }\left\vert \alpha _{\mu }\right\rangle \\
&=&e^{-\sum_{\mu }\left( \left\vert \alpha _{\mu }\right\vert ^{2}/2-\alpha
_{\mu }\bm{\psi}%
_{\mu }^{\dag }\right) }\left\vert \mathrm{vac}\right\rangle.
\label{General_coherent_state}
\end{eqnarray}%
Thus, the set of complex coefficients $\alpha _{\mu }$ plays the role of the
coefficients $f_{\mu }$ in Eq.~(\ref{fmus}). Somewhat surprisingly, this property of being a
coherent state remains true in \emph{every} wavelet basis (as can be
inferred from the quadratic form in the exponent of Eq. [\ref%
{General_coherent_state}]). 

The state of the line is in general not pure and must be described by a
density matrix $\rho _{\sigma \sigma ^{\prime }}$. This ultimate quantum
field description tool leads to the important notion of information
contained in the signal.

In general, in quantum mechanics, we can define for a system with a
finite-dimension Hilbert space, the Shannon$-$Von Neumann entropy%
\begin{equation}
S=-\mathrm{tr}\rho \ln \rho. 
\end{equation}%
The information contained in the system is then straightforwardly computed as%
\begin{equation}
\mathcal{I}=S\left( \rho _{\mathrm{mix}}\right) -S\left( \rho \right),
\end{equation}%
where $\rho _{\mathrm{mix}}$ is the fully mixed state in which all basis
states are equiprobable, with no off-diagonal correlations. The extension of
these ideas to a transmission line on which a signal propagates is not
trivial since the number of temporal modes is infinite and each temporal
mode has a Hilbert space with infinite dimensionality. Some constraints need
to be provided, for instance a fixed total energy for both $\rho $ and $\rho
_{\mathrm{mix}}$. We can also, in another instance, fix the maximum number
of excitation in each temporal mode. Supposing that the maximum number of
excitations is unity in the domain $(\left\vert m\right\vert ,p)\in \left\{
1,2,..,M\right\} \otimes \left\{ -P,...,+P\right\} $ and that other modes
are in the vacuum state, then, for a state of the line characterized by an
average photon number $\left\langle n_{\left\vert m\right\vert
p}\right\rangle $ per mode%
\begin{equation}
\mathcal{I}=\sum_{m=1}^{M}\sum_{p=-P}^{P}\mathcal{I}_{b}\left( \left\langle
n_{\left\vert m\right\vert p}\right\rangle -1/2\right),
\end{equation}%
where $\mathcal{I}_{b}\left( \left\langle X\right\rangle \right) $ the
information contained in a stochastic binary variable $X=\pm 1$:%
\begin{equation}
\mathcal{I}_{b}\left( x\right) =\log _{2}\left[ \sqrt{1-x^{2}}\left( \frac{%
1+\left\vert x\right\vert }{1-\left\vert x\right\vert }\right) ^{\frac{%
\left\vert x\right\vert }{2}}\right] .
\end{equation}
We refer the reader to Ref. \onlinecite{Holevo_2013} for a more complete description of the information carried by quantum signals. 

\subsection{Definition of traveling photon flux}
The dimension of the operators $a^{l}\left( t\right) $ is the inverse square root of time and it is tempting to interpret them as photon flux amplitudes. 
The propagating photon flux in terms of $a^{\rightleftarrows} \left[ \omega \right] $ is defined as:
\begin{widetext}
\begin{eqnarray}
\label{photflux1}
\langle a^{l}[\omega] a^{l'}[\omega']\rangle &=& \Big\{\bar{n}(\omega) + 1+2\pi P_a^{l}\delta(\omega-\omega_a)\Big\}\delta(\omega+\omega')\delta_{ll'},\ \omega>0,\nonumber \\&=& \Big\{\bar{n}(-\omega) +2\pi P_a^{l}\delta(\omega+\omega_a)\Big\}\delta(\omega+\omega')\delta_{ll'},\ \omega<0,
\end{eqnarray}
\end{widetext}
where $l,l'=\rightarrow, \leftarrow$. Here, $P_a^{\rightleftarrows}$ is the the photon flux of the incoming drive signal at angular frequency $\omega_a$ (in units of photons per unit time) and $\bar{n}(\omega)$ is the Bose occupation factor at the temperature of the electromagnetic excitations of the line. We have supposed that there is only one relevant signal frequency, and the generalization to the case with multiple frequencies is straightforward. Thus, in time-domain, the traveling photon-flux is given by
\begin{eqnarray}
\langle a^l(t)^\dagger a^{l'}(t)\rangle &=& \frac{\delta_{ll'}}{2\pi}\int_{-\infty}^\infty d\omega\int_{-\infty}^\infty d\omega'\langle a^{l}[\omega] a^{l'}[\omega']\rangle\nonumber\\&=&\delta_{ll'}\Big[\frac{1}{2\pi}\int_0^\infty\Big\{2\bar{n}(\omega) + 1)d\omega\Big\}+2P_a^l\Big]\nonumber,
\end{eqnarray}
where we have used Eq. \eqref{photflux1}. Thus, the first term is the contribution from the thermal and vacuum fluctuations, while the second term is the classical photon flux. 

\section{Quantum Langevin Equation and Input-Output theory}
\label{appQLE}
The main role of the previous section was to introduce the concept of
quantum electromagnetic fields propagating along a transmission line at
microwave frequencies. The elementary excitations of these fields, microwave
photons, can be seen as the carriers of the information transmitted by the
propagating field. In this section, we describe how these excitations of the transmission line interact with a localized signal processing device. 

\begin{figure}[H]
\centering
\includegraphics[width = 0.5\textwidth]{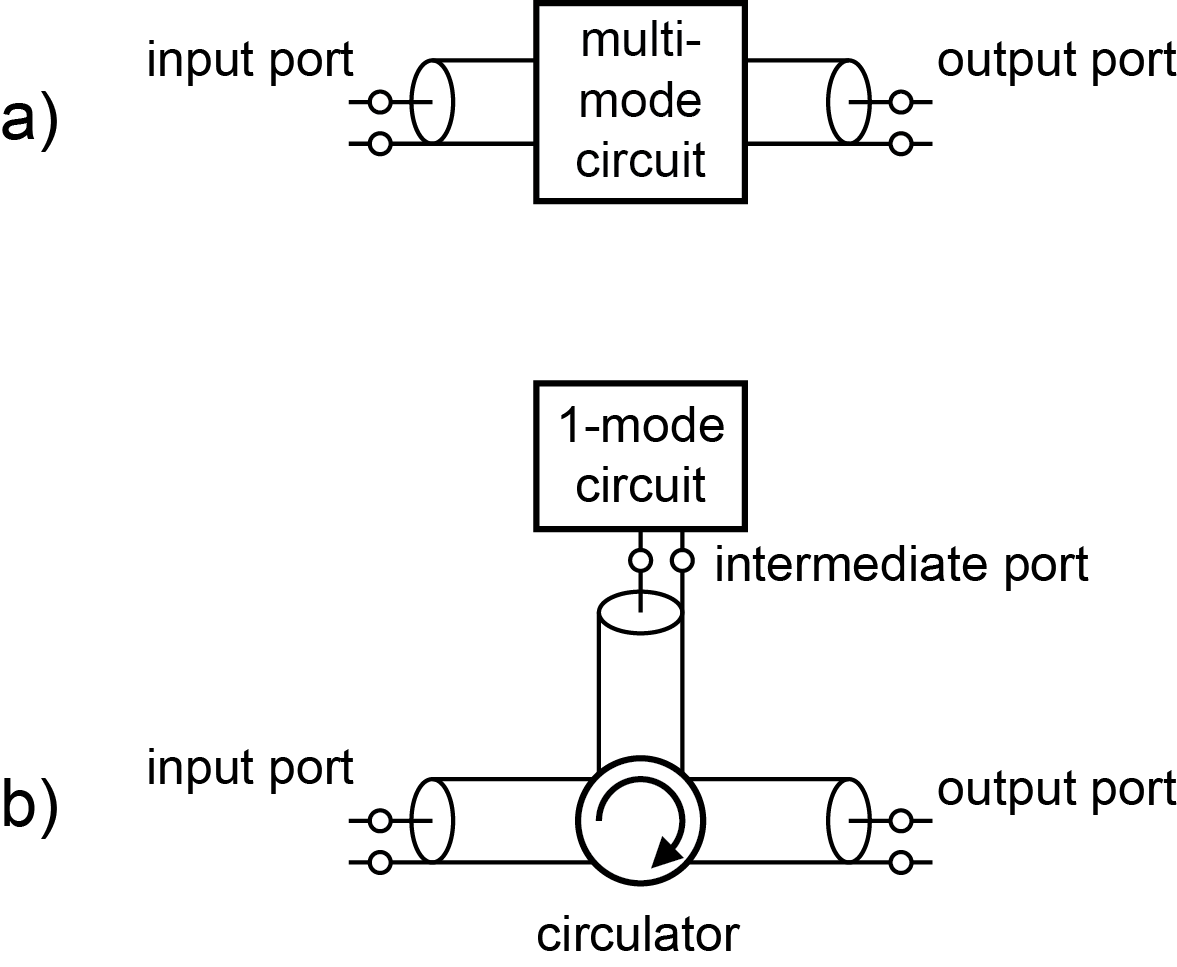}
\caption{\label{Fig_2}	Circuit modes and ports: a) We are interested
in quantum signal processing circuits in the lumped element regime. They 
possess, in general, several standing modes. Input lines and output lines are
attached to ports. In the simplest case b), a circuit with one standing mode communicates
with the outside through only one port, labeled here as``intermediate". An ideal circulator separates the
input from the output.}
\end{figure}

We consider a localized signal processing device as a lumped element circuit ($i.e.$, the spatial extent of the scatterer is much smaller than the wavelength of the radiation being scattered)
which is connected to two semi-infinite transmission lines [see Fig. \ref{Fig_2} (a)]. It
receives from the input line the propagating signal carrying the information
to be processed and re-emits into the output line another signal carrying the
result of the information processing. A crucial ingredient in the
description of the mapping of the input signal into the output signal, is
the coupling between the circuit, which houses standing electromagnetic
modes, and the transmission lines, which support propagating modes. This
coupling is dealt with theoretically using the QLE using input-output theory. 
For simplicity, we consider the situation in which a lumped element circuit with only one electromagnetic mode is coupled to only one
semi-infinite transmission line [Fig. \ref{Fig_2} (b)]. We will closely follow the form of input-output theory developed in Refs. \onlinecite{Gardiner_Collett_1985, Gardiner_Zoller_2004} (see also Chap. 3 of Ref. \onlinecite{Drummond_Ficek_2013}). In this so-called one-mode, one-port configuration, the signal processing occurs as a transformation of the incoming wave into the reflected outgoing wave.
However, in order to utilize the action of the one-mode, one-port circuit, a non-reciprocal linear device called a circulator has to be added in order to
separate the incoming and outgoing waves into two independent transmission lines [Fig. \ref{Fig_2} (b)]. This circuit configuration also provides a way to model directional, through amplifiers. \cite{Courty_Reynaud_1999}  We can leave the modeling of the circulator aside for the moment (it is a 3-port device), and derive the equation of motion for an operator of the lumped circuit element. 

\begin{figure}[H]
\centering
\includegraphics[width = 0.5\textwidth]{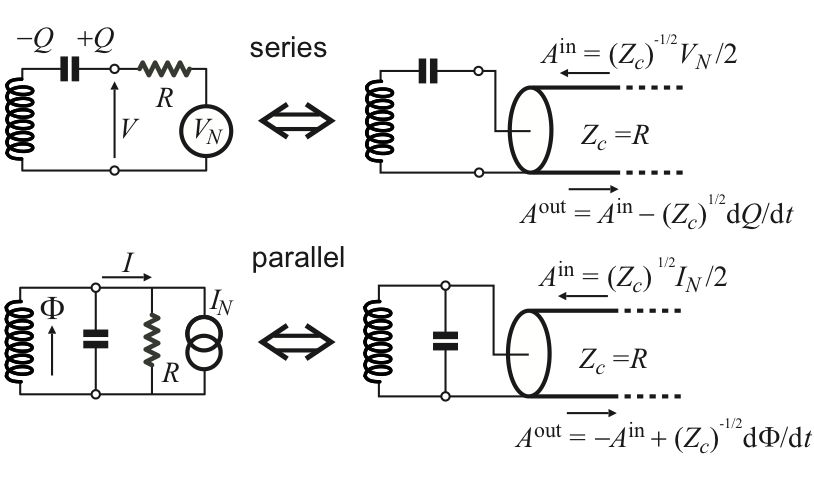}
\caption{\label{Fig_diss} The damping of a circuit by a resistance $Z_c=R$ can take
place in a series or parallel way, depending on whether the resistance
is placed in series with it or across a branch. The Nyquist model
represents the resistance by a transmission line with characteristic
impedance $Z_c=R$. In this model, the voltage or current source is replaced by an incoming wave. }
\end{figure}

The input-output formalism also provides a quantum-mechanical model for dissipation in these systems. In microwave circuits, the dissipation can be a resistance connected in series or in parallel (see Fig. \ref{Fig_diss}). In the following section, we derive the QLE for the case when the dissipation is caused by a resistor coupled in series with the circuit. In this case, a voltage source added in series with the resistor is responsible for coherent incident signals and fluctuations. The case when the dissipation is due to a resistor in parallel is treated afterwards. In the latter case, a current source in parallel with the resistor drives coherent incident signals and fluctuations. 

\subsection{Dissipation due to a resistor in series, together with a voltage source}
\subsubsection{Derivation of the Quantum Langevin Equation beyond RWA}
\label{appQLEa}
Unlike the previous section, we now have a semi-infinite transmission line with characteristic impedance $Z_c$, extending to $+\infty$, coupled to a scatterer (located at $x=0$), which we assume to be a lumped circuit with Lagrangian $L_{\rm{sys}}$. The total Lagrangian of the system plus transmission line is given by
\begin{align}
{\cal L} &= L_{\rm{sys}}+\int_0^\infty dx\bigg\{\frac{C_\ell}{2}\Big(\frac{\partial \Phi}{\partial t}\Big)^2- \frac{1}{2L_\ell}\Big(\frac{\partial \Phi}{\partial x}\Big)^2\bigg\} \nonumber \\&\qquad+ X\int_0^\infty dx \kappa(x) \frac{\partial \Phi}{\partial t},
\end{align}
where $C_\ell, L_\ell$ are defined as before and $X$ is a charge-like system operator that couples to the transmission line with the coupling constant $\kappa(x)$. We have chosen the coupling Hamiltonian so as to ensure that the line voltage $\frac{\partial \Phi}{\partial t}$ couples to X, as necessitated by the insertion of the transmission line in the circuit. The canonical conjugate momentum $\Pi$ now has a contribution from the system operator $X$:
\begin{align}
\frac{\partial{\cal L}}{\partial(\partial_t \Phi)} = C_\ell\frac{\partial \Phi}{\partial t} + X \kappa(x)=\Pi(x,t).
\end{align}
The operators $\Phi(x,t), \Pi(x,t)$ still obey the commutation relation: 
\begin{equation}
[\Phi(x,t), \Pi(x',t)] = i\hbar\delta(x-x').
\end{equation}
One can then write down the Hamiltonian for the system and transmission line:
\begin{align}
H&=H_{\rm{sys}} +\int_0^\infty \bigg[\frac{1}{2C_\ell}\big\{\Pi(x,t)-X \kappa(x)\big\}^2\nonumber\\&\qquad+ \frac{1}{2L_\ell}\Big(\frac{\partial \Phi}{\partial x}\Big)^2\bigg]. 
\end{align}
Next, we go to the Fourier domain, by defining the following:
\begin{eqnarray}
\label{trans_line_ft_defn_1}
q(\omega,t) &\equiv& \sqrt{\frac{2}{\pi v_p}}\int_{0}^\infty dx\ \Phi(x,t) \cos\frac{\omega x}{v_p},\\
p(\omega,t) &\equiv& \sqrt{\frac{2}{\pi v_p}}\int_{0}^\infty dx\ \Pi(x,t)\cos\frac{\omega x}{v_p},\\
\kappa(\omega) &\equiv& \sqrt{\frac{2}{\pi v_p}}\int_{0}^\infty dx\ \kappa(x) \cos\frac{\omega x}{v_p}
\end{eqnarray}
Here, we have involved only the Fourier cosine expansion. This is because for a lumped circuit element, the coupling $\kappa(x)$ is local and $\propto\delta(x)$. Only the Fourier cosine components  of  $\partial_t \Phi(x,t)$ and therefore, of $\Phi(x,t)$ couple to the system variable. The operators $q(\omega,t), p(\omega,t)$ satisfy the commutation relation
\begin{equation}
[q(\omega,t),p(\omega',t)] = i\hbar\delta(\omega-\omega').
\end{equation}
In the Fourier domain, the Hamiltonian of the system and the cosine part of the degrees of freedom of the line reads
\begin{align}
H &= H_{\rm{sys}} +\int_0^\infty d\omega \bigg[\frac{1}{2C_\ell}\big\{p(\omega,t)-X\kappa(\omega)\big\}^2\nonumber\\&\qquad+ \frac{\omega^2C_\ell}{2}q(\omega,t)^2\bigg].
\end{align}
The equations of motion for $q(\omega,t), p(\omega,t)$ are 
\begin{align}
\frac{dq(\omega,t)}{dt}&=\frac{p(\omega,t)-X\kappa(\omega)}{C_\ell}\\
\frac{dp(\omega,t)}{dt}&=-\omega^2C_\ell q(\omega,t).
\end{align}
We define annihilation operators for the different spatial modes of the transmission line as
\begin{equation}
\textgoth{a}(\omega,t) = \sqrt{\frac{\omega C_\ell}{2\hbar}}q(\omega,t) + \frac{i}{\sqrt{2\hbar\omega C_\ell}}p(\omega,t),
\end{equation}
which satisfy the commutation relation: \begin{equation}
[\textgoth{a}(\omega,t),\textgoth{a}(\omega',t)^\dagger] = \delta(\omega-\omega').
\end{equation}
Finally, the Heisenberg equation of motion of $a(\omega,t)$ is
\begin{equation}
\frac{d\textgoth{a}(\omega,t)}{dt}=-i\omega \textgoth{a}(\omega,t)-\sqrt{\frac{\omega }{2\hbar C_\ell}}\kappa(\omega) X.
\end{equation}
This can be solved to yield
\begin{align}
\textgoth{a}(\omega,t) &=\textgoth{a}(\omega,t_0)e^{-i\omega(t-t_0)}\nonumber\\&-\sqrt{\frac{\omega }{2\hbar C_\ell}}\int_{t_0}^t dt'X(t')e^{-i\omega(t'-t_0)},\\
\textgoth{a}(\omega,t) &=\textgoth{a}(\omega,t_1)e^{-i\omega(t-t_1)}\nonumber\\&+\sqrt{\frac{\omega }{2\hbar C_\ell}}\int_{t}^{t_1} dt'X(t')e^{-i\omega(t'-t_1)},
\end{align}
where $t_0<t$ is a time far in the past and $t_1>t$ is a time far in the future. Eventually, we will take $t_0$ and $t_1$ to $-\infty$ and $\infty$ respectively so that the dynamics of the circuit takes place entirely in the interval $[t_0,t_1]$. Now, we can rewrite $\Phi(x,t)$ in terms of the solutions in the past and the future:
\begin{align}
\Phi(x,t) &= \int_{0}^\infty d\omega \cos\Big(\frac{\omega x}{v_p}\Big)\sqrt{\frac{\hbar}{\pi\omega v_p C_\ell}}\big\{\textgoth{a}(\omega,t) + \textgoth{a}(\omega,t)^\dagger\big\},\nonumber
\end{align}
which leads to (after some algebra)
\begin{align}
\label{varphi1}\Phi(x,t) &= \Phi_{\rm{in}}\bigg(t-\frac{x}{v_p}\bigg)+\Phi_{\rm{in}}\bigg(t+\frac{x}{v_p}\bigg)\nonumber\\\qquad&-\frac{1}{2C_\ell}\int_{\frac{x}{v_p}-(t-t_0)}^{\frac{x}{v_p}+(t-t_0)}d\tau \kappa(v_p\tau)X\bigg(t-\bigg|\tau-\frac{x}{v_p}\bigg|\bigg)\\&=\Phi_{\rm{out}}\bigg(t-\frac{x}{v_p}\bigg)+\Phi_{\rm{out}}\bigg(t+\frac{x}{v_p}\bigg)\nonumber\\&\qquad+\frac{1}{2C_\ell}\int_{\frac{x}{v_p}+(t-t_1)}^{\frac{x}{v_p}-(t-t_1)}d\tau \kappa(v_p\tau)X\bigg(t+\bigg|\tau-\frac{x}{v_p}\bigg|\bigg),
\end{align}
where $\Phi_{\rm{in/out}}$ denote input and output fields in the past and the future, defined as
\begin{align}
\label{phiin}
\Phi_{\rm{in(out)}}\bigg(t\pm \frac{x}{v_p}\bigg) &= \frac{1}{2}\int_0^\infty d\omega \sqrt{\frac{\hbar}{\pi\omega v_p C_\ell}}\nonumber\\&\quad\Big\{\textgoth{a}(\omega,t_{0(1)})e^{i\omega t_{0(1)}}e^{-i\omega(t\pm\frac{x}{v_p})}+\rm{H.c.}\Big\}.
\end{align}
Next, we take $t_0, t_1$ to $-\infty, \infty$ respectively. For a lumped scatterer, the scattered field is outside the region where $\kappa(x)$ is nonzero and thus, $\tau-x/v_p<0$. Therefore,  
\begin{align}\label{Eqvarphi}
\Phi(x,t) &= \Phi^{\rm{in}}\bigg(t+\frac{x}{v_p}\bigg)+\Phi_{\rm{out}}\bigg(t-\frac{x}{v_p}\bigg),\\\label{Eqvarphi1}\Phi_{\rm{out}}\bigg(t\pm\frac{x}{v_p}\bigg)&=\Phi^{\rm{in}}\bigg(t\pm\frac{x}{v_p}\bigg)\nonumber\\&\quad-\frac{1}{2C_\ell}\int_{-\infty}^\infty d\tau \kappa(v_p\tau)X\bigg(t\pm\frac{x}{v_p}\mp\tau\bigg).
\end{align}
Eq. \eqref{Eqvarphi} shows that the field at any point $(x,t)$ is a sum of the input and output fields propagating in opposite directions. Eq. \eqref{Eqvarphi1} describes the output fields in terms of the input fields plus some system operator, convolved with the coupling constant. The equation of motion for a system operator $Y$ is given by
\begin{equation}
\frac{dY}{dt}=-\frac{i}{\hbar}[Y, H_{\rm{sys}}]-\frac{i}{2\hbar}\int_0^\infty dx \kappa(x)\bigg\{\frac{\partial \Phi}{\partial t},[X,Y]\bigg\}.
\end{equation}
Choosing 
\begin{equation}
\kappa(x) = 2\sqrt{Z_c v_pC_\ell}\delta(x),
\end{equation}
leads to
\begin{equation}
\label{eqY}
\frac{dY}{dt}=-\frac{i}{\hbar}[Y, H_{\rm{sys}}]-\frac{i\sqrt{Z_c v_p C_\ell}}{2\hbar}\bigg\{\frac{\partial \Phi}{\partial t}\bigg|_{x=0},[X,Y]\bigg\},
\end{equation}
where an extra factor of $1/2$ arises from the integral being from $0$ to $\infty$. From Eq. \eqref{varphi1}, setting $x=0$, we get
\begin{equation}
\label{varphi2}
\Phi(x=0,t) = 2\Phi^{\rm{in}}(t)-\sqrt{\frac{Z_c}{v_pC_\ell}}X(t).
\end{equation}
Inserting this in Eq. \eqref{eqY}, we arrive at
\begin{align}
\label{eqY1}
\frac{dY}{dt}&=-\frac{i}{\hbar}[Y, H_{\rm{sys}}]\nonumber\\&\quad+\frac{i}{2\hbar}\bigg\{Z_c \frac{dX}{dt}-2\sqrt{Z_c v_p C_\ell}\frac{\partial \Phi^{\rm{in}}}{\partial t}\bigg|_{x=0},[X,Y]\bigg\}.
\end{align}
Now, from Eq. \eqref{phiin}, we have
\begin{align}
\frac{\partial\Phi^{\rm{in}}}{\partial t}\bigg|_{x=0} &= \frac{-i}{2}\int_0^\infty d\omega \sqrt{\frac{\hbar\omega}{\pi v_pC_\ell}}\Big\{\textgoth{a}(\omega,t_{0})e^{-i\omega(t- t_{0})}\nonumber\\&\quad+\rm{H.c.}\Big\}.
\end{align} 
We define incoming and outgoing traveling photon field amplitudes (analogous to those in Sec. \ref{appQuantumSignals}) as $A^{\rm{in}}(t)$ and $A^{\rm{out}}(t)$ as
\begin{align}
 A^{\rm{in(out)}}(t)&=\sqrt{v_pC_\ell}\frac{\partial\Phi_{\rm{in(out)}}}{\partial t}\bigg|_{x=0}\nonumber\\&=\frac{-i}{\sqrt{2\pi}}\int_0^\infty d\omega \sqrt{\frac{\hbar\omega}{2}}\Big\{\textgoth{a}(\omega,t_{0(1)})e^{-i\omega(t- t_{0(1)})}\nonumber\\&\quad+\rm{H.c.}\Big\},
\end{align}
Thus, the QLE for the system operator $Y$ becomes:
\begin{widetext}
\begin{equation}
\label{eqY2}
\boxed{\frac{dY}{dt}=-\frac{i}{\hbar}[Y, H_{\rm{sys}}]+\frac{i}{2\hbar}\bigg\{Z_c \frac{dX}{dt}-2\sqrt{Z_c}A^{\rm{in}}(t),[X,Y]\bigg\}}.
\end{equation}
\end{widetext}
In deriving this equation, only the Markov approximation is made, which makes the QLE local in time. This is because we chose the coupling of the form $\kappa(x)\propto\delta(x)$. The RWA has not been made yet (see below). The output field can be computed from the incoming ones using Eq. \eqref{Eqvarphi}, \eqref{varphi2}:
\begin{align}
\label{eqAinAout1}
\boxed{A^{\rm{out}}(t) = A^{\rm{in}}(t)-\sqrt{Z_c}\frac{dX}{dt}}.
\end{align}
We point out that this form of the QLE beyond RWA is useful for analyzing problems such as the spin-boson problem\cite{Weiss1999} or the Kondo problem. \cite{Mahan2013, LeHur2012, Goldstein2013} Next, we perform the RWA on this QLE derived above.
 
\subsubsection{Derivation of the Quantum Langevin Equation under RWA}
\label{appQLEb}
For simplicity, we will derive the equation for an LC circuit (a harmonic oscillator). It is straightforward to generalize our results to include nonlinearities in our system Hamiltonian. Consider a series LC resonator with inductance $L$ and capacitance $C$ coupled in series with a resistor with resistance $Z_c$ and a voltage source $V_s$. The suitable choice for the position coordinate of the harmonic oscillator is the charge on the capacitor, denoted by $Q$ and the corresponding canonically conjugate momentum is the flux through the inductor, denoted by $\Phi$. They obey the commutation  relation: $[Q,\Phi]=i\hbar$. Then, the Hamiltonian for the system is given by 
\begin{equation}
H_{\rm{sys}} = \frac{Q^2}{2C}+\frac{\Phi^2}{2L}, 
\end{equation}
and the coupling operator $X$ of the previous section is $Q$. Then, the QLE for the operators $Q, \Phi$ are:
\begin{align}
\frac{dQ}{dt} &= \frac{\Phi}{L}\\
\frac{d\Phi}{dt} &= -\frac{Q}{C}-Z_c\frac{dQ}{dt}+2\sqrt{Z_c}A^{\rm{in}}(t).
\end{align}
Now, we define the annihilation operator for the harmonic oscillator as
\begin{align}
a = \sqrt{\frac{Z_0}{2\hbar}}Q + i\frac{1}{\sqrt{2\hbar Z_0}}\Phi,
\end{align}
where $Z_0=\sqrt{L/C}$ is the characteristic impedance of the LC oscillator. Then, the equation of motion of the operator $a$ is given by
\begin{align}
\frac{da}{dt} &= -i\omega_0 a-\frac{Z_c}{2L}(a-a^\dagger)\nonumber\\&\quad+\sqrt{\frac{Z_c}{2\hbar Z_0}}\int_0^\infty d\omega\sqrt{\frac{\hbar\omega}{\pi}}\{\textgoth{a}(\omega,t_0)e^{-i\omega(t-t_0)}+ \rm{H.c.}\},
\end{align}
where $\omega_0 =1/\sqrt{LC}$ is the resonant frequency of the oscillator. Since we are interested in frequencies close to the resonant frequency of the oscillator, we can drop the counter-rotating terms (terms with $a^\dagger$ and $\textgoth{a}(\omega,t_0)^\dagger$). Further, we can approximate the $\sqrt{\omega}$ under the integral with $\sqrt{\omega_0}$. These two approximations together comprise the RWA. The resultant equation is 
\begin{eqnarray}
\frac{da}{dt} &=& -i\omega_0 a-\frac{Z_c}{2L}a\nonumber\\&&\quad+\sqrt{\frac{Z_c}{L}}\frac{1}{\sqrt{2\pi}}\int_0^\infty d\omega\ \textgoth{a}(\omega,t_0)e^{-i\omega(t-t_0)}.
\end{eqnarray}
Finally, we identify $Z_c/L=\kappa$, which is the bandwidth of the resonator. This leaves us with the QLE: 
\begin{align}
\frac{da(t)}{dt} &= -i\omega_0 a(t)-\frac{\kappa}{2}a(t)+\sqrt{\kappa}a^{\rm{in}}(t),
\end{align}
where we have defined 
\begin{align}
a^{\rm{in}}(t)= \frac{1}{\sqrt{2\pi}}\int_0^\infty d\omega\ \textgoth{a}(\omega,t_0)e^{-i\omega(t-t_0)}.
\end{align}
Note that only positive frequencies appear in the definition of the incoming signal. In principle, one can integrate the QLE and express the circuit variable $%
a(t)$ in terms of the incoming field $a^{\rm{in}}\left( t\right) $. The output
field can be obtained from the input-output equation%
\begin{equation}
\label{input-output}
a^{\rm{out}}\left( t\right)=a^{\rm{in}}\left( t\right)-\sqrt{\kappa }a(t).
\end{equation}%
Again, the appearance of the simple coefficient $\sqrt{\kappa }$ in this
relation results from the Markov approximation. In the simple case of the harmonic oscillator, the elimination of $a$ between input and output can be
carried out fully at the analytical level and one obtains%
\begin{equation}
\left( \frac{d}{dt}+i\omega _{0}+\kappa /2\right) a^{\rm{out}}\left( t\right)
=\left( \frac{d}{dt}+i\omega _{0}-\kappa /2\right) a^{\rm{in}}\left( t\right).
\end{equation}%
Going to the Fourier domain, one obtains the reflection coefficient $r\left(
\omega \right) $

\begin{eqnarray}
a^{\rm{out}}\left[ \omega \right]  &=&r\left( \omega \right) a^{\rm{in}}\left[ \omega %
\right], \\
r_{\mathrm{RWA}}\left( \omega \right)  &=&\frac{\omega -\omega _{0}-i\kappa
/2}{\omega -\omega _{0}+i\kappa /2}.
\end{eqnarray}

The causality property of the circuit, which expresses the fact that it
cannot produce a response \textit{before} being submitted to a stimulus, is
implemented here by the analytic property of the complex function $r_{%
\mathrm{RWA}}(\omega )$: its pole is in the lower half complex plane while
its zero is in the upper half. On the other hand, the property of the
reflection coefficient to possess a single pole instead of a pair is an artifact of
RWA. As a matter of fact, when the circuit is linear as is the case here, one can
compute exactly the reflection coefficient using a more elaborate form of
QLE without RWA, while keeping the Markov approximation. One then obtains
the expression possessing the necessary pair of poles with values $\omega
_{\pm }=\left( -i\kappa \pm \sqrt{-\kappa ^{2}-4\omega _{0}^{2}}\right) /2$:

\begin{equation}
r\left( \omega \right) =\frac{\omega ^{2}-\omega _{0}^{2}-i\kappa \omega }{%
\omega ^{2}-\omega _{0}^{2}+i\kappa \omega }.
\end{equation}%
It is easy to see that in this last equation, $r\left( \omega \right) $
reduces to the single pole expression $r_{\mathrm{RWA}}\left( \omega \right) 
$ when $\omega $ is such that $\left\vert 1-\omega /\omega _{0}\right\vert
\ll 1$ and in the underdamped limit $\kappa /\omega _{0}\ll 1$.

It is straightforward to generalize our equation for arbitrary system Hamiltonian:
\begin{equation}
\boxed{\frac{da}{dt}\underset{%
\begin{array}{c}
\text{{\tiny Markov}} \\ 
\text{{\tiny RWA}}%
\end{array}%
}{=}\frac{i}{\hbar }\left[ H_{\rm{sys}},a\right] -\frac{\kappa }{2}a+\sqrt{\kappa }%
a^{\rm{in}}\left( t\right)},  \label{Quantum_Langevin}
\end{equation}%
with the boundary condition (for dissipation connected in series)
\begin{equation}
\label{input-output_1}
\boxed{a^{\rm{out}}\left( t\right)=a^{\rm{in}}\left( t\right)-\sqrt{\kappa }a(t)}.
\end{equation}
The remarkably simple form of the QLE is due to the two approximations: 
i) the Markov
approximation which considers that the coupling of the system with the
environment is ``ohmic": the density of modes of the environment can be
considered white across the set of circuit transition frequencies, as in an ideal resistance, ii) the
coupling is also supposed to be weak in the sense that $\kappa $ is much
smaller than any transition frequency between the energy levels of the
lumped circuit. Although approximations are made, the equation respects the
important commutation relation of the ladder operators:
\begin{equation}
\left[ a\left( t\right) ,a\left( t\right) ^{\dag }\right] =1
\end{equation}%
at all times $t$.

The incoming driving field has in general three components which are treated
on equal footing by the QLE: i) the deterministic signal to be processed,
ii) thermal or parasitic noise accompanying the information-carrying signal,
and iii) quantum noise, or, in other words, the zero-point fluctuations of
the field of the semi-infinite transmission line. The inclusion of this last
component is implemented implicitly in that the QLE is
an operator equation, in contrast with the Classical Langevin Equation which
is just a differential equation for a c-number function, albeit stochastic.
Note that the coefficient $\sqrt{\kappa }$ in front of the propagating field
amplitude embodies single-handedly the fluctuation-dissipation theorem: the
rate at which energy is radiated away from the circuit (the coefficient $%
\kappa $ of the second term) has to be tightly linked to the coupling
constant with which random radiation, emitted from the black body that the
line plays the role of, corrupt the purity of the state of the circuit. If one wonders
why $\kappa $ appears under a square root in this coupling coefficient,
one just needs to remember that while $a$ is a dimensionless standing photon
number amplitude, $a^{\rm{in}}\left( t\right) $ is the dimensioned amplitude
corresponding to a photon flux. Consequently, when the semi-infinite line is
in thermal equilibrium (input signal is only black-body noise with
temperature $T$), the following relation involving the anticommutator $%
\left\{ \square ,\square \right\} $ holds%
\begin{equation}
\left\langle \left\{ a^{\rm{in}}[\omega _{1}],a^{\rm{in}}[\omega _{2}]\right\}
\right\rangle _{T}\,=\coth \frac{\hbar \left(| \omega _{1}-\omega _{2}|\right) 
}{4k_{B}T}\delta \left( \omega _{1}+\omega _{2}\right),\nonumber
\end{equation}%
where $k_{B}$ is Boltzmann constant, $\left\{ A,B\right\} =AB+BA$ and $%
\left\langle ...\right\rangle _{T}$ the average in the thermal state. Given
an operating temperature around $20~\mathrm{mK}$, this expression shows that
the quantum fluctuations become fully dominant over thermal fluctuations at
frequencies above a GHz.

\subsection{Dissipation due to a resistor in parallel, together with a current source}
To compute the QLE for a resistor in parallel (see Fig. \ref{Fig_diss}, lower panel), the derivation presented in Sec. \ref{appQLEa}, \ref{appQLEb} can be used, with some modifications. The suitable variable for the semi-infinite transmission line (impedance $Z_c$ and extending to $+\infty$) is the charge at position $x$, denoted by $Q(x,t)$. The Lagrangian for the total system plus transmission is now given by:
\begin{align}
{\cal L} &= L_{\rm{sys}}+\int_0^\infty dx\bigg\{\frac{L_\ell}{2}\Big(\frac{\partial Q}{\partial t}\Big)^2- \frac{1}{2C_\ell}\Big(\frac{\partial Q}{\partial x}\Big)^2\bigg\} \nonumber \\&\qquad+ X\int_0^\infty dx \kappa(x) \frac{\partial Q}{\partial t},
\end{align}
where $C_\ell, L_\ell$ are defined as before, $L_{\rm{sys}}$ is the Lagrangian of the system and $X$ is now a flux-like system operator that couples to the transmission line with the coupling constant $\kappa(x)$. Then, all the manipulations in Sec. \ref{appQLEa} can be done with the substitutions:
\begin{equation}
\Phi(x,t)\rightarrow Q(x,t), \quad L_\ell\leftrightarrow C_\ell. 
\end{equation}
For brevity, we skip the steps and present only the results. As before $q(\omega,t), p(\omega,t)$ and $\kappa(\omega)$ denote the Fourier cosine-transforms of $Q(x,t), \Pi(x,t), \kappa(x)$ respectively where $\Pi(x,t)$ is the canonically conjugate momentum of $Q(x,t)$. Note that only in this subsection, $q(\omega, t), p(\omega,t)$ correspond to the Fourier transforms of the charge and flux operators of the semi-infinite transmission line. The annihilation operator is defined as 
\begin{equation}
\textgoth{a}(\omega,t) = \sqrt{\frac{\omega L_\ell}{2\hbar}}q(\omega,t) + \frac{i}{\sqrt{2\hbar\omega L_\ell}}p(\omega,t),
\end{equation}
Solving the Heisenberg equation of motion for the annihilation operator and plugging it into the expression for $Q(x,t)$, we get
\begin{align}
\label{Q1}Q(x,t) &= Q_{\rm{in}}\bigg(t-\frac{x}{v_p}\bigg)+Q_{\rm{in}}\bigg(t+\frac{x}{v_p}\bigg)\nonumber\\\qquad&-\frac{1}{2L_\ell}\int_{\frac{x}{v_p}-(t-t_0)}^{\frac{x}{v_p}+(t-t_0)}d\tau \kappa(v_p\tau)X\bigg(t-\bigg|\tau-\frac{x}{v_p}\bigg|\bigg)\\&=Q_{\rm{out}}\bigg(t-\frac{x}{v_p}\bigg)+Q_{\rm{out}}\bigg(t+\frac{x}{v_p}\bigg)\nonumber\\&\qquad+\frac{1}{2L_\ell}\int_{\frac{x}{v_p}+(t-t_1)}^{\frac{x}{v_p}-(t-t_1)}d\tau \kappa(v_p\tau)X\bigg(t+\bigg|\tau-\frac{x}{v_p}\bigg|\bigg),
\end{align}
where $Q_{\rm{in/out}}$ denote input and output fields in the past and the future, defined as
\begin{align}
\label{Qin}
Q_{\rm{in(out)}}\bigg(t\pm \frac{x}{v_p}\bigg) &= \frac{1}{2}\int_0^\infty d\omega \sqrt{\frac{\hbar}{\pi\omega v_p L_\ell}}\nonumber\\&\quad\Big\{\textgoth{a}(\omega,t_{0(1)})e^{i\omega t_{0(1)}}e^{-i\omega(t-\frac{x}{v_p})}+\rm{H.c.}\Big\}.
\end{align}
The equation of motion for a system operator $Y$ is given by
\begin{equation}
\frac{dY}{dt}=-\frac{i}{\hbar}[Y, H_{\rm{sys}}]-\frac{i}{2\hbar}\int_0^\infty dx \kappa(x)\bigg\{\frac{\partial Q}{\partial t},[X,Y]\bigg\}.
\end{equation}
Choosing 
\begin{equation}
\kappa(x) = 2\sqrt{\frac{v_pL_\ell}{Z_c}}\delta(x),
\end{equation}
and noting that 
\begin{equation}
\sqrt{v_pL_\ell}\frac{\partial Q_{\rm{in}}}{\partial t}\bigg|_{x=0} = A^{\rm{in}}(t),
\end{equation}
the QLE for the system operator $Y$ becomes:
\begin{widetext}
\begin{equation}
\label{eqY22}
\boxed{\frac{dY}{dt}=-\frac{i}{\hbar}[Y, H_{\rm{sys}}]+\frac{i}{2\hbar}\bigg\{\frac{1}{Z_c} \frac{dX}{dt}-\frac{2}{\sqrt{Z_c}}A^{\rm{in}}(t),[X,Y]\bigg\}}.
\end{equation}
\end{widetext}
This is the QLE for a system coupled to a resistor in parallel, together with a current source where only the Markov approximation has been made. The output field for this case is given by 
\begin{align}
\label{eqAinAout2}
\boxed{A^{\rm{out}}(t) = -A^{\rm{in}}(t)+\frac{1}{\sqrt{Z_c}}\frac{dX}{dt}}.
\end{align}
Note that Eqs. (\ref{eqY22}, \ref{eqAinAout2}) are the parallel connection counterparts of Eqs. (\ref{eqY2}, \ref{eqAinAout1}) given for series connection. 
Finally, one can perform identical manipulations as in Sec. \ref{appQLEb} to arrive at the same QLE under RWA given in Eq. \eqref{Quantum_Langevin}:
\begin{equation}
\boxed{\frac{da}{dt}\underset{%
\begin{array}{c}
\text{{\tiny Markov}} \\ 
\text{{\tiny RWA}}%
\end{array}%
}{=}\frac{i}{\hbar }\left[ H_{\rm{sys}},a\right] -\frac{\kappa }{2}a+\sqrt{\kappa }%
a^{\rm{in}}\left( t\right)} ,  \label{Quantum_Langevin_1}
\end{equation}
with the modified boundary condition corresponding to parallel dissipation: 
\begin{equation}
\label{input-output_2}
\boxed{a^{\rm{out}}\left( t\right)=-a^{\rm{in}}\left( t\right)+\sqrt{\kappa }a(t)}.
\end{equation}
Here, Eqs. (\ref{Quantum_Langevin_1}, \ref{input-output_2}) are the parallel counterparts of Eqs. (\ref{Quantum_Langevin}, \ref{input-output_1}) given for series coupling. 

It is important to note that while the overall form of the QLE under RWA is the same for both kinds of dissipation, both the damping rates and the input-output boundary relations are different. In the case of parallel LC resonator (inductance $L$ and capacitance $C$) coupled to a resistor in parallel, the dissipation rate is $\kappa = 1/(Z_c C)$. This should be compared with the dissipation rate derived in the series configuration where $\kappa$ was equal to $Z_c/L$. As expected, the dissipation for a series (parallel) configuration increases (decreases) with increasing resistance.

\subsection{Generalized QLE for multi-port devices}
Continuing to work in the framework of both RWA and the Markov
approximation, one can easily deal with more than one circuit mode and more
than one semi-infinite line. Denoting by $M$ the circuit standing mode index
and $P$ the port index, one obtains the multi-mode, multi-port generalized
QLE:%
\begin{equation}
\frac{d}{dt}a_{M}=\frac{i}{\hbar }\left[ H,a_{M}\right] +\sum_{P}\left[ -%
\frac{\kappa _{MP}}{2}a_{M}+\varepsilon _{MP}\sqrt{\kappa _{MP}}%
a_{P}^{\rm{in}}\left( t\right) \right].
\end{equation}

\begin{figure}[H]
\centering
\includegraphics[width = 0.5\textwidth]{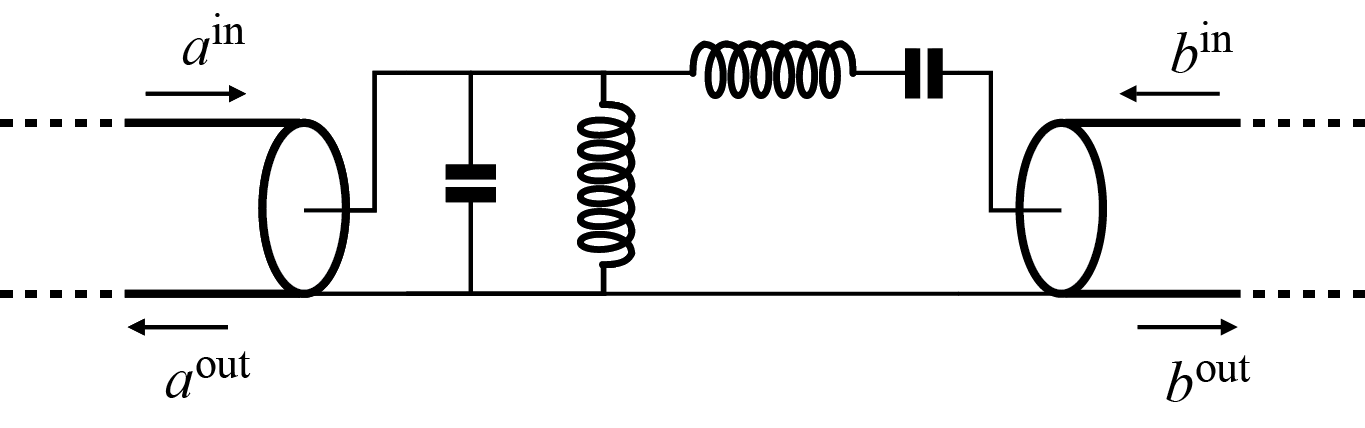}
\caption{\label{Fig_4} Example of a two-mode,
two-port circuit in which care must be taken in the amplitude factors of the
Quantum Langevin Equation.}
\end{figure}

Apart from a simple extension of the number of variables, this new equation
contains the rectangular matrix $\varepsilon _{MP}$ whose 
coefficients are $\varepsilon _{MP}=\pm1$.
This matrix can be computed from the details of the coupling of the lines to
particular elements of the circuit (capacitances or inductances, series or
parallel connections). A simple example of a situation where the $%
\varepsilon _{MP}$ cannot be set to unity by a re-definition of the mode
amplitude $a_{M}$ is presented in Fig. \ref{Fig_4}. The general Input-Output Equation
takes the form%
\begin{equation}
a_{M}=\sum_{P}\frac{1}{\sqrt{\kappa _{MP}}}\left[ \varepsilon
_{MP}a_{M}^{\rm{in}}\left( t\right) +(\varepsilon^{-1})_{MP}a_{P}^{\rm{out}}\left(
t\right) \right] .
\end{equation}%

\section{Calculation of total output power in parametric amplifiers}
\label{output_power_calc}
\subsection{Degenerate case}
In this appendix, we outline the calculation of the output power of a degenerate parametric amplifier. We will treat the gain $G$ as a parameter, and thus, our results apply to both stiff and depleted calculations. The scattering matrix for this system is given in Eq. \eqref{jba_scatmat} of Sec. \ref{sec_4}. The output power is defined as
\begin{align}
P_a^{\rm{out}} &= \langle a^{\rm{out}}(t)^\dagger a^{\rm{out}}(t)\rangle\\&=\frac{1}{2\pi}\int_0^\infty d\omega\int_0^\infty d\omega' \langle a^{\rm{out}}[-\omega]a^{\rm{out}}[\omega']\rangle e^{i(\omega-\omega')t}\nonumber.
\end{align}
In the next step, we substitute the output fields in terms of the input fields using Eq. \eqref{jba_scatmat}. Thus, we arrive at 
\begin{widetext}
\begin{equation}
\label{pout1}
P_a^{\rm{out}} =\frac{1}{2\pi}\int_0^\infty d\omega\int_0^\infty d\omega'  e^{i(\omega-\omega')t}\bigg\langle \bigg\{\frac{M_1(\omega)}{D(\omega)^*}a^{\rm{in}}[-\omega]+\frac{M_2^*}{D(\omega)^*}a^{\rm{in}}[2\omega_a-\omega]\bigg\}\bigg\{\frac{M_1(\omega')}{D(\omega')}a^{\rm{in}}[\omega']+\frac{M_2}{D(\omega')}a^{\rm{in}}[\omega'-2\omega_a]\bigg\}\bigg\rangle.\nonumber
\end{equation}
\end{widetext}
Here, $M_1(\omega)=|\chi _{a}^{-1}\left( \omega _{S}\right)| ^{2}+\rho
_{\rm{aa}}^{2}$, $M_2=-2\rho _{\rm{aa}}e^{-i\theta }$ and $D(\omega)=\chi _{a}^{-2}\left( \omega _{S}\right)-\rho_{\rm{aa}}^{2}$.
For a coherent tone incident on resonance on the $a$-mode, from Eq. \eqref{photflux1}, 
\begin{align}
\label{ain1}
\langle a^{\rm{in}}[\omega]a^{\rm{in}}[\omega']\rangle &= \delta(\omega+\omega')\big\{\theta(\omega)[1+2\pi P_a^{\rm{in}}\delta(\omega-\omega_a)]\nonumber\\&\quad +\theta(-\omega)2\pi P_a^{\rm{in}}\delta(\omega+\omega_a)\big\},
\end{align}
where we have neglected the thermal photon population. Using Eq. \eqref{pout1} and keeping only non-vanishing terms, 
\begin{widetext}
\begin{align}
\label{pout2}
P_a^{\rm{out}}&=\frac{1}{2\pi}\int_0^\infty d\omega\int_0^\infty d\omega'  e^{i(\omega-\omega')t}\delta(\omega-\omega')\bigg\{\frac{M_1(\omega)M_1(\omega')}{D(\omega)^*D(\omega')}2\pi P_a^{\rm{in}}\delta(\omega-\omega_a)\nonumber\\&\quad+\frac{|M_2|^2}{D(\omega)^*D(\omega')}\theta(2\omega_a-\omega)[1+2\pi P_a^{\rm{in}}\delta(\omega-\omega_a)]\bigg\}
\end{align}
\end{widetext}
where we have used that $\omega_a\gg \kappa_a$. Simplifying the above equation, we get 
\begin{equation}
P_a^{\rm{out}} =(2G-1)P_a^{\rm{in}} + \frac{1}{2\pi}\int_0^{2\omega_a}d\omega\bigg|\frac{M_2}{D(\omega)}\bigg|^2,
\end{equation}
which leads to 
\begin{equation}
\label{eqpout-1}
P_a^{\rm{out}}=(2G-1)P_a^{\rm{in}} +\frac{\kappa_a}{\sqrt{G}}(G-1)\frac{1+\rho_{\rm{aa}}^2}{8},
\end{equation}
where 
\begin{equation}
G = \bigg(\frac{1+\rho_{\rm{aa}}^2}{1-\rho_{\rm{aa}}^2}\bigg)^2.
\end{equation}
In performing the integral, we have again used $\omega_a\gg\kappa_a$. In Eq. \eqref{eqpout-1}, the first term is the amplified classical power. The second term in the output power arises from vacuum fluctuations, amplified by a factor $G-1$, integrated over the effective bandwidth $\sim\kappa_a/\sqrt{G}$. 
\subsection{Non-degenerate case}
The output power for the non-degenerate case proceeds analogously and in the case when only the signal mode is driven, one arrives at
\begin{align}
\label{eqpout-2}
P_a^{\rm{out}} &=GP_a^{\rm{in}} +\frac{\kappa_a}{\sqrt{G}}(G-1)\frac{1+\rho_{\rm{ab}}^2}{8},
\end{align}
where 
\begin{equation}
G = \bigg(\frac{1+\rho_{\rm{ab}}^2}{1-\rho_{\rm{ab}}^2}\bigg)^2.
\end{equation}

\section{Dynamic range results for the degenerate parametric amplifier}
\label{dyn_range_deg_paramp}
In this section, we present the results of the calculations of the gain, output power and shift of threshold for the parametric oscillation in the degenerate parametric amplifier. The behaviors of these quantities are completely analogous to the non-degenerate case presented in the main text. 

\begin{figure}[H]
\centering
\includegraphics[width=0.5\textwidth]{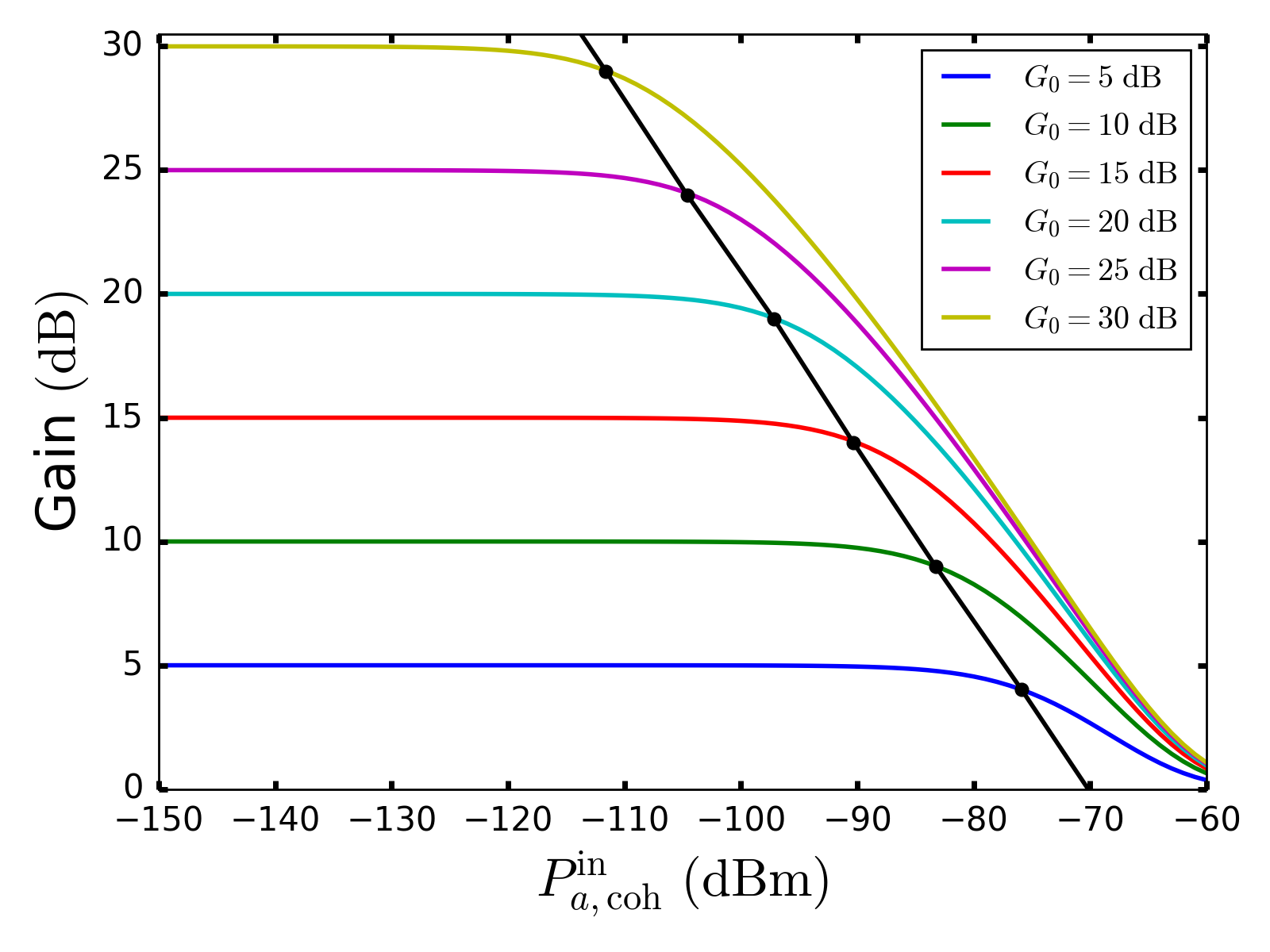}
\caption{\label{jba_gain} Gain of degenerate parametric amplifier as a function coherent incident signal power. The different solid lines correspond to un-depleted gain of $5$ to $30$ dB, in steps of $5$ dB. For realistic system parameters, we have chosen $\omega_a/2\pi = 10$ GHz, $\omega_c/2\pi = 20$ GHz, $\kappa_a/2\pi = 100$ MHz, $\kappa_c/2\pi = 600$ MHz and $g_2/2\pi=0.1$ MHz. The black dots on each curve correspond to the 1dB compression point, where the gain of the amplifier drops by 1dB. These dots lie on a straight line (the black line in the figure), whose slope in the given plot is $\sim -0.7$. As in the non-degenerate case, the asymptotic value of the slope in the limit of high gain is $-2/3$ (see Ref. \onlinecite{Abdo_Devoret_2013_b}). }
\end{figure}

\begin{figure}[H]
\centering
\includegraphics[width=0.5\textwidth]{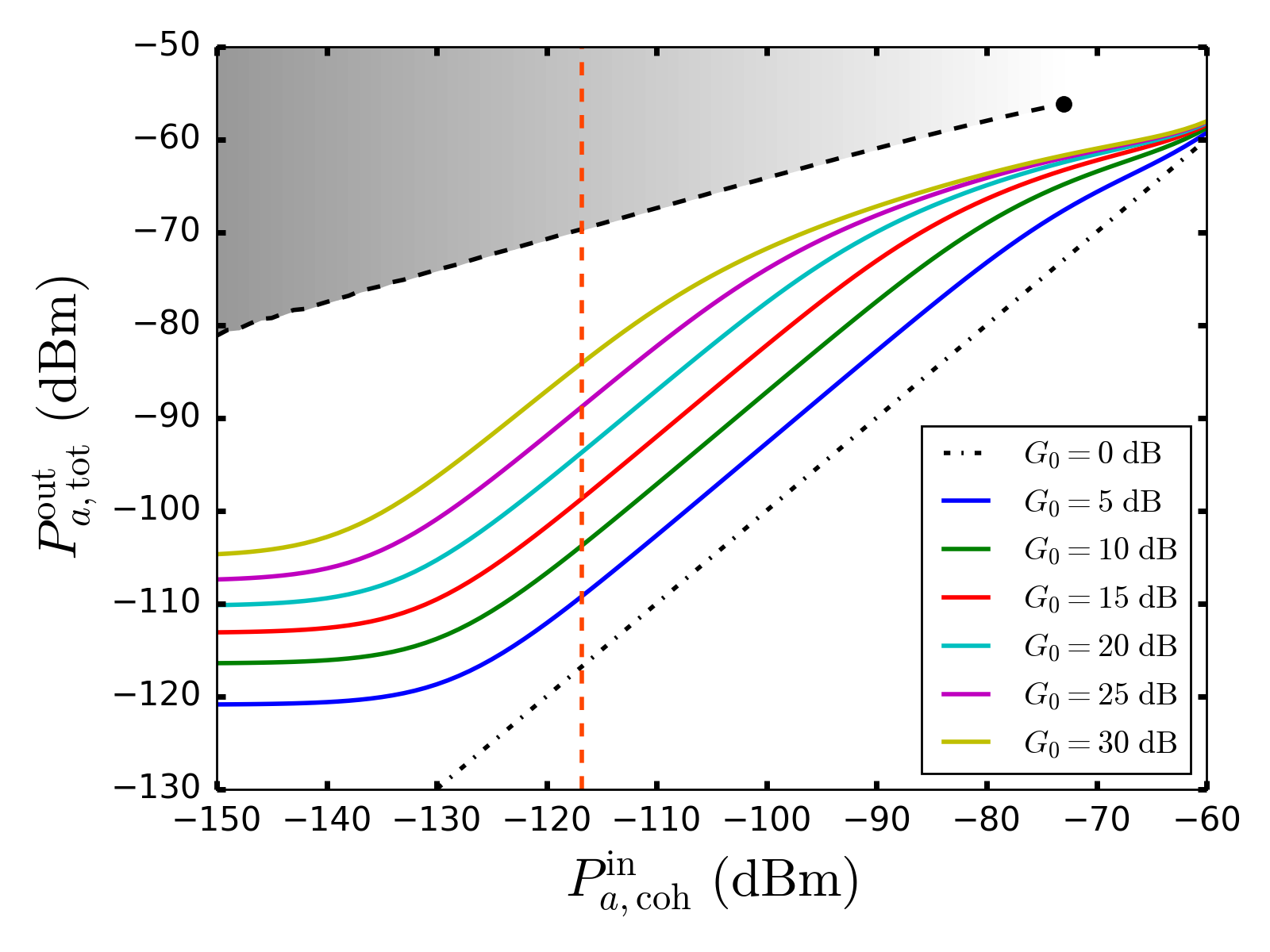}
\caption{\label{jba_pout} Total output signal power as a function of coherent incident signal power for the degenerate paramp.  The different solid lines correspond to un-depleted gain of $5$ to $30$ dB, in steps of $5$ dB. The system parameters are chosen as in Fig. \ref{jba_gain}. As the  incident coherent signal power goes to zero, the output power tends to a constant value corresponding to amplified vacuum fluctuations. 
The black dot-dashed line correspond to when the pump tone is switched off. The black dashed line corresponds to the maximum output signal power, before the onset of spontaneous oscillation and for the device to function as an amplifier, one needs to operate below the corresponding pump powers (see Fig. \ref{jba_thresh} for an estimate).  Sufficient increase of the incident signal power removes the spontaneous oscillation, indicated by the black circle. In the shaded region, the system shows parametric oscillation. The gray color gradient schematically indicates the difference between the two possible classical amplitudes of the output signal in the region of parametric oscillation. This difference goes to zero when the incident power is large enough for the system to stop showing parametric oscillation. The vertical orange line corresponds to the power of half a photon of noise incident on the signal port.}
\end{figure}

\begin{figure}[H]
\centering
\includegraphics[width=0.5\textwidth]{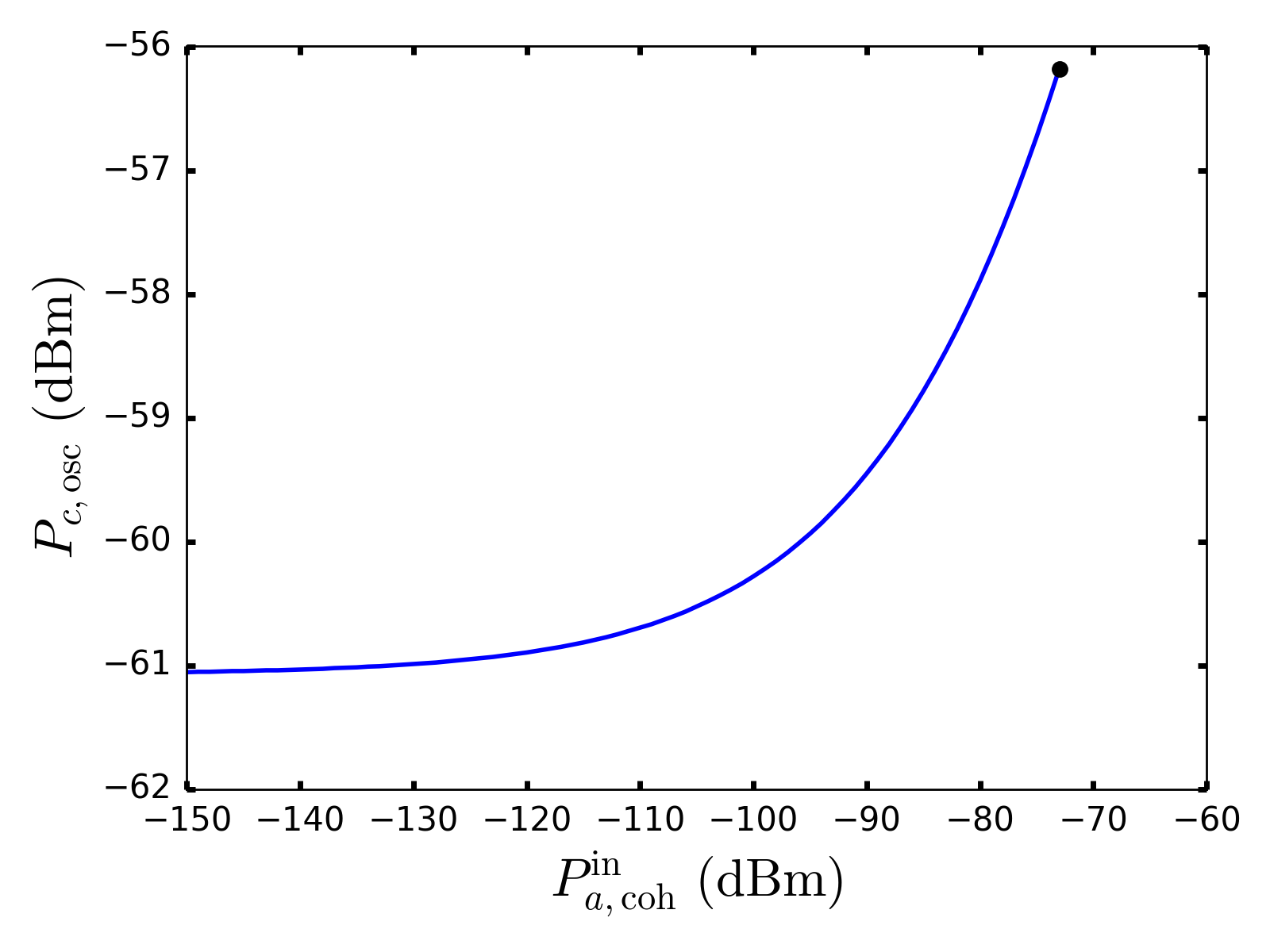}
\caption{\label{jba_thresh} Shift of threshold of spontaneous oscillation upon increase of coherent signal power. The system parameters are chosen as in Fig. \ref{jba_gain}. }
\end{figure}

Fig. \ref{jba_gain} shows the resultant gain of the device as a function of coherent incident signal power ($P_{\rm{a, coh}}^{\rm{in}}$). The different curves correspond to un-depleted gain of $5$ to $30$ dB, in steps of $5$ dB. 
In Fig. \ref{jba_pout}, the total output signal power is plotted as a function of $P_{\rm{a, coh}}^{\rm{in}}$. The solid curves denote the output power for un-depleted gain of $5$ to $30$ dB, in steps of $5$ dB. As the incident coherent signal power goes to zero, the output power saturates and corresponds to amplified vacuum fluctuations incident on the signal port. The black dot-dashed line correspond to zero un-depleted gain when the pump tone is switched off. Finally, the dashed black line correspond the maximum output power that the device can produce before the onset of spontaneous parametric oscillation. 
As incident signal power is increased, the pump power needed for onset of oscillation, $i.e.$ the threshold pump power, increases. This is shown in Fig. \ref{jba_thresh}. For sufficiently high incident signal power, the system ceases to exhibit parametric oscillation (denoted by the black circle in Figs. \ref{jba_pout}, \ref{jba_thresh}). 

\bibliography{library_1}

\end{document}